\documentclass[a4paper,11pt]{article}
\pdfoutput=1
\usepackage{jcappub}

\usepackage{subfigure,float,epstopdf,hyperref,multirow}

\title{Reconstruction of a direction-dependent primordial power spectrum from Planck CMB data}

\author[a]{Amel Durakovic,}
\author[b]{Paul Hunt,}
\author[c,d,e,f]{Suvodip Mukherjee,}
\author[a,b]{Subir Sarkar,}
\author[f]{Tarun Souradeep}
\emailAdd{amel@nbi.dk}
\emailAdd{smukherjee@flatironinstitute.org}
\emailAdd{s.sarkar@physics.ox.ac.uk}
\emailAdd{tarun@iucaa.in}

\affiliation[a]{Niels Bohr Institute, Blegdamsvej 17, DK-2100 Copenhagen, Denmark}
\affiliation[b]{Rudolf Peierls Centre for Theoretical Physics, 1 Keble Road, Oxford OX1 3NP, UK}
\affiliation[c]{Center for Computational Astrophysics, Flatiron Institute, 162 5th Avenue, New York, NY 10010, USA}
\affiliation[d]{Institut d'astrophysique de Paris, 98 bis Bld Arago, 75014 Paris, France}
\affiliation[e]{Institut Lagrange de Paris, Sorbonne Universit\'es, 98 bis Bld Arago, 75014 Paris, France}
\affiliation[f]{Inter-University Centre for Astronomy and Astrophysics, Post Bag 4, Ganeshkind, Pune 411007, India}

\abstract{We consider the possibility that the primordial curvature perturbation is direction-dependent. To first order this is parameterised by a quadrupolar modulation of the power spectrum and results in statistical anisotropy of the CMB, which can be quantified using `bipolar spherical harmonics'. We compute these for the Planck DR2--2015 SMICA  map and estimate the noise covariance from Planck Full Focal Plane 9 simulations. A constant quadrupolar modulation is detected with $2.2 \sigma$ significance, dropping to $2\sigma$ when the primordial power is assumed to scale with wave number $k$ as a power law. Going beyond previous work we now allow the spectrum to have arbitrary scale-dependence. Our non-parametric reconstruction then suggests several spectral features, the most prominent at $k \sim 0.006~\mathrm{Mpc}^{-1}$. When a constant quadrupolar modulation is fitted to data in the range $0.005 \leq k/\mathrm{Mpc}^{-1} \leq 0.008$, its preferred directions are found to be related to the cosmic hemispherical asymmetry and the CMB dipole. To determine the significance we apply two test statistics to our reconstructions of the quadrupolar modulation from data, against reconstructions of realisations of noise only. With a test statistic sensitive only to the amplitude of the modulation, the reconstructions from the multipole range $30 \leq \ell \leq 1200$ are unusual with $2.1\sigma$ significance. With the second test statistic, sensitive also to the direction, the significance rises to $6.9\sigma$. Our approach is easily generalised to include other data sets such as polarisation, large-scale structure and forthcoming 21-cm line observations which will enable these anomalies to be investigated further.} 


\begin{document}
\maketitle
\flushbottom

\section{Introduction}
The observed temperature fluctuations of the cosmic microwave background (CMB) radiation are believed to be due to curvature perturbations generated at an early stage in the evolution of the universe. The most discussed mechanism for this is the quantum fluctuations of a scalar field, the vacuum energy of which drives a period of accelerated expansion known as inflation. In the simplest scenario where the inflaton field evolves slowly down an almost flat potential, the (dimensionless) primordial power spectrum, which is the variance of the Fourier components of the curvature perturbation, is nearly scale-invariant, with slightly more power on large scales. Provided that the primordial power spectrum is independent of the direction (and the background cosmology is homogeneous and isotropic), the CMB fluctuations will also be statistically isotropic.

We consider here the possibility of a direction-dependent primordial power spectrum resulting in  \emph{statistically anisotropic} CMB fluctuations.  In order for this to occur, some field must single out a particular direction $u^{\mu}$ and break rotational symmetry, and so the agent should either carry a space-time index or have a spatial gradient present in the initial field configuration. The first such model \cite{Ford:1989me} considered a vector field $A^{\mu}$ driving inflation via the effective scalar $\xi = A_{\mu}A^{\mu}$ with a sufficiently flat potential $V(\xi)$. When the vector field singles out a direction, its energy-momentum tensor is no longer isotropic, so its direction-dependence can play a role gravitationally; in this case, inflation is followed by anisotropic expansion. Other work, motivated by some observational CMB anomalies, found that an anisotropy in the initial expansion rate following inflation can cause the fluctuations to inherit the anisotropy \cite{Gumrukcuoglu:2006xj}. A concrete realisation of the anisotropy, with a vector field responsible, was subsequently proposed \cite{Ackerman:2007nb}. Such vector field models were, however, found to be unstable \cite{Himmetoglu:2008zp,Himmetoglu:2008hx}. It is important that the anisotropic field should \emph{not} contribute significantly to the energy density, otherwise the expansion itself will be anisotropic to a degree in conflict with observations. Another possibility is that the vector field couples to the inflaton and thereby transfers the anisotropy, as in models with a coupling $f(\phi)^2 F_{\mu \nu} F^{\mu \nu}$ where $f(\phi)$ is a function of the inflaton field and $F_{\mu \nu} = \partial_{\mu} A_{\nu} - \partial_{\nu} A_{\mu}$ is the field strength of the vector field $A_\mu$ \cite{Watanabe:2009ct,Bartolo:2012sd}.


All such models produce an angular modulation \cite{Pullen:2007tu} such that the full (dimensionless) primordial power spectrum is:
\begin{align}
\mathcal{P} ( \mathbf{k} ) &= \mathcal{P} \left(k\right) + \sqrt{4 \pi} \sum_{L M} g_{L M}(k) Y_{L M}( \mathbf{\hat{k}}).
\label{eq:glm}
\end{align}
We adopt a data-driven approach, not committing to any specific theoretical model, and focus on the quadrupole modulation:
\begin{align}
\mathcal{P} ( \mathbf{k} ) &= \mathcal{P} \left(k\right) + \sqrt{4 \pi} \sum_{M=-2}^{2} g_{2 M}(k) Y_{2 M}( \mathbf{\hat{k}}), 
\label{eq:split}
\end{align}
where $\mathcal{P}(k)$ is the usual isotropic power spectrum, $Y_{2 M}(\mathbf{\hat{k}})$ are spherical harmonics that carry the direction-dependence and $g_{2 M}(k)$ are the associated harmonic space coefficients which may vary \emph{freely} with the wave number $k$ and be either positive or negative. As the coefficients $g_{2 M}(k)$ determine the modulation, these are our main focus for reconstruction from data. 

Note that we have chosen the split (\ref{eq:split}) in order to remain agnostic about the isotropic part of the primordial power spectrum, unlike the alternative parameterisation
\begin{align}
\mathcal{P}(\mathbf{k})  &= \mathcal{P}(k) (1 + g(k) (\mathbf{\hat{k}} \cdot \mathbf{\hat{n}})^2),
\label{eq:ca}
\end{align}
which reconstructs the shape of the quadrupole modulation relative to the isotropic primordial power spectrum. By contrast we reconstruct the \emph{absolute} quadrupole modulation. Also, eq.\eqref{eq:split} refers to the most general quadrupole modulation while the form of eq.\eqref{eq:ca} is a special case where the two directions that describe a general quadrupole coincide. Relations between $g_{2M}$ and the two directions of a quadrupole are derived in Appendix~\ref{sec:genquadr}.

While $g(k)$ is scale-invariant (up to logarithmic corrections) in most models, there is no fundamental reason for the quadrupole modulation to be scale-free. For example, in a study of perturbations in a Bianchi I universe which isotropises as inflation proceeds, $g_{2M}(k)$ was found to have strong oscillatory features on large scales \cite{Pitrou:2008gk}.\footnote{A limited duration of inflation was postulated so that the effects of the primordial anisotropic phase do not extend beyond the current horizon.} Therefore we allow arbitrary scale-dependence of $g_{2M}(k)$ and attempt to determine this from the data. 

Symmetry considerations dictate that the quadrupole is the first modulation that can appear \cite{Pullen:2007tu}. The next contribution would be the hexadecapole $L=4$ as only even $L$ are allowed (see \S~\ref{sec:app}). Since the $L=2$ modulation is the first one that can appear and also has some theoretical motivation we will  consider only this case although our formalism readily allows for the inclusion of higher order modulations.

The quadrupole modulation of the primordial power spectrum will manifest itself in the CMB temperature fluctuations which will now be statistically anisotropic. The temperature fluctuations $\Delta T$ are conventionally expanded in spherical harmonics such that
\begin{align}
\Delta T (\theta,\phi) &= \sum_{\ell m} a_{\ell m} Y_{\ell m}(\theta, \phi).
\end{align}
For statistical isotropy, the correlation function is:
\begin{align}
\langle a_{\ell m} a^{\ast}_{\ell' m'} \rangle &= C_\ell \delta_{\ell \ell'} \delta_{m m'},
\end{align}
where $C_\ell$ is the (isotropic) angular power spectrum and the brackets indicate ensemble averages. In the statistically \emph{anisotropic} case, this no longer holds and there are additional terms with `bipolar spherical harmonics' (BipoSH) coefficients \cite{Hajian:2003qq,Hajian:2005jh}. Those associated with a given multipole modulation can readily be calculated (see \S~\ref{sec:app}) and in turn given the BipoSH coefficients of a CMB map, it is possible to reconstruct the direction-dependent primordial power spectrum. Previous attempts by the Planck collaboration to extract this from their CMB maps have either assumed a constant value of the quadrupole modulation \cite{Ade:2015hxq} or a smooth power-law scale dependence \cite{Ade:2015lrj}. In addition to these exercises, we will reconstruct the full spectrum from Planck data, allowing the quadrupole modulation to vary \emph{freely} with wave number. We follow an approach to reconstruction based on `Tikhonov regularisation' that has been demonstrated to work well for isotropic primordial power spectra \cite{Hunt:2013bha,Hunt:2015iua}.

Potential features will be checked against two regions of interest: the hemispherical asymmetry \cite{Eriksen:2003db} and the CMB dipole. The former is a dipolar modulation $A \hat{\mathbf{p}} \cdot \hat{\mathbf{n}}$ of an otherwise statistically isotropic sky $T_\mathrm{iso}(\hat{\mathbf{n}})$ such that the observed sky
\begin{align}
		T(\hat{\mathbf{n}}) = T_{\mathrm{iso}}(\hat{\mathbf{n}}) (1+A \hat{\mathbf{p}} \cdot \hat{\mathbf{n}})
\end{align}
where $A$ is the amplitude of the modulation and $\hat{\mathbf{p}}$ is a preferred direction.

This paper is organised as follows. In Section \ref{sec:app}, the detailed formalism is presented: BipoSH coefficients are defined and related to quadrupole modulations and Tikhonov regularisation is described. In Section \ref{sec:set}, estimates of uncertainties are made using the Planck Full Focal Plane (FFP9) simulations \cite{Ade:2015via} and used to construct a simplified likelihood for binned data which is then tested. The reconstruction is performed on benchmark spectra adopting the estimated uncertainties to check how well it performs. In Section \ref{sec:res}, the data is presented. The best-fit constant and power-law quadrupole modulations are calculated, followed by our main result: the non-parametric reconstructions. We check here if the directions are related to those of the hemispherical asymmetry and the CMB dipole. In Section \ref{sec:statsig}, the statistical significance of possible spectral features is discussed. We summarise in Section \ref{sec:sum}.

\section{Formalism}
\label{sec:app}

The general two-point function of spherical harmonics coefficients can be written
\begin{align}
\langle a_{\ell m} a^{\ast}_{\ell' m'} \rangle &= C_\ell \delta_{\ell \ell'} \delta_{m m'} + \sum_{L M} \left(-1\right)^{m'}C^{L M}_{\ell m \ell' -m'} A'^{L M}_{\ell \ell'},
\end{align}
where $ C^{L M}_{\ell m \ell' -m'} $ are Clebsch-Gordan coefficients and $A'^{L M}_{\ell \ell'}$ are BipoSH coefficients \cite{Hajian:2003qq,Hajian:2005jh}. These are associated with the \emph{bipolar} spherical harmonics which form an orthonormal basis for functions of two directions. This is most evident when considering the temperature correlation function in real space
\begin{align}
C(\mathbf{\hat{n}},\mathbf{\hat{n}}') &= \langle \Delta T(\mathbf{\hat{n}}) \Delta T(\mathbf{\hat{n}}')  \rangle = \sum_{\ell \ell' L M} A'^{LM}_{\ell \ell'} \left\{ Y_\ell(\mathbf{\hat{n}}) \otimes Y_{\ell'}(\mathbf{\hat{n}}') \right\}_{L M} \nonumber\\ &= \sum_{\ell} \frac{2 \ell + 1}{4 \pi} C_\ell P_{\ell}(\mathbf{\hat{n}} \cdot \mathbf{\hat{n}}') + \sum_{\ell \ell' L >0, M} A'^{L M}_{\ell \ell'} \left\{ Y_{\ell}(\mathbf{\hat{n}}) \otimes Y_{\ell'}(\mathbf{\hat{n}}') \right\}_{L M} ,
\end{align}
where,
\begin{align}
\left\{ Y_\ell (\mathbf{\hat{n}}) \otimes Y_{\ell'} \left(\mathbf{\hat{n}}'\right) \right\}_{LM}&= \sum_{m m'} C^{L M}_{ \ell m \ell' -m'} Y_{\ell m}(\mathbf{\hat{n}}) Y_{\ell'm'}(\mathbf{\hat{n}}').
\end{align}
Just as the spherical harmonics coefficients can be calculated from a map by projection onto the basis functions, $a_{\ell m} = \int \mathrm{d} \Omega \, Y^{\ast}_{\ell m}(\mathbf{\hat{n}}) T(\mathbf{\hat{n}})$, the BipoSH coefficients can be similarly computed as
\begin{align}
A'^{LM}_{\ell \ell'} &= \int \mathrm{d}\Omega \int \mathrm{d}\Omega' \, C(\mathbf{\hat{n}},\mathbf{\hat{n}}') \{ Y_{\ell}(\mathbf{\hat{n}}) \otimes Y_{\ell'}(\mathbf{\hat{n}}') \}_{L M}^{\ast}.
\end{align}
This can be written more straightforwardly in terms of the spherical harmonics as
\begin{align}
A'^{LM}_{\ell \ell'} &= \sum_{mm'} \langle a_{\ell m} a^{\ast}_{\ell' m'}\rangle (-1)^{m'} C^{L M}_{\ell m \ell' -m'},
\label{eq:almforbip}
\end{align}
which can have both even and odd parity \cite{Book:2011na}. Only the \emph{even} parity BipoSH coefficients are non-zero for a quadrupolar modulation of the power spectrum and these are related to the even parity BipoSH spectra as: 
\begin{align}
 A'^{LM}_{\ell \ell'} =  \sqrt{\frac{(2 \ell +1)(2\ell'+1) }{2L + 1}} C^{L0}_{\ell 0 \ell' 0} A^{LM}_{\ell \ell'}. \label{eq:convfa}
\end{align}
We adopt this definition in our work, following the WMAP collaboration \cite{Bennett:2010jb}.

The dimensionful primordial power spectrum $P(\mathbf{k})$ is the variance of the curvature perturbation $\mathcal{R}(\mathbf{k})$ such that
\begin{align}
\langle \mathcal{R}(\mathbf{k}) \mathcal{R}(\mathbf{k}') \rangle = (2 \pi)^3 \delta^{(3)}(\mathbf{k}+\mathbf{k}') P(\mathbf{k})
\label{eq:rrcov}
\end{align}
which suffices to describe a Gaussian field and where it is furthermore assumed that different modes $\mathbf{k}$ and $\mathbf{k}'$ are independent.
A reflection of the vectors on both sides of eq.\eqref{eq:rrcov} implies that $P(-\mathbf{k}) = P(\mathbf{k})$. Since the spherical harmonics are inside $P(\mathbf{k})$, only those spherical harmonics which equal themselves upon reflection are admitted, and this holds for the even $L$ only. It is useful to note that since $Y_{LM}^{\ast}(\mathbf{\hat{k}}) = (-1)^{M} Y_{L,-M}(\mathbf{\hat{k}})$, in order for $P(\mathbf{k})$ to be real-valued, it must be the case that $g_{LM}^{\ast}(k) = (-1)^{M} g_{L,-M}(k)$. This means that it is only necessary to reconstruct $g_{LM}$ for non-negative $M$.

The central relation is that between the direction-dependent modulations of the primordial power spectrum and the induced BipoSH coefficients associated with the temperature anisotropies. The coefficients are:
 \begin{align}
 A_{\ell \ell'}^{L M} &= 4 \pi (-i)^{\ell - \ell'} \int_{0}^{\infty} \mathrm{d} \log k \, g_{L M}(k) \Delta_\ell (k) \Delta_{\ell'}(k),
 \label{eq:relate}
 \end{align}
 where $\Delta_\ell (k)$ are temperature transfer functions relating curvature perturbations $\mathcal{R}(k)$ to multipoles of the temperature perturbations. This relation is derived in Appendix~\ref{sec:relate}.

 It is useful to introduce another variable $d$ that counts the distance from $\ell$ to $\ell'\equiv\ell+d$. It will only be necessary to compute for $d=0$ and $d=2$ since the BipoSH coefficient $A^{LM}_{\ell \ell'}$ appears with the Clebsch-Gordan coefficient $C^{LM}_{\ell m \ell' -m'}$ which for $L=2$ is non-zero only when $\ell$ and $\ell'$ are equal or differ by two. It will not be necessary to calculate for $d=-2$ as eq.\eqref{eq:relate} is symmetric in $\ell$ and $\ell'$ so that $A^{2 M}_{\ell~\ell-2}$ is equal to $A^{2 M}_{\ell-2~\ell}$. Similarly the hexadecapole would require the calculation for $d=0,2,4$.

\begin{figure}[ht]
\centering
\subfigure{\includegraphics[scale=0.48]{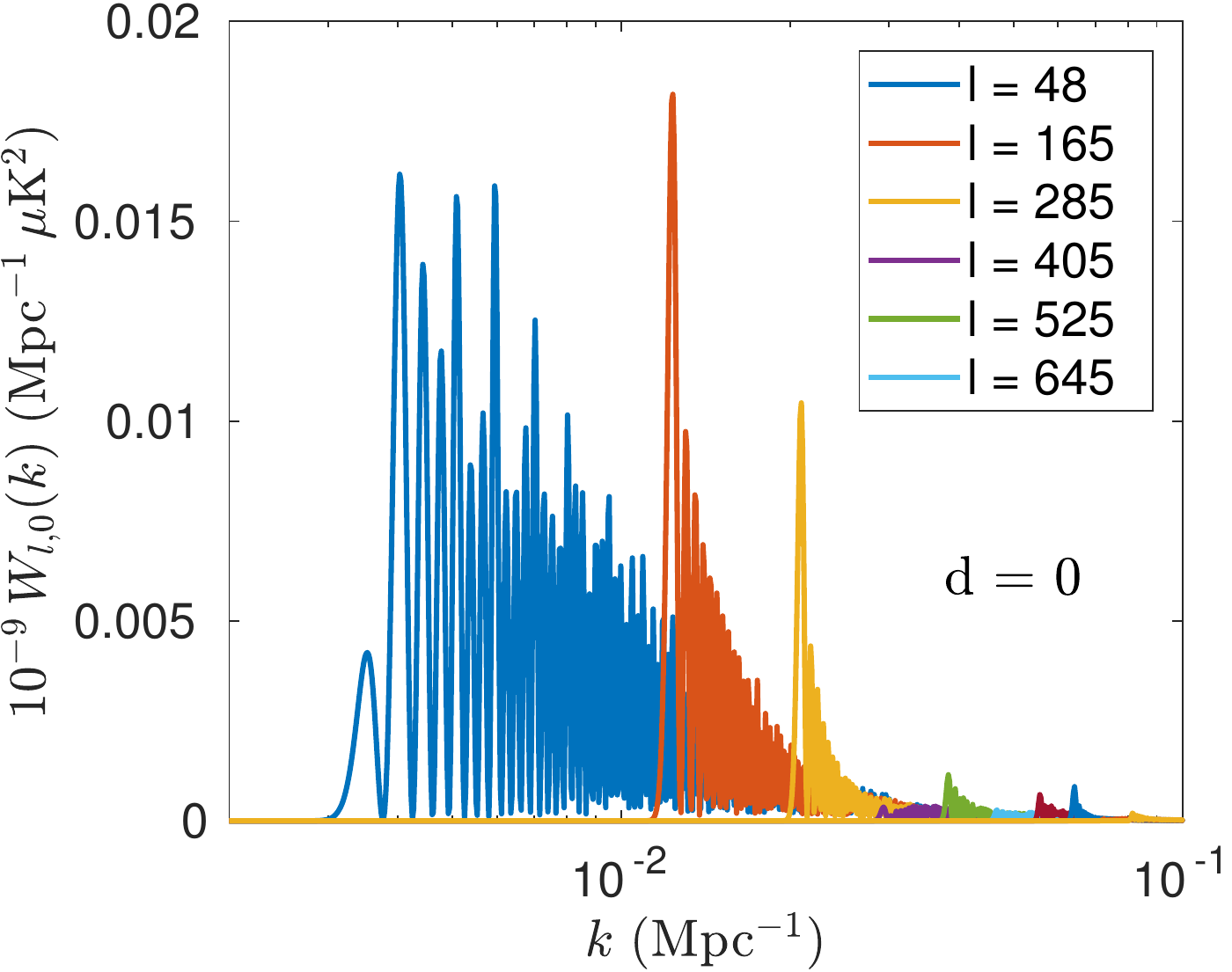}
\label{fig:transa}
}
\subfigure{\includegraphics[scale=0.48]{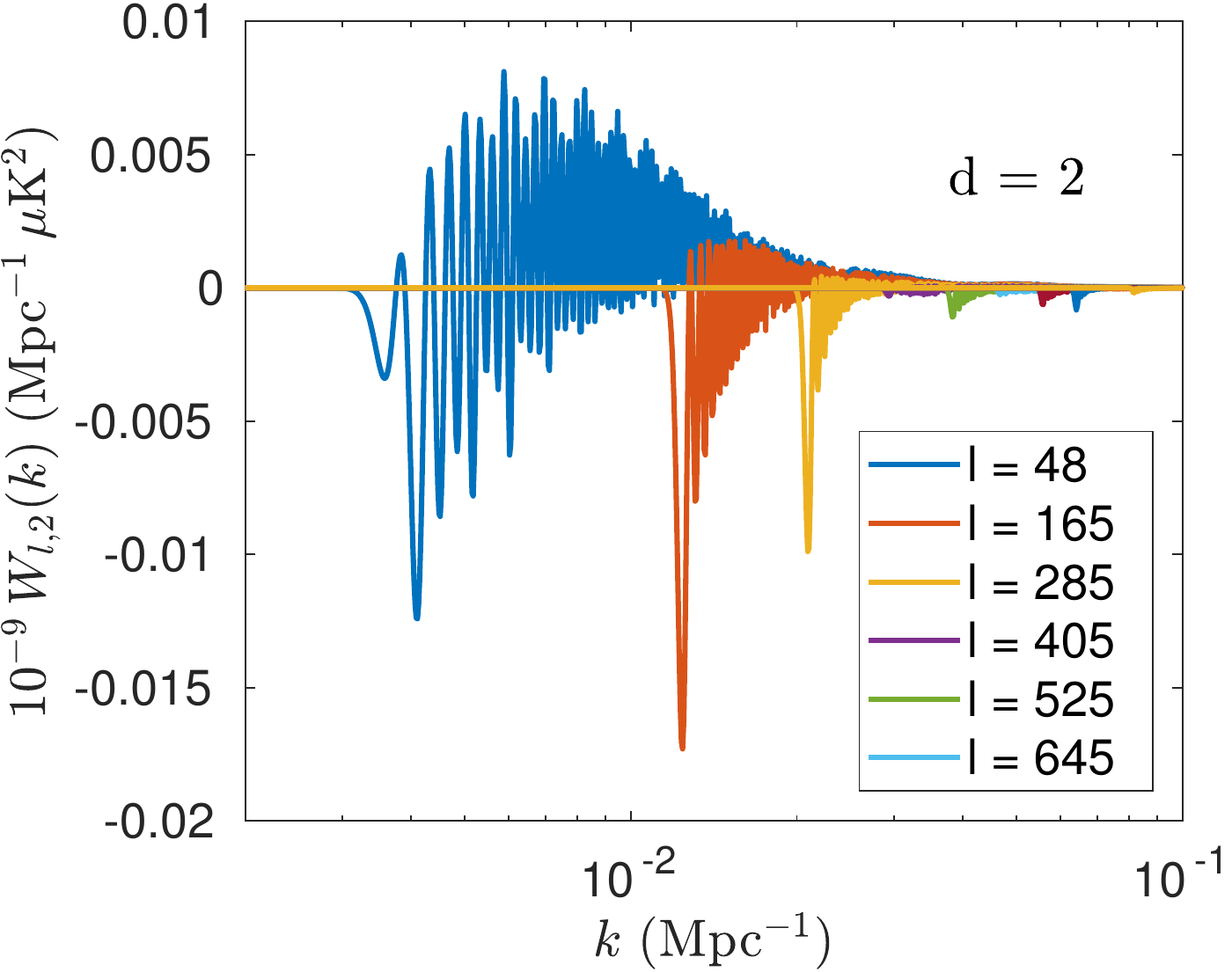}
\label{fig:transb}
}
\caption{The transfer functions $W_{p k}$ relating $g^{LM}_k$ to $A^{L M}_{\ell d}$ for $d = 0$ (left panel) and $d = 2$ (right panel) and representative choices of the multipole $\ell$. As a rule of thumb, the maxima are near $k = \ell \eta_0^{-1} \approx \ell/14000 \,\mathrm{ Mpc^{-1}}$, where $\eta_0$ is the conformal time today.}
\label{fig:transf}
\end{figure}

 In order to make the problem of reconstruction amenable to numerical analysis, an evenly spaced grid in $\log k$ space at positions $k_i$ is introduced and made sufficiently fine that its discretisation does not matter. The modulation $g_{LM}(k)$ is then written in terms of functions $\phi_j(k)$ that are equal to unity in the space between the grid points $k_j$ and $k_{j+1}$ and zero otherwise, such that \begin{align} g_{LM}(k)&= \sum_i g^{LM}_i \phi_i(k), \end{align} where $g^{LM}_i$ are now coefficients. Upon introducing $p$ as a collective variable for $\ell$ and $d$, eq.\eqref{eq:relate} can now be written as a matrix equation \begin{align} A^{L M}_p &= \sum_{i} W_{p j} g^{LM}_j \Leftrightarrow \mathbf{A}^{LM} = \mathbf{W} \mathbf{g}^{LM}, \label{eq:meq} \end{align} where $\mathbf{A}^{LM}$ and $\mathbf{g}^{LM}$ are vectors and the matrix $\mathbf{W}$ is given by
 \begin{align}
W_{p j} &= 4 \pi i^{d} \int_{0}^{\infty} \mathrm{d} \log k\, \phi_j (k) \Delta_\ell (k) \Delta_{\ell+d}(k).
 \label{eq:wmat}
\end{align}
Furthermore, we combine all components of $\mathbf{g}^{LM}$, real and imaginary, into a single vector $\mathbf{g}$, with imaginary parts of the components $M=1,2$ following the real parts of $M=0,1,2$:
\begin{align}
		\mathbf{g} = \begin{pmatrix}
		  \mathbf{g}^{20} \\ \mathrm{Re} \, \mathbf{g}^{21} \\ \mathrm{Re}\, \mathbf{g}^{22} \\ \mathrm{Im} \, \mathbf{g}^{21} \\ \mathrm{Im} \, \mathbf{g}^{22}
		\end{pmatrix}, \quad
		\mathbf{A} = \begin{pmatrix}
		  \mathbf{A}^{20} \\ \mathrm{Re} \, \mathbf{A}^{21} \\ \mathrm{Re}\, \mathbf{A}^{22} \\ \mathrm{Im} \, \mathbf{A}^{21} \\ \mathrm{Im} \, \mathbf{A}^{22}
		\end{pmatrix},
		\mathbf{W}' = \mathbf{I}_{5\times 5} \otimes \mathbf{W} = \begin{pmatrix} \mathbf{W} & & & & \\  & \mathbf{W} & & & \\  & & \mathbf{W} & & \\  & & & \mathbf{W} & \\ & & & & \mathbf{W} \end{pmatrix}
\end{align}
where we have also grouped $\mathbf{A}^{LM}$ into $\mathbf{A}$ and we have made a block matrix out of $\mathbf{W}$ so that it still holds that
\begin{align}
		\mathbf{A} = \mathbf{W}' \mathbf{g}.
\end{align}
The matrix $\mathbf{W}$, which relates a given modulation $\mathbf{g}^{LM}$ to the BipoSH coefficients $\mathbf{A}^{LM}$, has a non-trivial kernel, so the matrix equation \eqref{eq:meq} \emph{cannot} be inverted to obtain the required $\mathbf{g}^{LM}$. Physically, this is due to projection effects wherein one mode contributes to many BipoSH coefficients so it is impossible to invert the relation without additional assumptions. This is illustrated in Fig.~\ref{fig:transf} where different rows of $\mathbf{W}$ are plotted, relating the contribution of a mode $k$ to the BipoSH coefficient associated with one of the curves.

In order to proceed further, this deconvolution problem needs regularisation. We opt to choose that solution $\hat{\mathbf{g}}$ amongst the multitude of otherwise equally likely solutions which has the least `roughness' $R(\mathbf{g})$. This is defined, in the continuous case, as
\begin{align}
R ({g}) =\sum_{LM} \left( \int \mathrm{d}k \left( \frac{\mathrm{d} \, \mathrm{Re} \, g_{LM} (k)} {\mathrm{d} \log k} \right)^2 + \int \mathrm{d}k \left( \frac{\mathrm{d} \, \mathrm{Im} \, g_{LM} (k)} {\mathrm{d} \log k} \right)^2 \right),
\end{align}
while, in the discrete case, it is 
\begin{align}
R (\mathbf{g}) &= \mathbf{g}^{T} \mathbf{\Gamma} \mathbf{g},
\end{align}
where
\begin{align}
\mathbf{\Gamma} = \mathbf{I}_{5\times 5} \otimes \mathbf{\tilde{\Gamma}} = \begin{pmatrix} \mathbf{\tilde{\Gamma}} & & & & \\  & \mathbf{\tilde{\Gamma}} & & & \\  & & \mathbf{\tilde{\Gamma}} & & \\  & & & \mathbf{\tilde{\Gamma}} & \\ & & & & \mathbf{\tilde{\Gamma}} \end{pmatrix}
 \textrm{ and }  \mathbf{\tilde{\Gamma}} &= \mathbf{D}^{T} \mathbf{D} =
\begin{pmatrix}
  1 & -1  \\
  -1 & 2 & -1 \\
 & \ddots & \ddots & \ddots\\
 &  & -1 & 2 & -1 \\
 &  &   & -1 & 1
\end{pmatrix}.
\end{align}
Here $\mathbf{D}$ is the first-order derivative operator and these are the matrix elements of $\tilde{\mathbf{\Gamma}}$ when the grid is evenly spaced in $\log k$. Note that the roughness penalty function does not couple modulations $g_{LM}(k)$ with different $L,M$ values.

The degree to which the solution should be penalised for this roughness is quantified by the \emph{regularisation parameter} $\lambda$.
The solution will be that which maximises the likelihood (or, equivalently, minimises the negative log likelihood) subject to the constraint that the roughness has the specific value determined by $\lambda$ through:
\begin{align}
\hat{\mathbf{g}} &=  \min_{\mathbf{g}} \, \{  -2 \log \mathcal{L}(\mathbf{g},\tilde{\mathbf{A}} ) + \lambda \mathbf{g}^{T} \mathbf{\Gamma} \mathbf{g} \, \}.
\label{eq:soln}
\end{align}
Here $\mathcal{L}$ denotes the likelihood and $\tilde{\mathbf{A}}$ denotes the observed values extracted from the CMB map. The likelihood has been multiplied by $-2$ to match the standard $\chi^2$ statistic for the case of a Gaussian likelihood.

This approach to deconvolution known as `Tikhonov regularisation' has been widely used in image analysis and can be interpreted as Bayesian inference. In the case with a likelihood $\mathcal{L}$, the solution eq.\eqref{eq:soln} can be seen as the one that maximises the posterior probability $P(\mathbf{g}| \tilde{\mathbf{A}} )$ when the prior probability is given by $P(\mathbf{g} ) = \exp (- \lambda R(\mathbf{g})/2)$. If the likelihood is Gaussian and the roughness function a quadratic form, then the posterior probability distribution too will be Gaussian. The inverse covariance matrix of the posterior gives the Hessian which in this case is
\begin{align}
H_{i j}(\lambda) = -\frac{\partial^2}{\partial g_i \partial g_j}\log \mathcal{L} + \frac{\lambda}{2} \Gamma_{ij},
\end{align}
so its inverse provides the covariance matrix of the posterior, from which credible intervals may be derived.

Note that the regularisation parameter $\lambda$ appears in the prior and can therefore only come from knowledge about the roughness of $\mathbf{g}$. As this is lacking \emph{a priori}, we simply consider a range of values for $\lambda$ and report the results. If the likelihood is Gaussian and the roughness function has a quadratic form then the reconstruction can be performed analytically without resorting to numerical minimisation schemes (e.g. `BFGS' which would be necessary for more complicated likelihoods).

Consider the penalised log-likelihood
\begin{align}
Q(\lambda) &= ( \mathbf{W}'  \mathbf{g} - \tilde{\mathbf{A}} )^{T} {\mathbf{\Sigma}}^{-1} ( \mathbf{W}'  \mathbf{g} - \tilde{\mathbf{A}} ) + \lambda {\mathbf{g}}^{T} \mathbf{\Gamma} \mathbf{g}\textrm{.}
\end{align}
The penalised likelihood is maximised for
\begin{align}
\hat{\mathbf{g}} &= ( \mathbf{W}'^{T} {\mathbf{\Sigma}}^{-1} \mathbf{W}' + \lambda \mathbf{\Gamma} )^{-1} \mathbf{W}'^{T} {\mathbf{\Sigma}}^{-1} \tilde{\mathbf{A}},
\label{eq:recon}
\end{align}
and $\mathbf{H} = \mathbf{W}'^{T} \, {\mathbf{\Sigma}}^{-1} \mathbf{W}' + \lambda \mathbf{\Gamma} $ is the Hessian.

It is also useful to consider the frequentist covariance matrix. It is simply the error in $\hat{\mathbf{g}}$ that is induced by the error in the data $\mathbf{\tilde{A}}$. Since $\hat{\mathbf{g}}$ and $\mathbf{\tilde{A}}$, are related by a linear transformation \eqref{eq:recon}, the (frequentist) covariance matrix of $\mathbf{g}$ \emph{viz.} $\mathbf{\Sigma}_{\mathrm{F}}$ is related to the data covariance matrix $\mathbf{\Sigma}$ by a similarity transformation
\begin{equation}
\mathbf{\Sigma}_\mathrm{F} = \mathbf{M} \mathbf{\Sigma} \mathbf{M}^{T},
\label{eq:freqcov}
\end{equation}
where $\mathbf{M} = ( \mathbf{W}'^{T} {\mathbf{\Sigma}}^{-1} \mathbf{W}' + \lambda \mathbf{\Gamma} )^{-1} \mathbf{W}'^{T} {\mathbf{\Sigma}}^{-1}$. The confidence intervals (frequentist) given by the square root of the diagonal values of $\mathbf{\Sigma}_{\mathrm{F}}$ will in general be \emph{different} from the (Bayesian) credible intervals given by the square root of the diagonal values of $\mathbf{H}^{-1}$. To illutrate this we will provide both in what follows.

\section{Setup and tests} \label{sec:set}
\subsection{Data and covariance matrix setup}
The transfer functions in eq.\eqref{eq:wmat} were calculated using a customised version of \verb+CAMB+ \cite{Lewis:2002ah} adopting the Planck best-fit cosmological model. A total of 2316 bins, evenly distributed in $\log k$ space, were used to approximate $g_{2 M}(k)$ in the range from $k_{\mathrm{min}} \sim 7 \times 10^{-6}\,\mathrm{Mpc}^{-1}$ to $k_{\mathrm{max}} \sim 0.4 \,\mathrm{Mpc}^{-1}$. 

\begin{figure}[ht]
\centering
\includegraphics[width=1\textwidth]{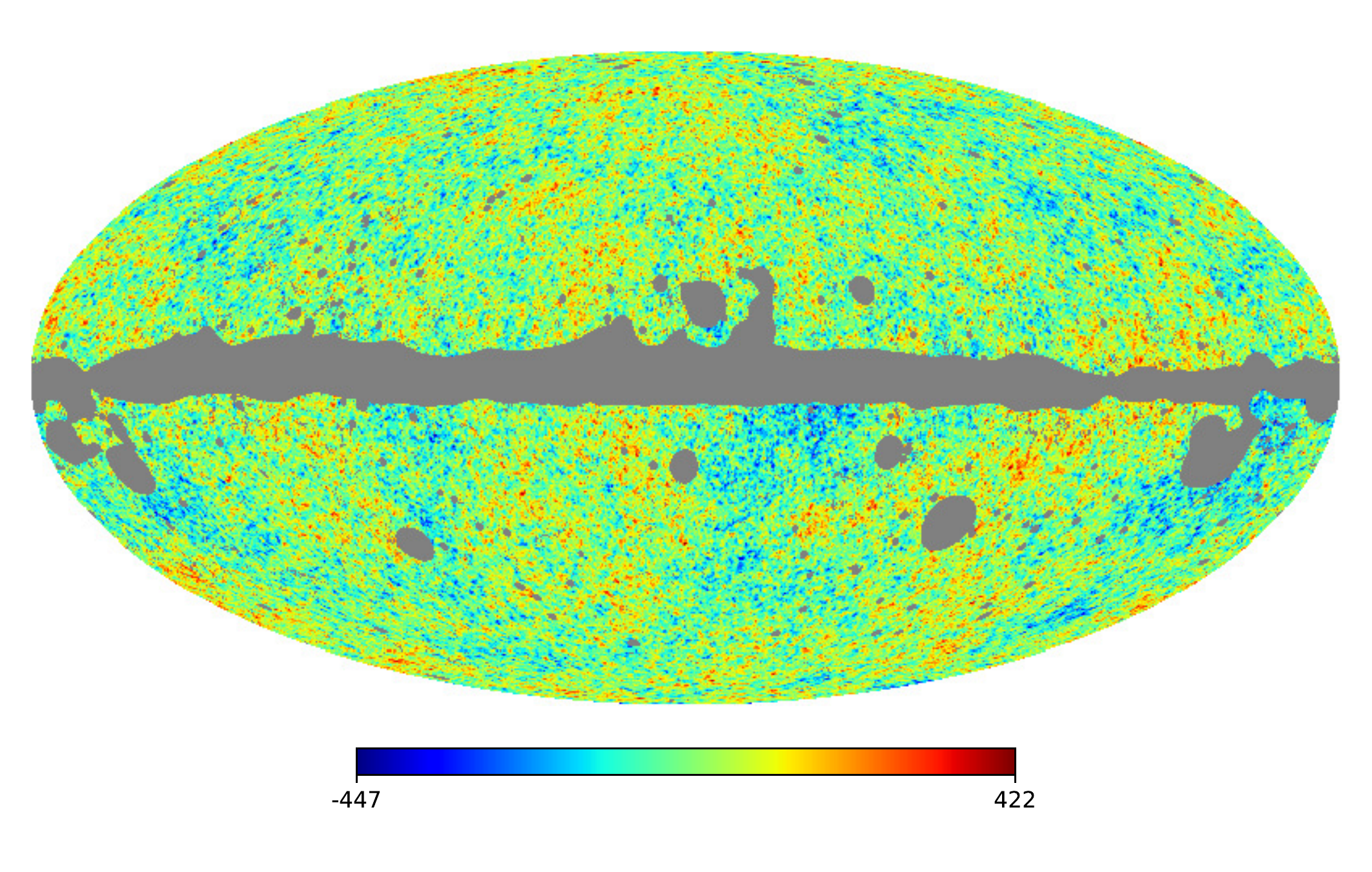}
\caption{The masked PR2--2015 SMICA temperature map in units of $\mu$K.}
\label{fig:maskedsmica}
\end{figure}

Since statistical anisotropies can also be induced by the instrument itself, due to e.g. non-circular beam effects, pointing errors \emph{etc.}, or by astrophysical foregrounds (see e.g. refs.\cite{Das:2014awa,Kim:2013gka}), we used the most realistic simulations of the instrument and sky, \emph{viz.} 999 Planck Full Focal Plane (FFP9) simulations \cite{Ade:2015via} to estimate the uncertainties on the extracted BipoSH coefficients. The simulated sky maps were masked using the Planck SMICA mask (COM\_Mask\_CMB-confidence-Tmask-IQU-smica\_1024\_R2.02-full with $N_\mathrm{side}=1024$) \cite{pla} shown in grey in Fig.~\ref{fig:maskedsmica}.

From the spherical harmonics of the cut sky map $b_{\ell m}$, using eq.\eqref{eq:almforbip} and eq.\eqref{eq:convfa}, we then compute the BipoSH coefficients 
in the following way:
\begin{align}
 		\hat{A}^{L M}_{\ell \ell'} = \frac{1}{C^{L 0}_{\ell 0 \ell' 0} } \sqrt{\frac{2L+1 }{(2\ell+1)(2 \ell'+1)}} \sum_{mm'} \hat{b}_{\ell m} \hat{b}^{\ast}_{\ell' m'} (-1)^{m'} C^{L M}_{\ell m \ell' -m'}. \label{eq:cutsky}
\end{align} 
Lost power was accounted for by multiplying the coefficients by
$1/f_\mathrm{sky} \sim 1.176$, where $f_\mathrm{sky}$ is the available sky fraction
$\sim 85 \%$.

The BipoSH coefficients $A^{2 M}_{\ell~\ell+d}$ were binned with bin size $\Delta \ell = 30$ and weighted by $\ell (\ell+1)$, yielding averaged BipoSH coefficients $A^{2M}_{\ell_b~\ell_b + d}$ in effective multipoles $\ell_b$. Specifically, a matrix $\mathbf{B}$ was defined with elements
 \begin{align}
 B_{\ell_b \ell} &= \frac{\ell (\ell + 1)}{\sum_{\ell \in b} \ell (\ell + 1)},
 \end{align}
where $b$ is the set of multipoles in the bin. The effective multipoles are then $\boldsymbol \ell_b = \mathbf{B} \boldsymbol \ell$ with effective BipoSH coefficients $A^{2 M}_{\ell_b~\ell_b + d} = \sum_{\ell} B_{\ell_b \ell} A^{2 M}_{\ell~\ell+d} $.

We remind the reader of the collective variable $q$ which compresses the full index structure that would otherwise comprise of $M$, $\ell_b$, $d$ and whether the real or imaginary part of the BipoSH coefficient is considered. The ordering of $q$ is such that $d$ is iterated over first, then over $\ell_b$, then over $M$, with real parts first and then imaginary parts such that the ordering of the outermost structure is
\begin{align}
		(M=0, \mathrm{Re}\, M=1, \mathrm{Re}\, M=2, \mathrm{Im}\, \mathrm{M}=1, \mathrm{Im}\, \mathrm{M}=2 )
\end{align}
and within each of the five cells, $A^{2M}_{\ell_b,\ell_b}$ precedes $A^{2M}_{\ell_b,\ell_b+2}$. 

We compute the mean BipoSH coefficient $\bar{A}_{q}^{2 M} = N_\mathrm{sim}^{-1} \sum_{n=1}^{N_\mathrm{sim}} A^{2 M}_{q}(n)$ where $N_\mathrm{sim}=999$ in this case and the number in the parentheses labels the realisation. Next we compute the covariance matrix:
\begin{align}
\Sigma_{q q'} &= \langle ( A_{q} - \bar{A}_{q}  ) ( A_{q'} - \bar{A}_{q'} )  \rangle =  \frac{1}{N_\mathrm{sim}-1} \sum_{n=1}^{N_\mathrm{sim}} ( A_{q}(n) - \bar{A}_{q}  ) ( A_{q'}(n) - \bar{A}_{q'} )\mathrm{,} \label{eq:sigmamat}
\end{align}
which, when expanded in the individual components, reads
\begin{align}
				\mathbf{\Sigma} = \begin{pmatrix}
		\langle (\mathbf{A}^{20} - \bar{\mathbf{A}}^{20}) ( {\mathbf{A}^{20}} - \bar{\mathbf{A}}^{20} )^{T} \rangle & & \cdots & & \langle (\mathbf{A}^{20} - \bar{\mathbf{A}}^{20} ) (\mathrm{Im} \, {\mathbf{A}^{22}} - \mathrm{Im} \, {\bar{\mathbf{A}}^{22}}  )^{T} \rangle  \\
		  \langle (\mathrm{Re}\, \mathbf{A}^{21} - \mathrm{Re}\, \bar{\mathbf{A}}^{21} ) ( {\mathbf{A}^{20}} - \bar{\mathbf{A}}^{20} )^{T} \rangle & \ddots  &  & & \\
		  \vdots & & & & \langle (\mathrm{Im} \, \mathbf{A}^{22} - \mathrm{Im} \, \bar{\mathbf{A}}^{22} ) (\mathrm{Im} \, {\mathbf{A}^{22}} - \mathrm{Im} \, {\bar{\mathbf{A}}^{22}} )^{T} \rangle
		\end{pmatrix} 
\end{align}
and invert this to obtain the inverse covariance matrix ${\mathbf{\Sigma}}^{-1}$ which was used to finally construct the likelihood:
\begin{align}-2 \log \mathcal{L}(\mathbf{g}, \tilde{\mathbf{A}}) &= ( \mathbf{W}' \mathbf{g} - \tilde{\mathbf{A}} + \bar{\mathbf{A}} )^{T} {\mathbf{\Sigma}}^{-1} (\mathbf{W}'  \mathbf{g} -  \tilde{\mathbf{A}} + \bar{\mathbf{A}}). 
\label{eq:likf}
\end{align}

\begin{figure}[ht]
\centering
\includegraphics[width=\textwidth]{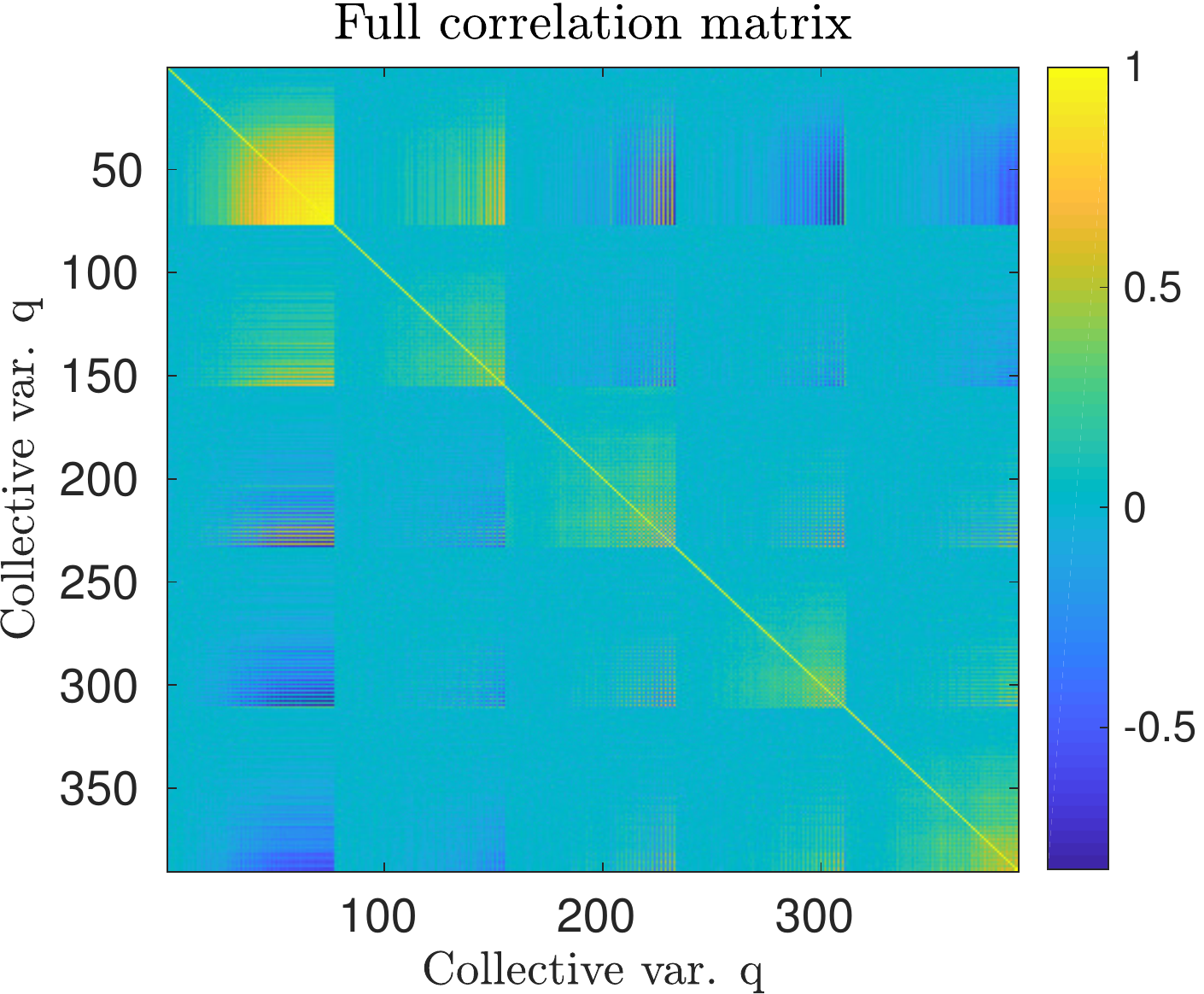}
\caption{Full correlation matrix. Collective variable $q$ now also includes $M=0,1,2$ and real and imaginary parts, listing first the real parts and then the imaginary parts: real pt. $M=0$, real pt. $M=1$, real pt. $M=2$, imaginary pt. $M=1$, imaginary pt. $M=2$.}
\label{fig:fcor}
\end{figure}

The covariance matrix $\mathbf{\Sigma}$ was found to have non-neglible off-diagonal values despite the $\Delta \ell = 30$ binning. For $\ell_b > 600$ and in particular in the subspace corresponding to the covariance of $M=0$, the bins are strongly correlated. 
The \emph{full} correlation matrix is shown in Fig.~\ref{fig:fcor}. There are strong correlations as well as anticorrelations for different $M$, most pronounced for correlations with $M=0$. It is therefore necessary to include the full covariance matrix in the likelihood.
\subsection{Gaussianity test}
Since the likelihood is the probability of observing the data $\tilde{\mathbf{A}}$ in the Planck best-fit $\Lambda$CDM cosmology 
 with an instrument that has the characteristics modelled by the FFP simulations and its foregrounds, eq.\eqref{eq:likf} is valid only if that is indeed what the model predicts. However since that likelihood would be very involved given the complex nature of the FFP simulations, we instead inspected the distributions of the BipoSH coefficients of the 999 realisations.

\begin{figure}[ht]
\centering
\subfigure{\includegraphics[width=0.48\textwidth]{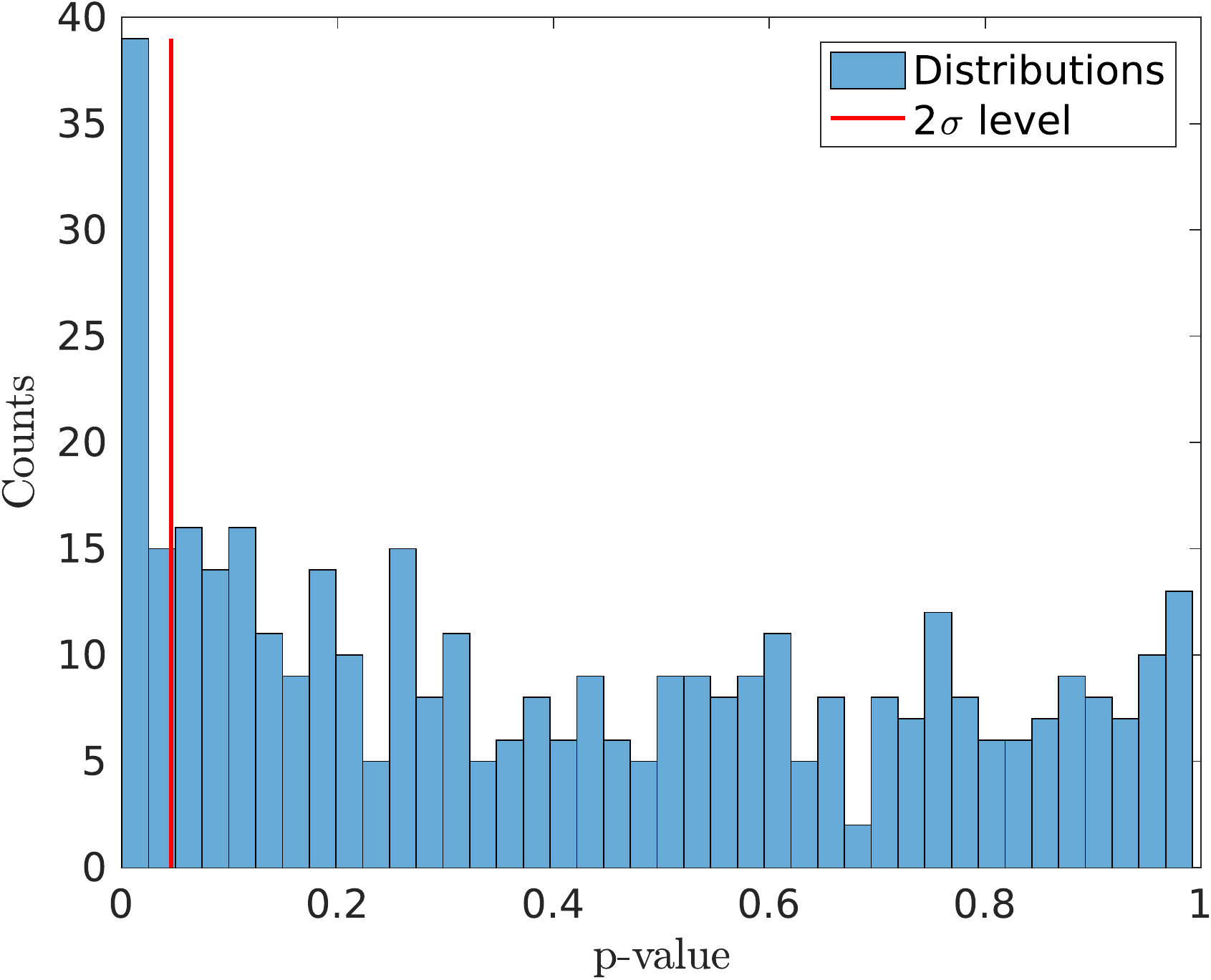}
\label{fig:pval}
}
\subfigure{\includegraphics[width=0.48\textwidth]{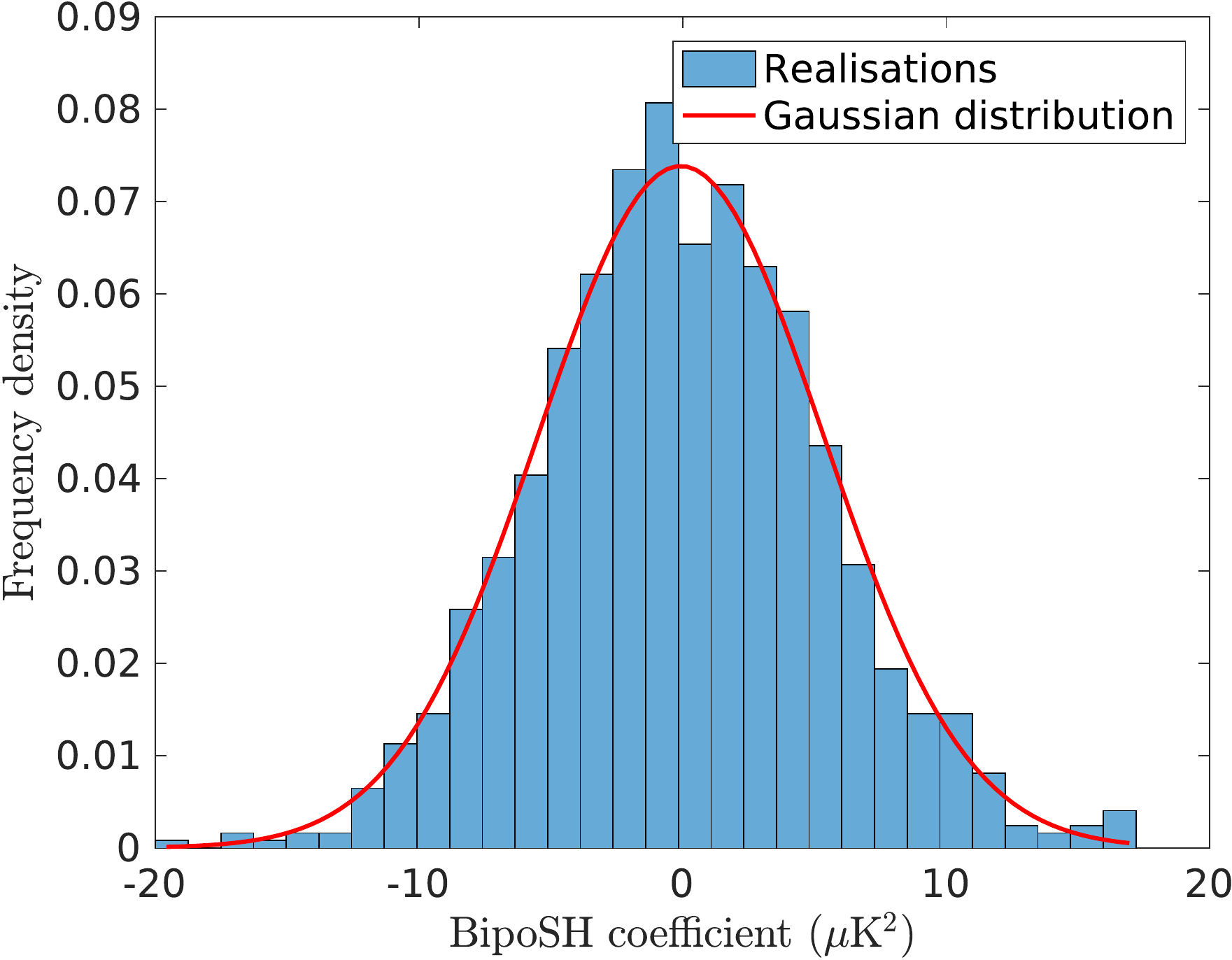}
\label{fig:adist}
}
\caption{Tests of normality of distributions of the BipoSH coefficients $A^{2M}_{\ell_b~\ell_b + d}$. In the left panel, a histogram of p-values of Anderson-Darling tests on the distributions is shown with a vertical (red) line marking the $2 \sigma$ significance level. The right panel shows a histogram of 999 realisations of a particular BipoSH coefficient $A^{2,0}_{76~76}$, with a Gaussian distribution superimposed.}
\label{fig:nortest}
\end{figure}

To test the normality of the distribution of BipoSH coefficients from the FFP simulations, an Anderson-Darling test \cite{adarling} was performed on each distribution of $A^{2 M}_{\ell_b~\ell_b + d}$. The details of this test, which is sensitive to the tail of a distribution, can be found in Appendix~\ref{sec:adtest}. The resulting p-values are shown in Fig.~\ref{fig:nortest} which also shows a typical distribution.

\begin{figure}[ht]
\centering
\includegraphics[scale=0.5]{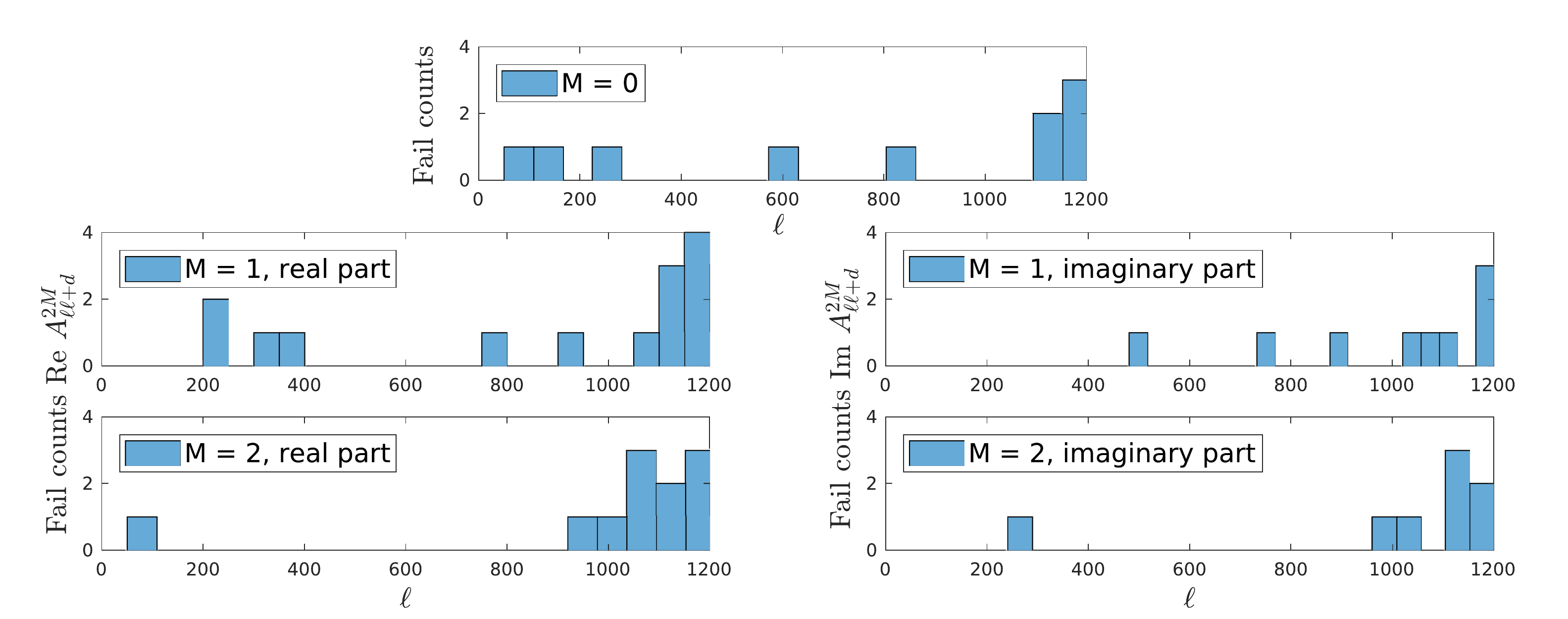}
\caption{Counts of distributions of the BipoSH coefficients $A^{2 M}_{\ell~\ell+d}$ failing the Anderson-Darling test at $2 \sigma$ significance level. The distributions of the real and imaginary part are shown, respectively, in the left and right panel. Note the larger count for $\ell > 1000$.}
\label{fig:reimtest}
\end{figure}

It is seen that at $2 \sigma$ significance, 52 out of 390 \emph{i.e.} 13.3\% of the distributions did not pass the Anderson-Darling test. Of these, 41 come from multipoles $\ell > 700$ as seen in Fig.~\ref{fig:reimtest}.

\begin{figure}[ht]
\centering
\subfigure{\includegraphics[width=0.48\textwidth]{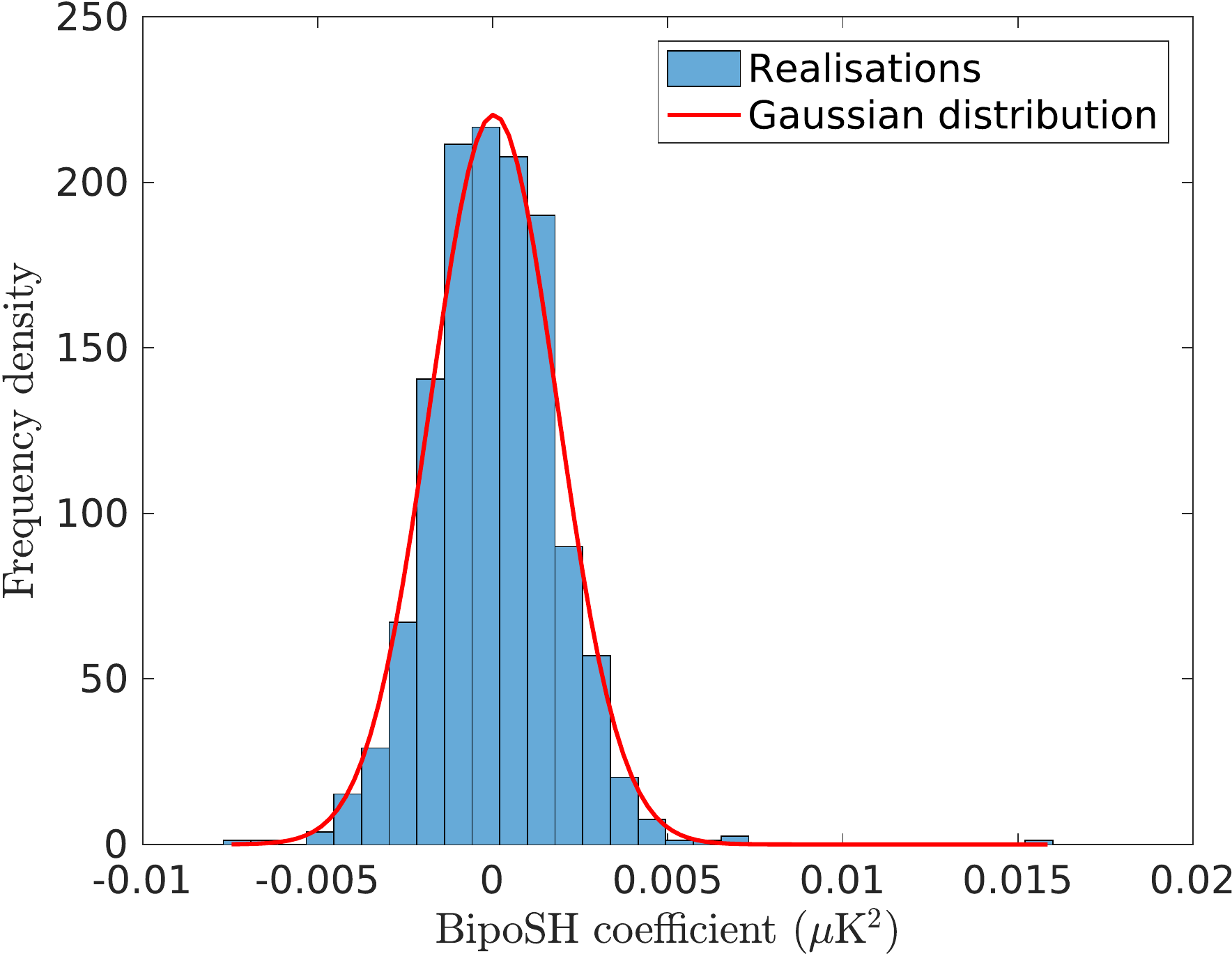}
\label{fig:thefailure}
}
\subfigure{\includegraphics[width=0.48\textwidth]{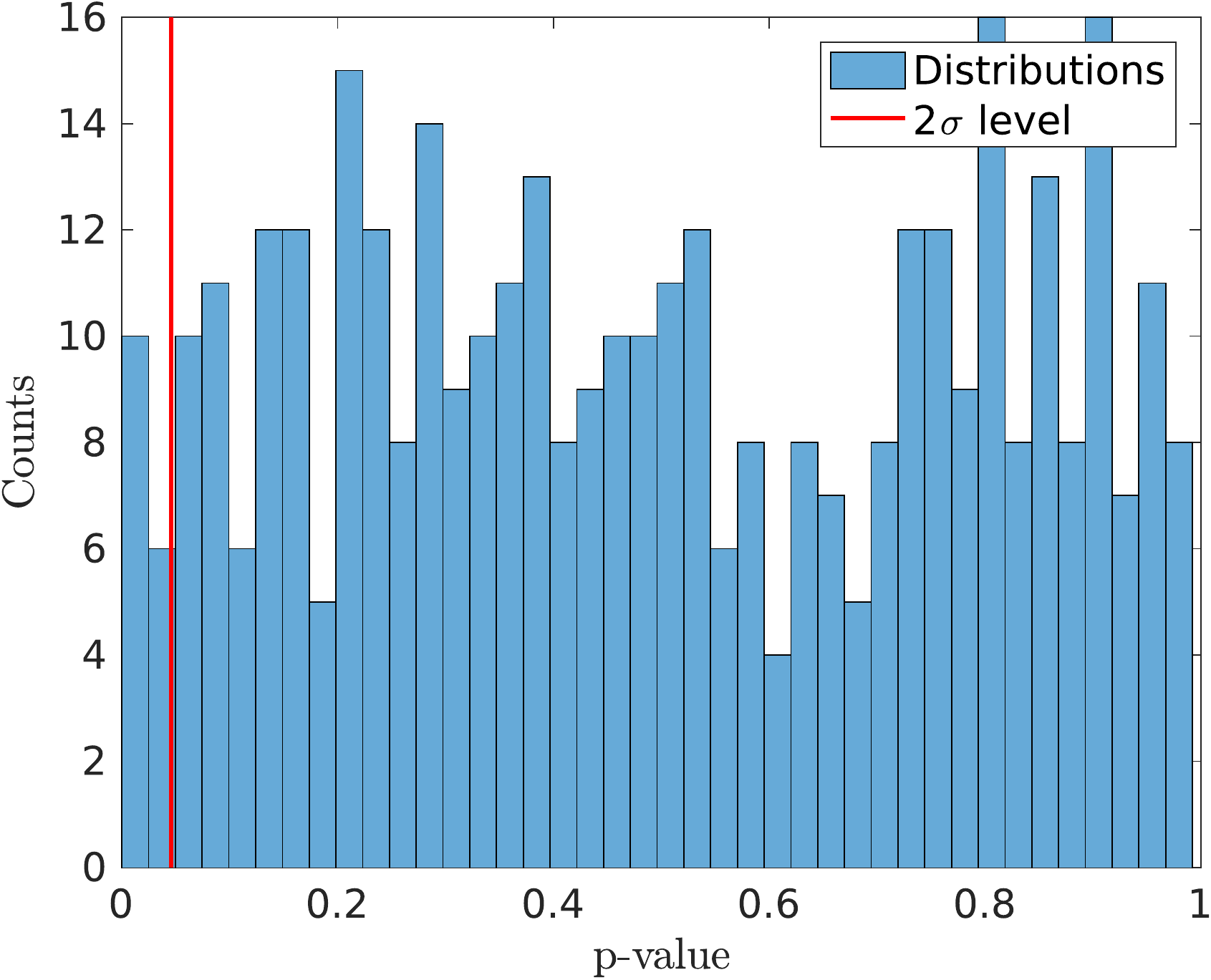}
\label{fig:umr}
}
\caption{The left panel shows a distribution of realisations of the BipoSH coefficient $A^{2,2}_{1185~1185}$ which fail the Anderson-Darling test due to an outlier at $\sim0.015~\mu$K$^2$. In the right panel the normality test is repeated on unmasked realisations --- now only 16 distributions fail the Anderson-Darling test, compared to 52  when the mask is used.}
\label{fig:mixed}
\end{figure}

An example of distribution that failed the test is shown in Fig.~\ref{fig:mixed}. An outlier is present that causes the distribution, which would otherwise be well modelled by a normal distribution, to fail the test. It turns out that the mask used on the skymap is responsible for most test failures. This is demonstrated by the fact that when  the Anderson-Darling test is run on \emph{unmasked} sky realisations, only 16 distributions failed, compared to 52 for the masked sky. (Of course, a certain number is expected to fail in any case as only a finite number of simulations have been run.) Despite this problem of outliers for $\ell > 700$, the Gaussian model of the likelihood seems justified and we adopt it as our likelihood. 

\subsection{Mock data tests}
\begin{figure}[ht]
\centering
\subfigure{\includegraphics[width=0.48\textwidth]{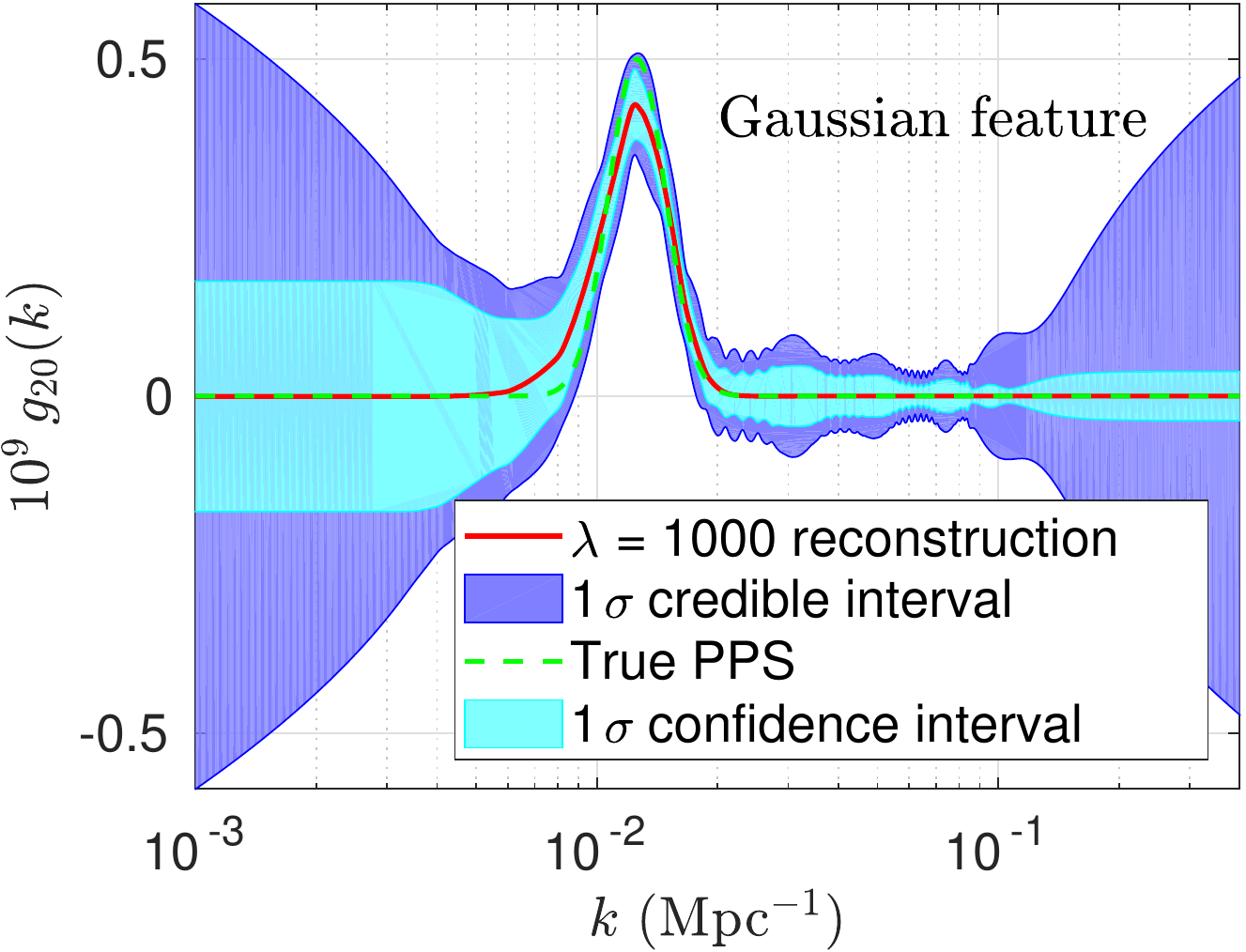}
\label{fig:bump}
}
\subfigure{\includegraphics[width=0.48\textwidth]{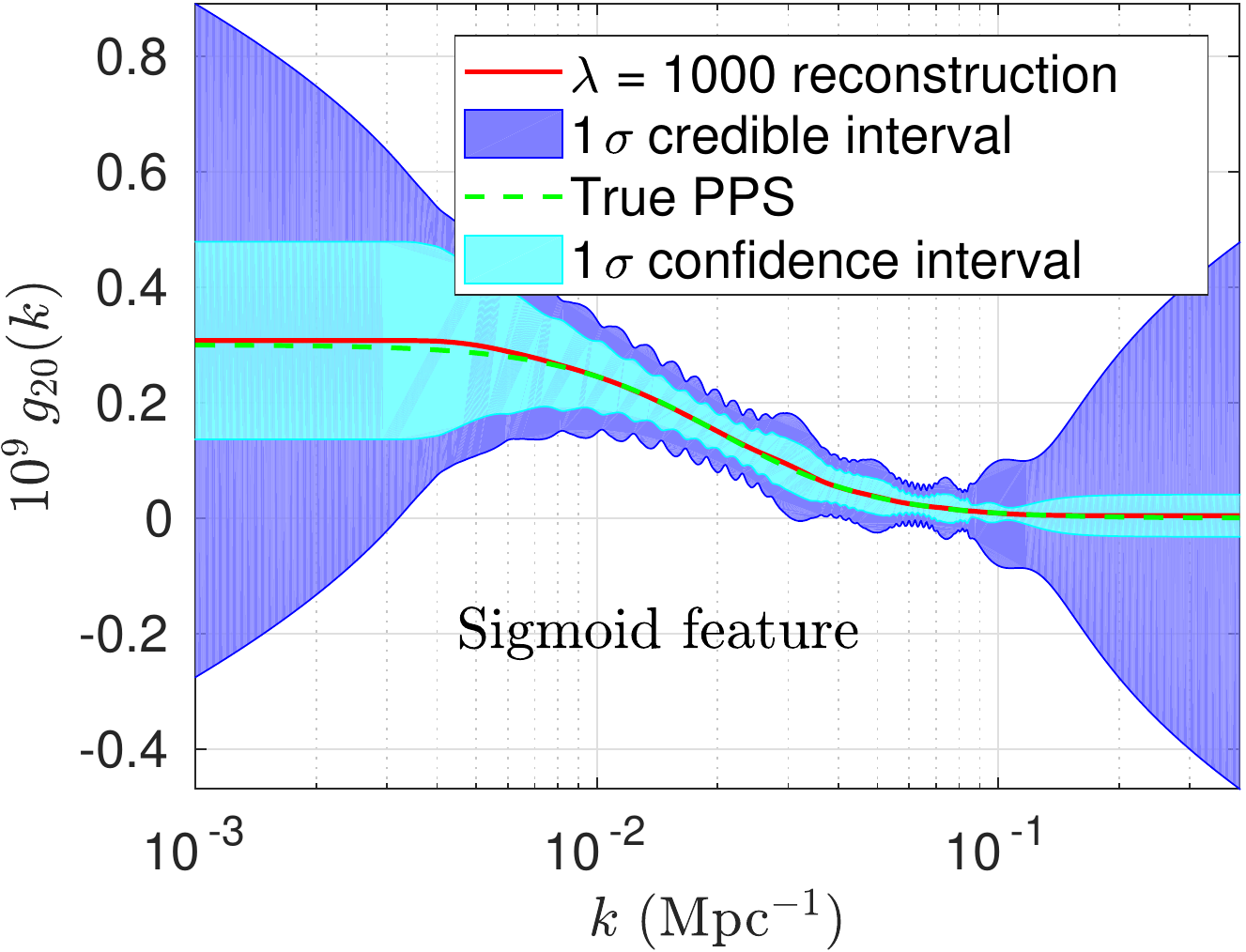}
\label{fig:drop}
}
\caption{Reconstruction of noise-free test spectra with regularisation parameter $\lambda = 1000$. Purple bands indicate the uncertainty in the posterior probability which reflects the uncertainties associated with the experiment --- cosmic variance as well as the choice of prior. Cyan bands indicate the scatter in the reconstruction due to scatter in the data. The reconstruction (full red line) matches well the true spectrum (dotted green line) and is clearly distinguishable from no power. The left and right panels show the reconstruction of a Gaussian (\ref{gaussian}) and sigmoid (\ref{sigmoid}) feature respectively.}
\label{fig:mods}
\end{figure}

To test our approach, the BipoSH coefficients of known quadrupole modulations were calculated. The quadrupole modulations without noise were then reconstructed from these BipoSH coefficients using eq.\eqref{eq:recon}. The uncertainties associated with the reconstructions were taken to be those expected from an experiment that is properly modelled by the FFP simulations, including cosmic variance. The results for two different assumed modulations of $g_{20}(k)$ are shown in Fig.~\ref{fig:mods}. One is a Gaussian
\begin{equation}
g_{20}(k) = a \exp(- (\log_{10}(k/k_0) )^2 / q^2 ),
\label{gaussian}
\end{equation}
with $a=0.5$ , $k_0= 1.26 \times 10^{-2} \, \mathrm{Mpc}^{-1}$ and $q = 0.1$, while the other is a sigmoid
\begin{equation}
g_{20}(k) = a \exp(-(\log_{10}(k/k_0) )/q) / (1+ \exp(-(\log_{10}(k/k_0))/q ) ),
\label{sigmoid}
\end{equation}
with $k_0=2 \times 10^{-2} \, \mathrm{Mpc}^{-1}$, $q = 0.2$ and $a=0.3$. Although slight biases towards smoother spectra are visible, our reconstructions capture the features well and the credible intervals are seen to be quite distinct from zero power.

\begin{figure}[ht]
\centering
\includegraphics[width=0.48\textwidth]{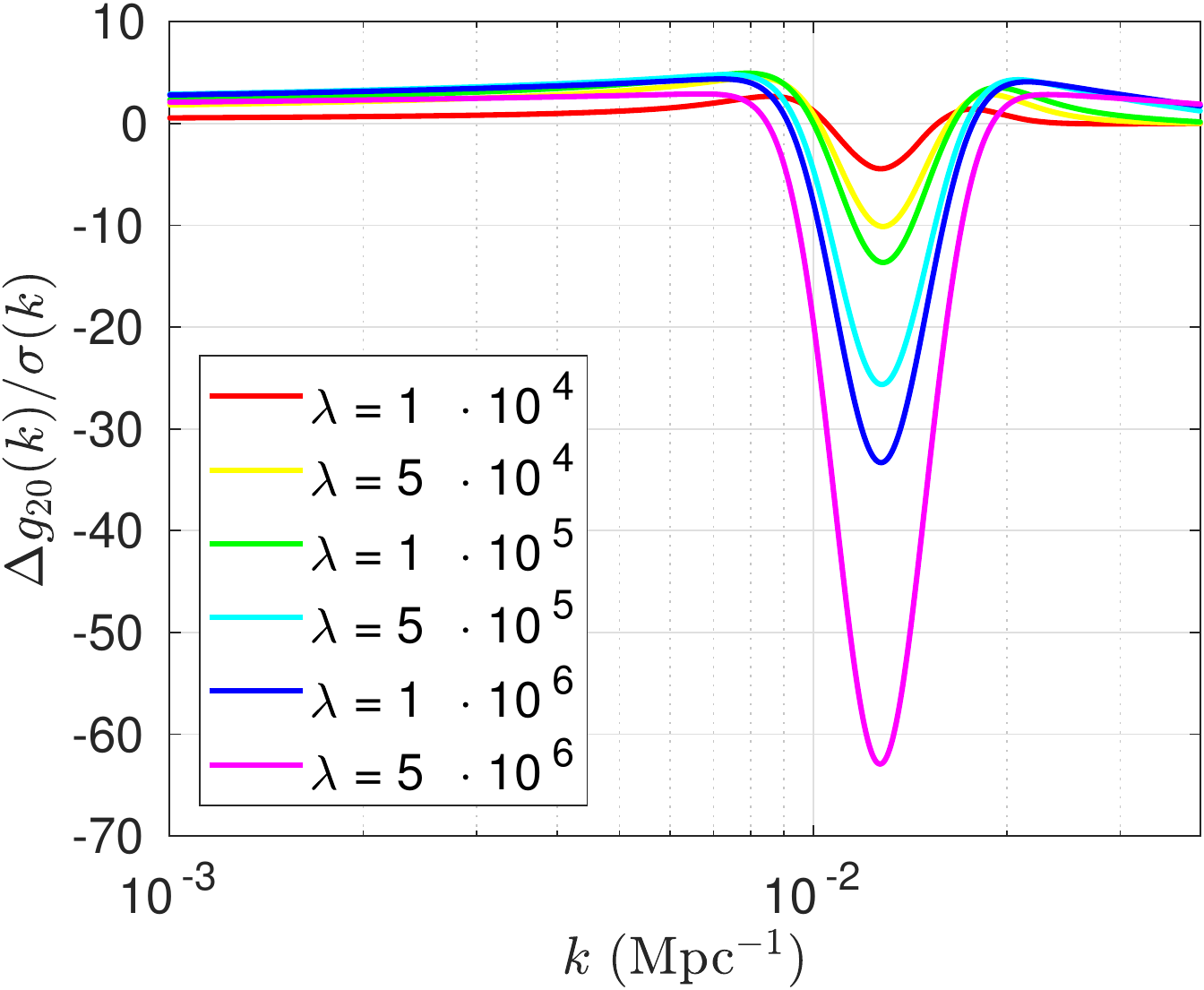}
\caption{Bias in reconstruction of a Gaussian feature: Starting with $\lambda = 10^4$ (red) and ending with $\lambda = 5 \times 10^6$ (blue), a bias is observed with the lowest $\lambda$ being least biased. The plot shows the difference between the true and the reconstructed spectrum relative to the uncertainty.}
\label{fig:thebias}
\end{figure}

The reconstruction parameter will be biased against features if a high value of the regularisation parameter is chosen, as is evident in the Bayesian view of Tikhonov regularisation where $\lambda$ controls the roughness of the prior. The bias relative to the uncertainty is demonstrated in Fig.~\ref{fig:thebias} where the Gaussian feature of Fig.~\ref{fig:mods} is reconstructed with different choices of $\lambda$. The least biased curve is the one with the smallest $\lambda$. Lowering $\lambda$ increases the uncertainty however, so that too small a value makes the reconstruction uninformative. Clearly this determines an \emph{optimal} value of $\lambda$ for reconstruction, as has been discussed in ref.\cite{Hunt:2013bha}.

\section{Results} 
\label{sec:res}

\begin{figure}[ht]
\centering
\subfigure{\includegraphics[width=0.48\textwidth]{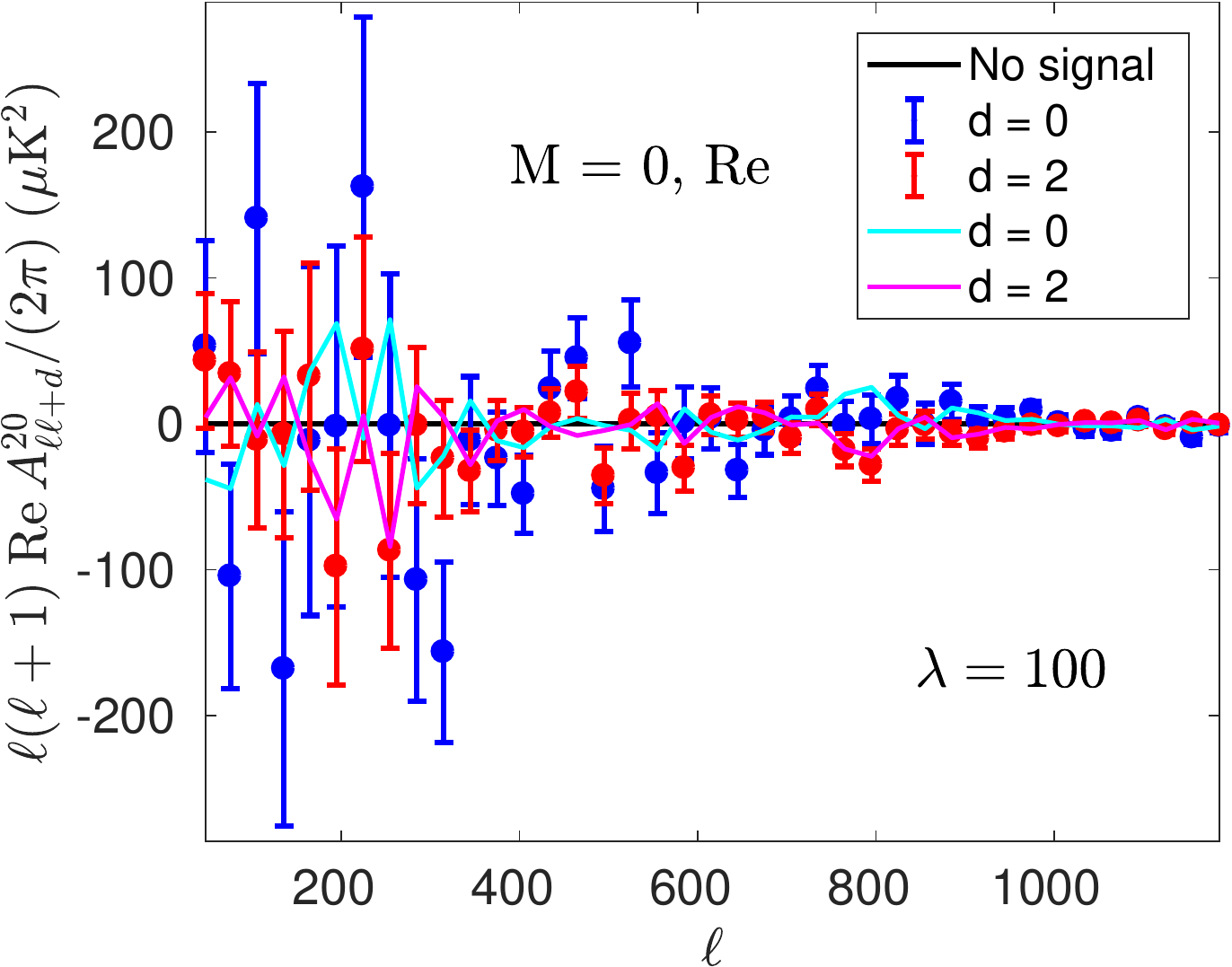}} \\
\subfigure{\includegraphics[width=0.48\textwidth]{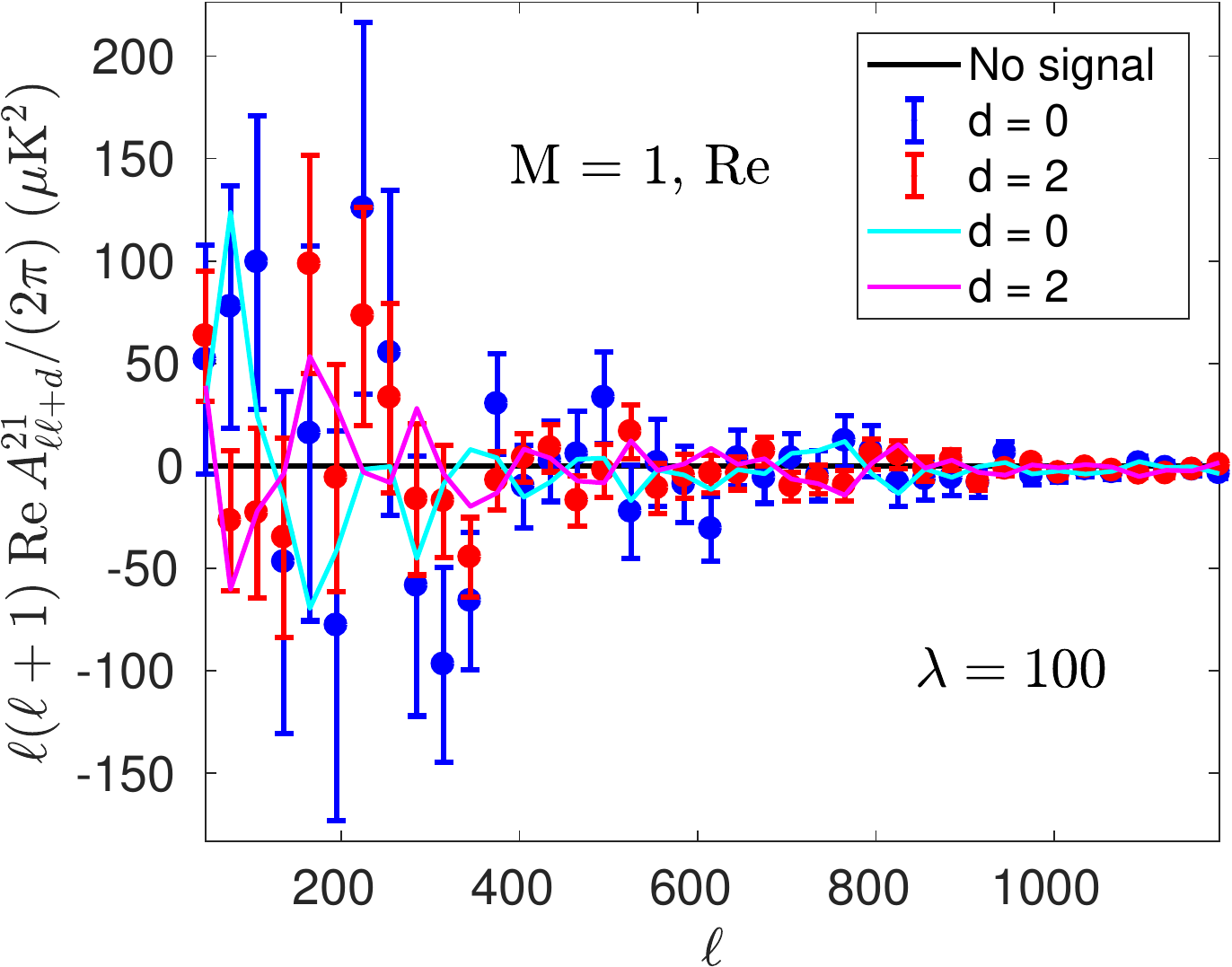}}
\subfigure{\includegraphics[width=0.48\textwidth]{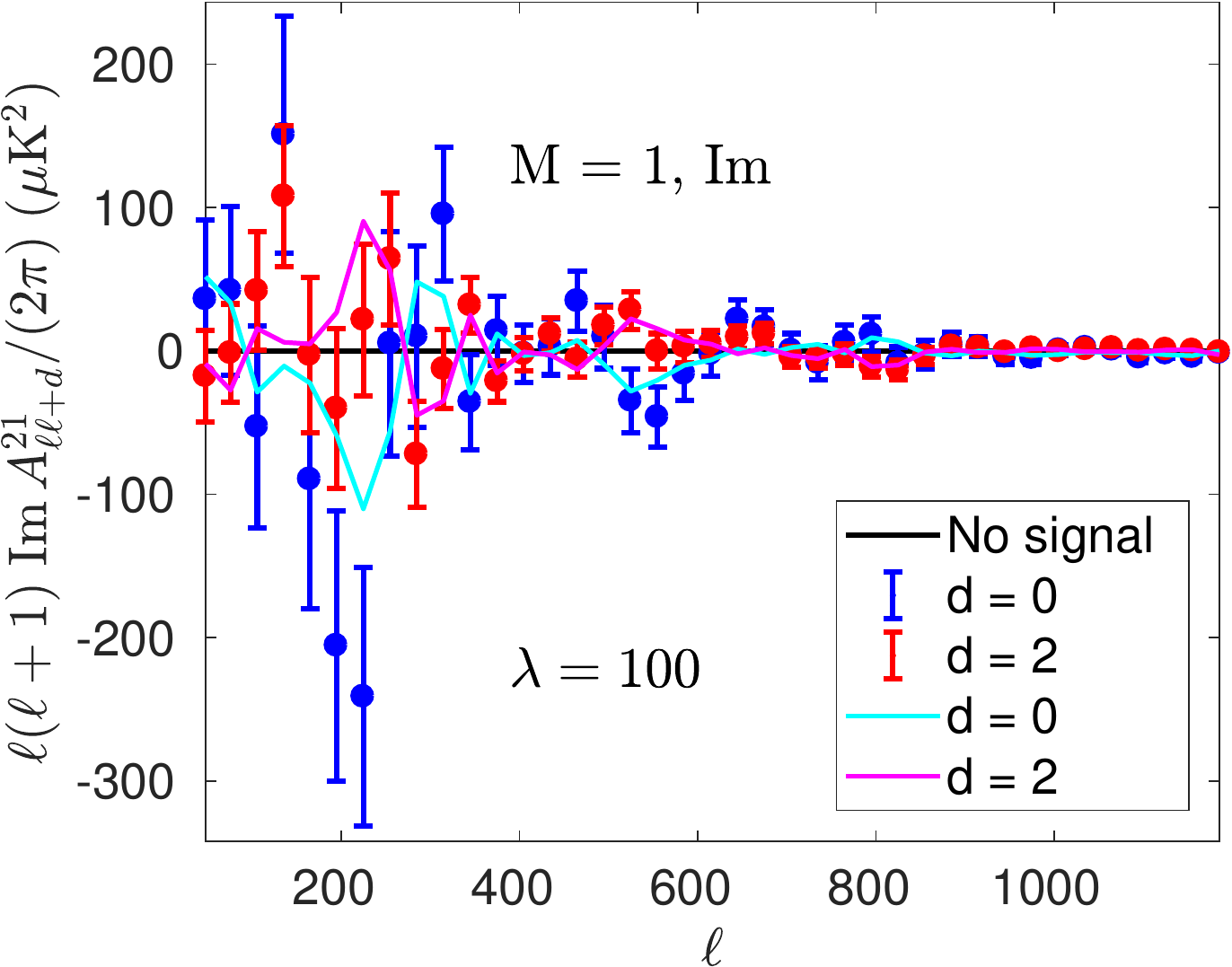}}
\subfigure{\includegraphics[width=0.48\textwidth]{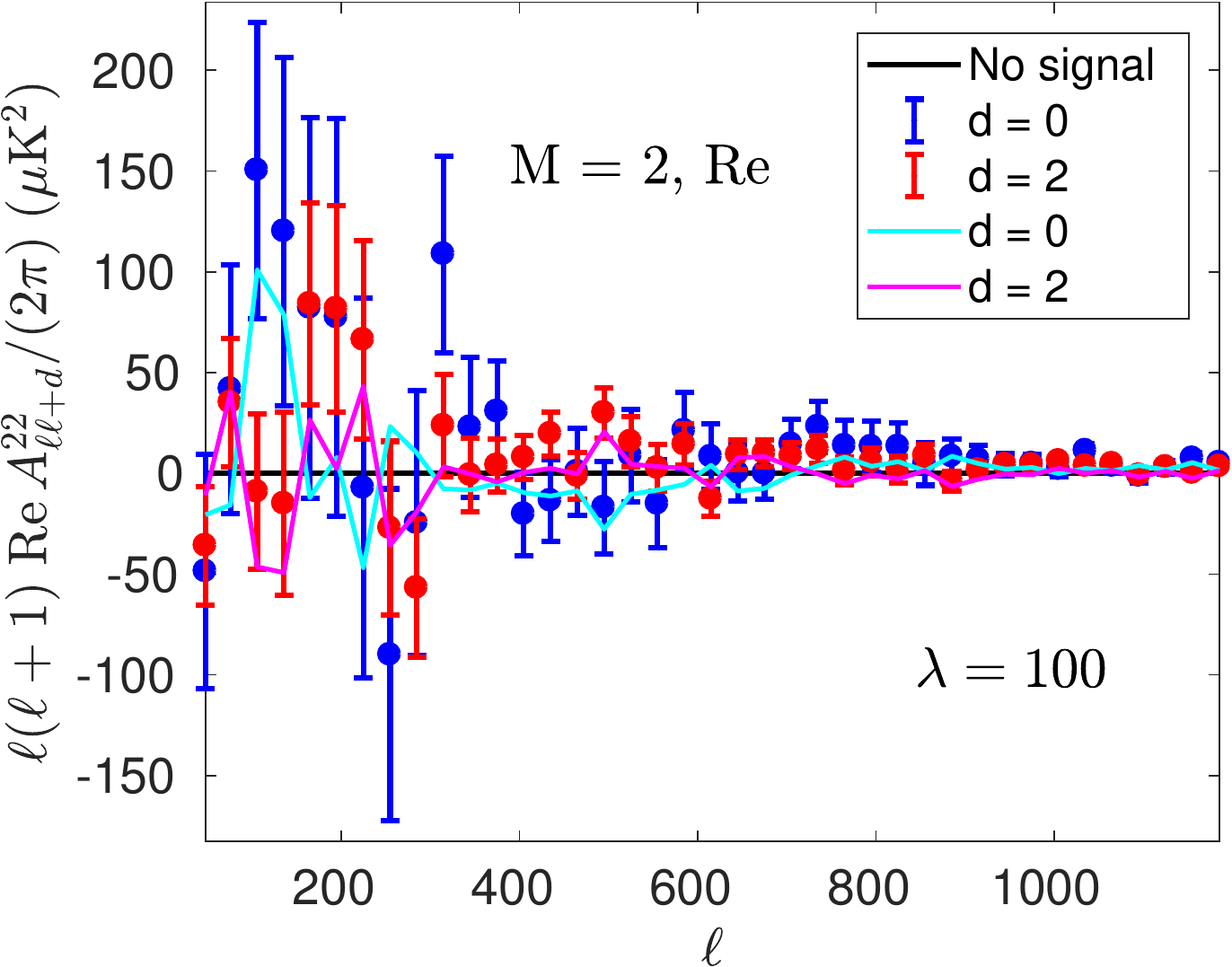}}
\subfigure{\includegraphics[width=0.48\textwidth]{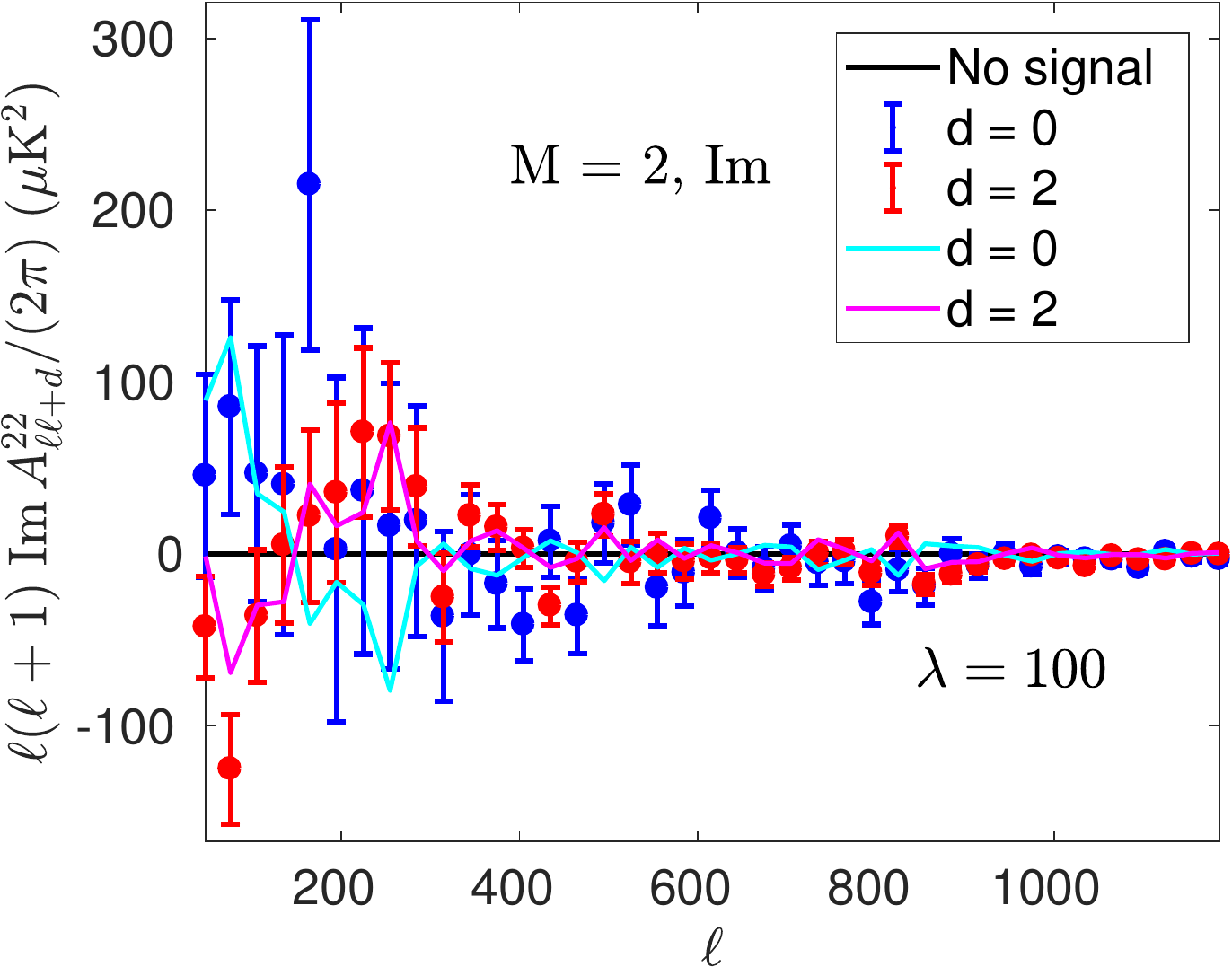}}
\caption{Real and imaginary parts of the corrected BipoSH coefficients $A^{2M}_{\ell, \ell+d}$ of the masked PR2--2015 SMICA map in bins of $\Delta \ell = 30$, 
with $A^{2M}_{\ell, \ell}$ ($d=0$) in blue, and $A^{2M}_{\ell, \ell+2}$ ($d=2$) in red. The BipoSH coefficients associated with the quadrupole modulation reconstructed ($\lambda = 100$) from data (magenta lines for $d=0$ and cyan lines for $d=2$) are superimposed.}
\label{fig:bip}
\end{figure}

We look for a modulation in the masked PR2--2015 SMICA intensity map\footnote{The full name is COM\_CMB\_IQU-smica\_1024\_R2.01\_full with $N_{\mathrm{side}}=1024$ \cite{pla}.} shown in Fig.~\ref{fig:maskedsmica}. The BipoSH coefficients $\tilde{\mathbf{A}}^{2M}$ are calculated using eq.\eqref{eq:cutsky} for $M=0,1,2$ and $d=0,2$ in the range $30\leq \ell \leq 1200$. There is only a real part for $M=0$ but $M=1, 2$ also have imaginary parts. Fig.~\ref{fig:bip} shows the bias-corrected data points $\tilde{\mathbf{A}}^{2M} - \bar{\mathbf{A}}^{2M}$. 

\subsection{Scale-independent quadrupole modulation}

The best-fit \emph{constant} spectrum $g_{2M \star}$ can be defined via the equation:\footnote{The numerical factor cancels the $\sqrt{4 \pi}$ in eq.\eqref{eq:split} and enables a direct comparison with related work \cite{Ade:2015hxq}.}
\begin{equation}
g_{2M}(k) = (4 \pi)^{-1/2} g_{2M \star} \mathcal{P}(k) ,
\label{eq:best-fit} 
\end{equation}
where the best-fit isotropic spectrum is taken to be $\mathcal{P}(k) = A (k/k_{\ast})^{n_s-1}$ with $k_{\ast} = 0.05 \, \mathrm{Mpc^{-1}}$, $A = 2.198$ and $n_\mathrm{s} = 0.9655$ (best-fit parameters from Planck TT+lowP) \cite{Ade:2015xua}.
We look for the $g_{2M \star}$, collectively denoted $ \boldsymbol \alpha$, which maximises the likelihood \eqref{eq:likf} which is in this case,
\begin{align}
		- 2 \log \mathcal{L} = ( (4 \pi)^{-1/2} \, \mathbf{W}' \mathbf{Q} \boldsymbol \alpha  - \tilde{\mathbf{A}} + \bar{\mathbf{A}}  )^{T} \mathbf{\Sigma}^{-1} ( (4 \pi)^{-1/2} \, \mathbf{W}' \mathbf{Q} \boldsymbol \alpha   - \tilde{\mathbf{A}} + \bar{\mathbf{A}} )
\end{align}
where
\begin{align}
\mathbf{Q} = \mathbf{I}_{5\times 5}\otimes \mathbf{p} = \begin{pmatrix}
		 \mathbf{p} & & \\
		 & \ddots & \\ & & \mathbf{p}
		\end{pmatrix}		
\end{align}
is a block matrix made of the isotropic primordial power spectrum $\mathbf{p}$ stacked five times.
The maximum likelihood estimate is obtained by differentiating with respect to $\boldsymbol \alpha$, setting the expression equal to zero and isolating $\boldsymbol \alpha$:
\begin{align}
	\hat{g}_{2M \star} & \equiv  \boldsymbol \alpha =\sqrt{4 \pi} \mathbf{J}^{-1} \mathbf{Q}^{T} \mathbf{W}'^{T} \mathbf{\Sigma}^{-1} (\tilde{\mathbf{A}} - \bar{\mathbf{A}} )  \label{eq:bfc}
\end{align}
where
\begin{align}
		\mathbf{J} = \mathbf{Q}^{T} (\mathbf{W}')^{T} \mathbf{\Sigma}^{-1} \mathbf{W}' \mathbf{Q}.
\end{align}
The uncertainty of $g_{2M \star}$ is just the inverse Hessian, \emph{i.e.}, the inverse of the double derivative of $-\log \mathcal{L}$ with respect to $g_{2M \star}$, which in this case is:
\begin{align}
\mathbf{H}^{-1} &= 4 \pi \mathbf{J}^{-1}. \label{eq:bfce}
\end{align}


When we quote the 1D uncertainties they correspond to the square root of the diagonal entries of the inverse Hessian:
\begin{align}
		\sigma_{g_{2M \star}} \equiv \sigma_{i} = \sqrt{4 \pi (J^{-1})_{ii}}
\end{align}
where $J_{ii}$ are the diagonal values of $\mathbf{J}$ and $i$ is the index running from $1$ ($\mathrm{Re} \,  g_{20 \star}$) to $5$ ($\mathrm{Im} \, g_{22 \star}$).

Our results are summarised in Table~\ref{tab:main}. Both $\mathrm{Re} \, g_{20 \star} $ and $\mathrm{Re} \, g_{22 \star}$ appear to be higher than expected, over twice the expected value. We perform the p-value test later.  


\begin{table}[ht]
\begin{center}
\begin{tabular}{|l|r|r|}
\hline
$M$ & $ 10^2 \times \mathrm{Re} \, {g}_{2M \star}$ & $10^2 \times \mathrm{Im} \, {g}_{2M \star}$ \\
\hline
$0$ & $0.76\pm 0.34$ &  \\
\hline
$1$ & $0.05\pm 0.20$ & $-0.14 \pm 0.20$  \\
\hline
$2$ & $0.48 \pm 0.21$ & $0.09 \pm 0.20$ \\
\hline
\end{tabular}
\caption{Results for the best-fit constant quadrupole modulation $g_{2M \star}$, eq.\eqref{eq:best-fit}.}
\label{tab:main}
\end{center}
\end{table}

\subsection{Power-law quadrupole modulations}

We also test for a power-law quadrupole modulation of the form
\begin{equation}
	g_{2M}(k) = \frac{4 \sqrt{\pi} g_{\ast}}{15} \left( \frac{k}{k_{\ast}}\right)^{q} \mathcal{P}(k) Y^{\ast}_{2M}(\theta,\phi) ,
\label{eq:newform}
\end{equation}
where $\mathcal{P}(k)$ is the best fit isotropic spectrum and the coefficient is chosen to ensure a direct comparison with the Planck collaboration's results \cite{Ade:2015lrj}. In real space, this corresponds to a quadrupole modulation $\propto (\hat{\mathbf{k}} \cdot \hat{\mathbf{n}})^2$ where $\hat{\mathbf{n}} = (\theta,\phi)$ is a unit vector in the direction of the hot spot. The posterior distribution of the parameters $(g_{\ast},\theta,\phi)$ is estimated using the Metropolis-Hastings algorithm with flat priors on the angles and a flat prior on $g_{\ast}$ in the range $-10 \leq 10^2 g_{\ast} \leq 10$. 
The results are shown in Table~\ref{tab:astpowlaws}.

\begin{figure}[ht]
\centering
\subfigure{\includegraphics[width=0.48\textwidth]{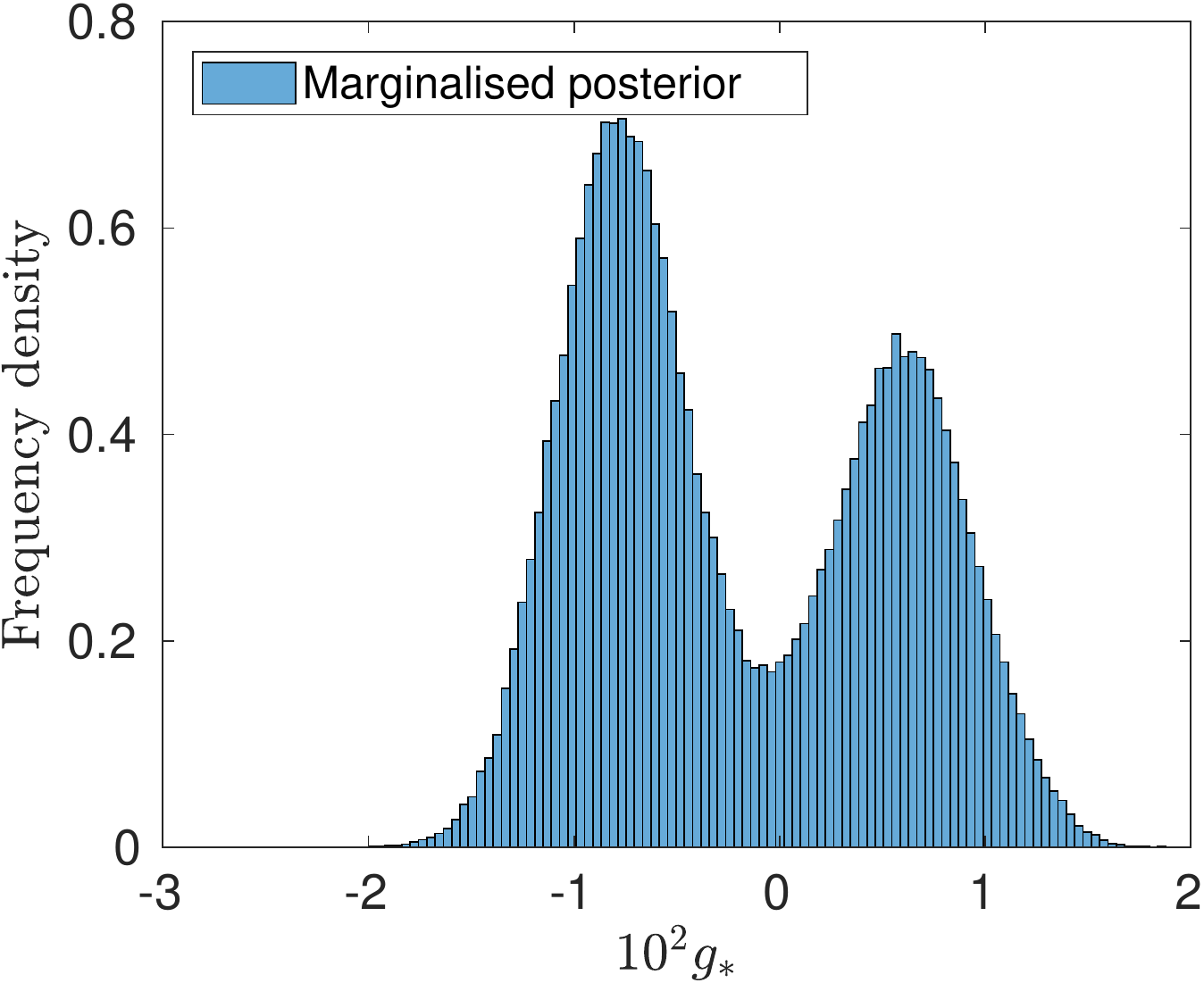}}
\subfigure{\raisebox{2.25 em}{\includegraphics[width=0.48\textwidth]{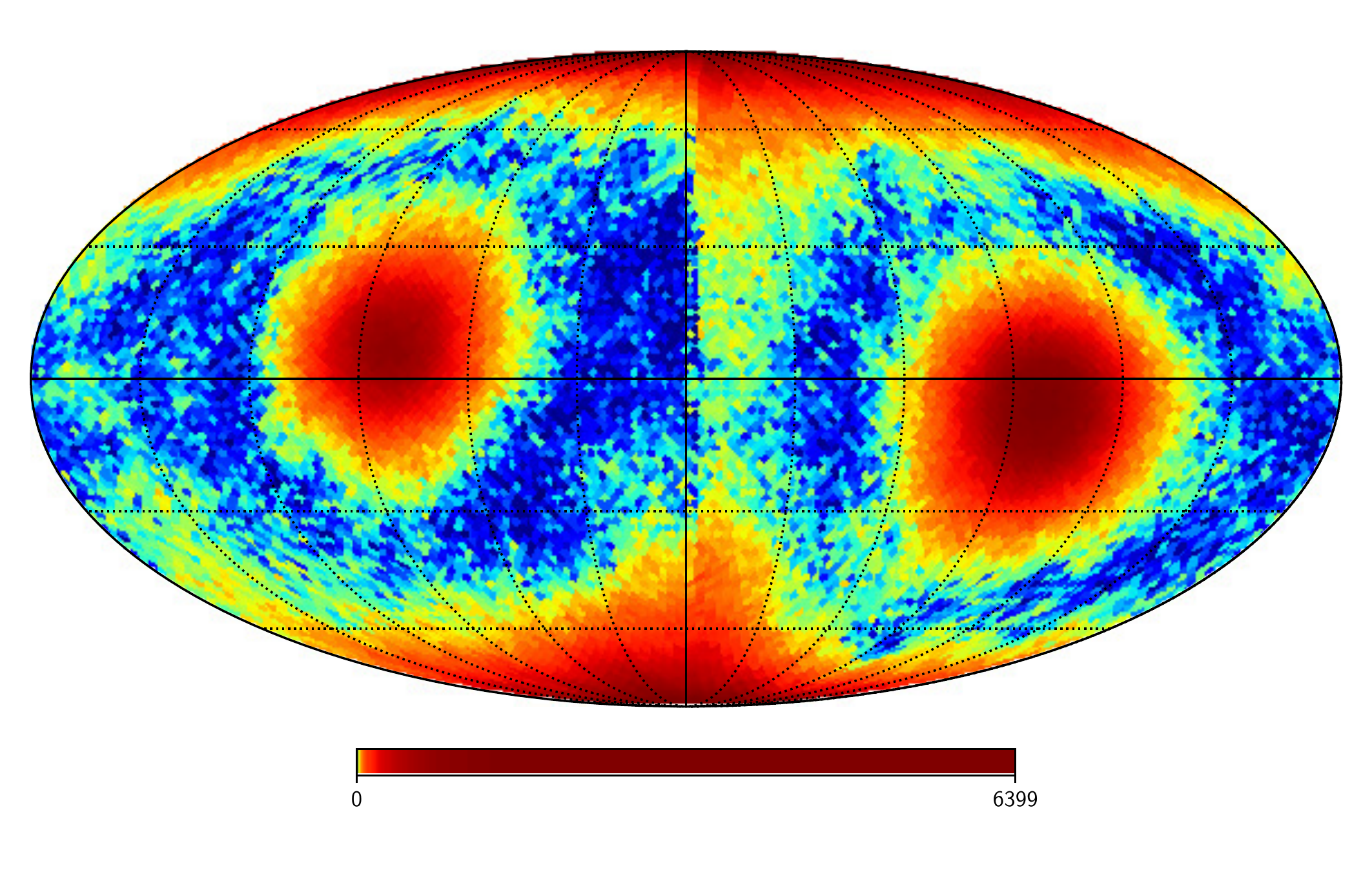}}}
\caption{The left panel shows the Metropolis-Hastings sampled posterior distribution of $10^2 g_{\ast}$ for the scale-independent case $q=0$ assuming a flat prior on $-10 \leq 10^2 g_{\ast} \leq 10$ and marginalising over the angles $\theta,\phi$. The posterior distribution is bimodal. The right panel shows the Mollweide projection of the posterior distribution of directions $(\theta,\phi)$.}
\label{fig:mcmchist}
\end{figure}

\begin{table}[ht]
\begin{center}
\begin{tabular}{|r|r|r|r|r|r|}
\hline
$q$ & $-2$ & $-1$ & $0$ & $1$& $2$ \\
\hline
 $ 10^2 {g}_{\ast}<0$ & \multirow{2}{*}{$-0.00\pm 0.13$} & $-0.56^{+0.40}_{-0.39}$ & $-0.74^{+0.34}_{-0.33}$ & $-0.47^{+0.22}_{-0.21}$ & $-0.26^{+0.13}_{-0.12}$ \\
 \cline{0-0}\cline{3-6}
 $ 10^2 {g}_{\ast}>0$ & & $0.82^{+0.38}_{-0.39}$ & $0.63^{+0.33}_{-0.35}$ & $0.27^{+0.17}_{-0.18}$ & $0.14^{+0.10}_{-0.10}$ \\
\hline
\end{tabular}
\end{center}
\caption{Results for the best-fit power-law quadrupole modulations of the form of eq.\eqref{eq:newform} for different power law indices.}
\label{tab:astpowlaws}
\end{table}
The credible intervals are such that $68.2\%$ of the probability is contained in the quoted ranges. The posterior distribution of $g_{\ast}$ found after marginalising over the directions $\theta,\phi$ is shown in Fig.~\ref{fig:mcmchist} for the case of $q=0$. It is bimodal. A Mollweide projection of the \emph{posterior distribution} of the direction $(\theta,\phi)$ is also shown.
Nearly half the total posterior probability ($49\%$) is focussed in the direction $(l,b)=({84^{\circ}}^{+13}_{-15},{7^{\circ}}^{+13}_{-12})$ with amplitude $10^2 g_{\ast} = -0.74^{+0.34}_{-0.33}$ which correspond to the two lobes. The best-fit positive amplitude is $10^{2} g_{\ast} = 0.63^{+0.33}_{-0.35}$. The probability distribution of the direction of the positive amplitude modulation $g_{\ast}>0$ is almost uniform in the azimuthal direction, but not in the polar direction. The best-fit here points perpendicular to the Galactic plane $b={1.92^{\circ}}^{+21}_{-21}$. This is suggestive of possible contamination of the SMICA map by Galactic foregrounds. 

\begin{figure}[ht]
\centering
\subfigure{\includegraphics[width=0.48\textwidth]{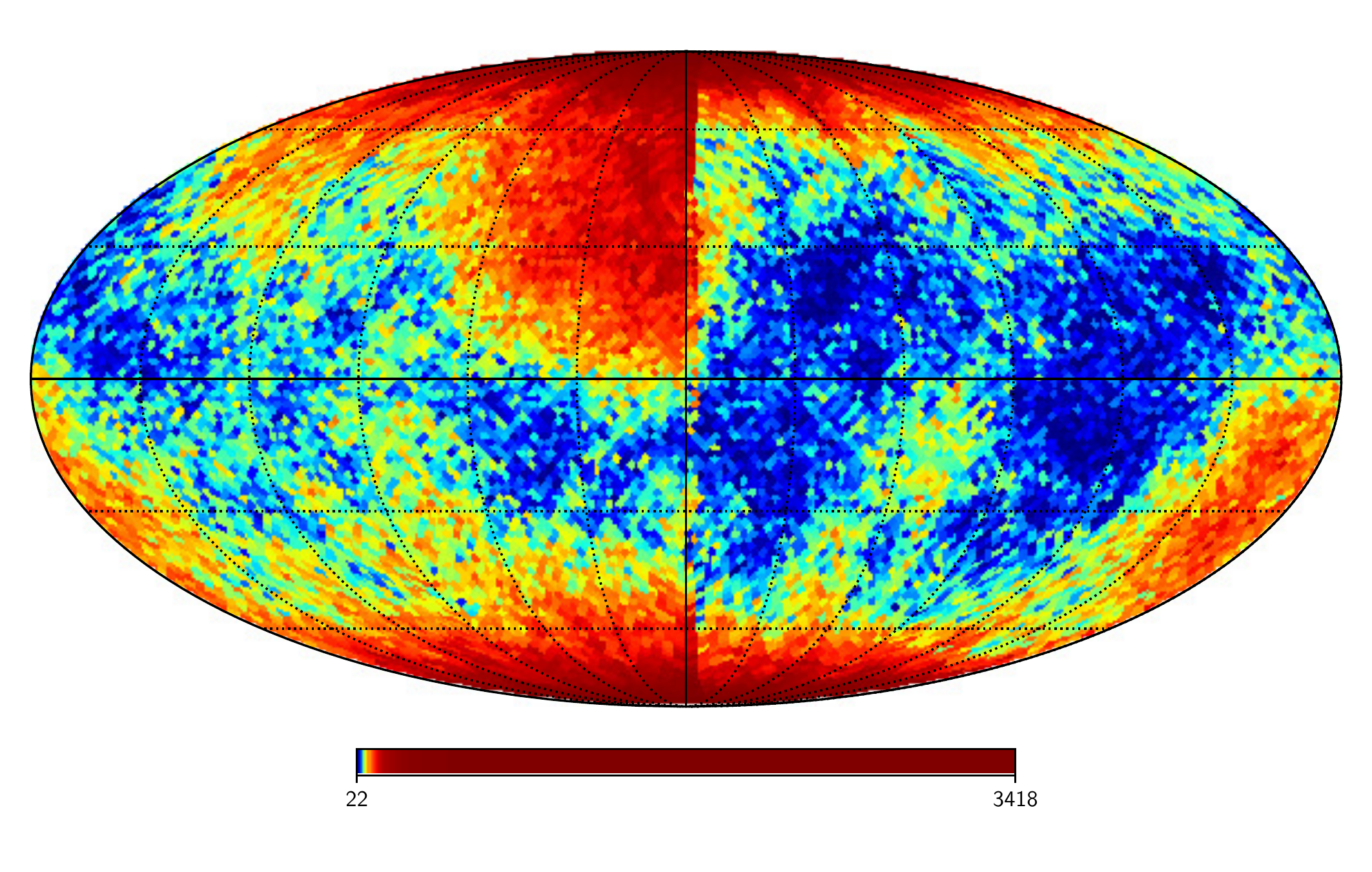}}
\subfigure{\includegraphics[width=0.48\textwidth]{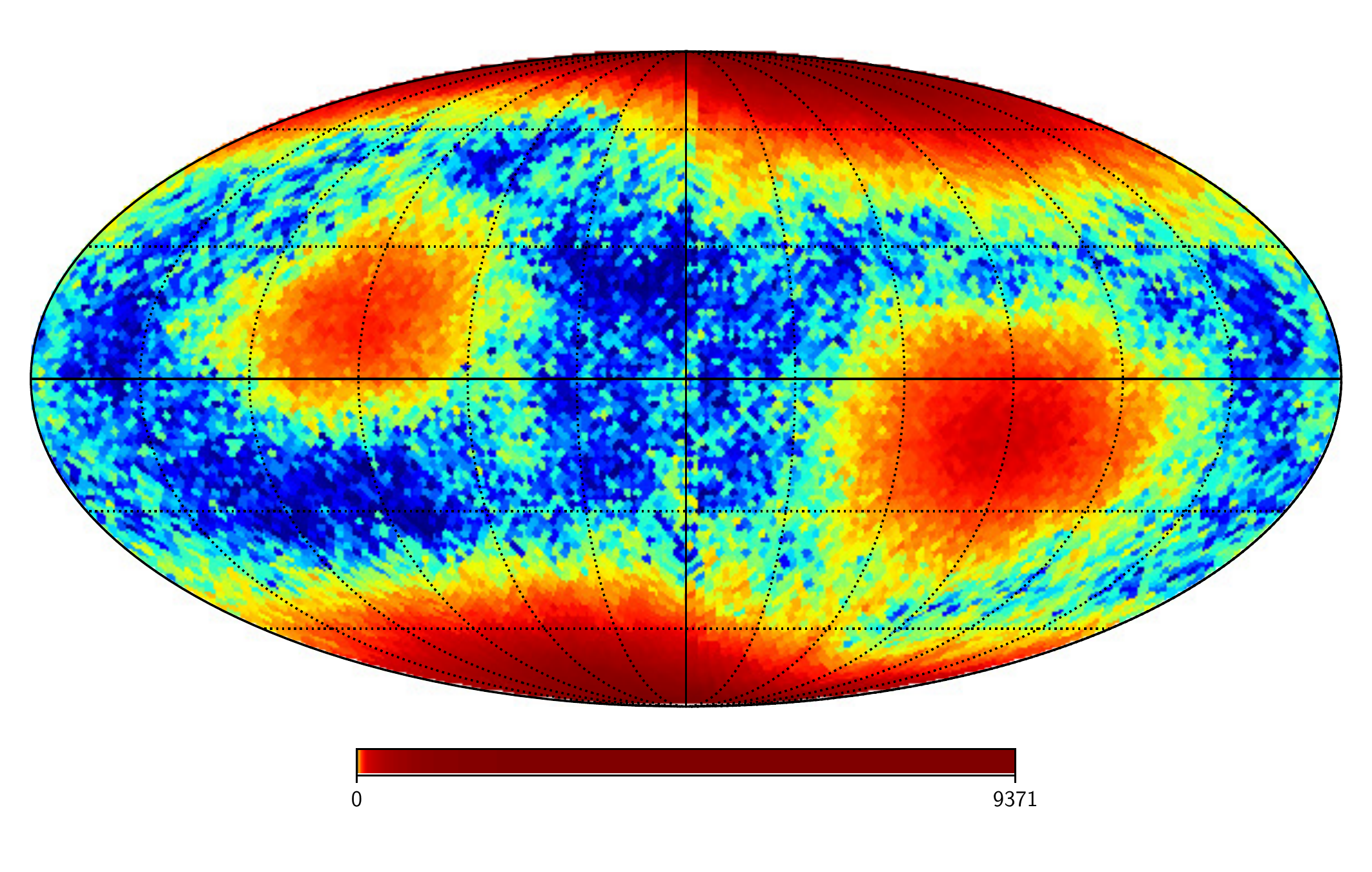}}
\subfigure{\includegraphics[width=0.48\textwidth]{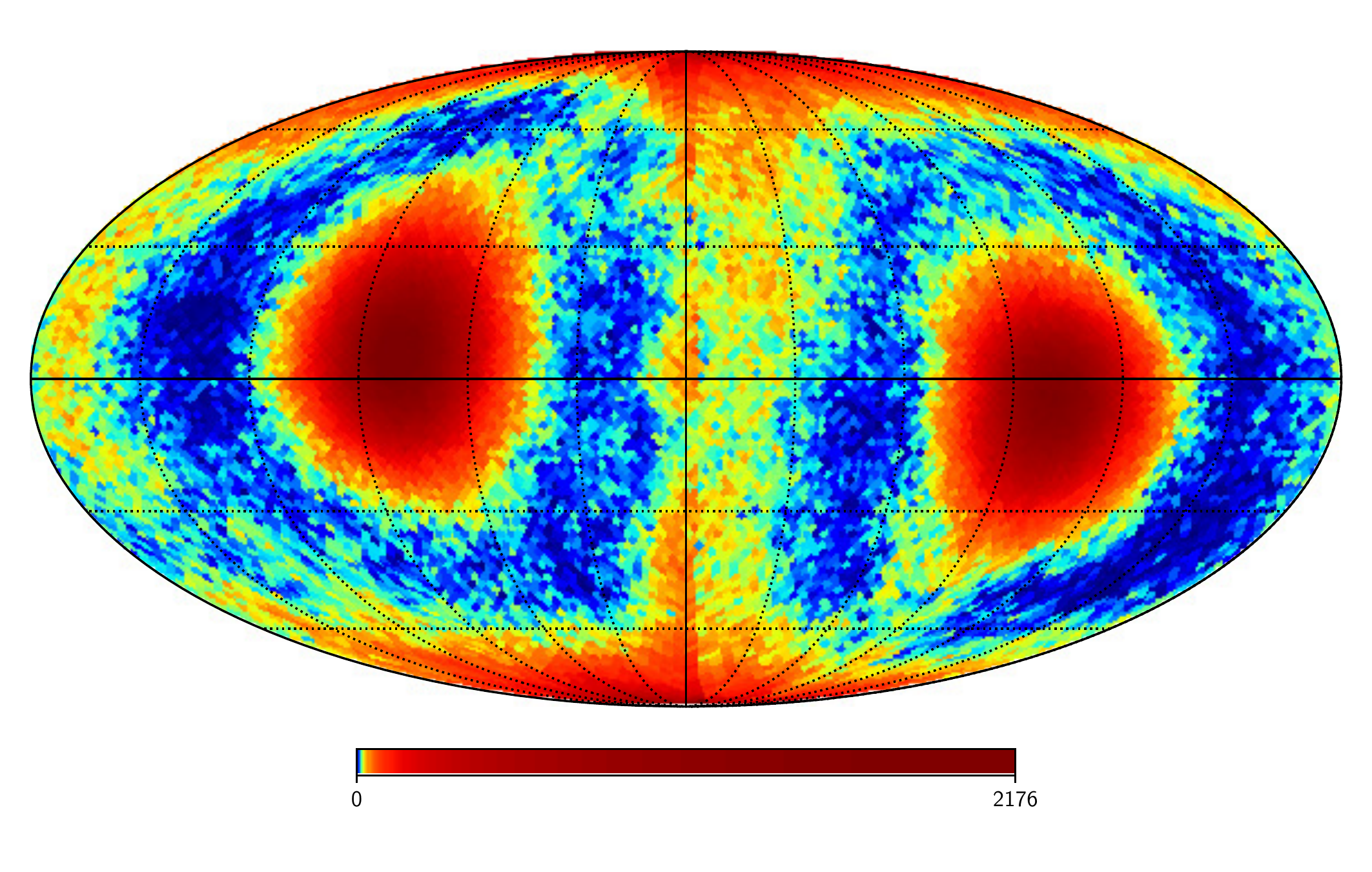}}
\subfigure{\includegraphics[width=0.48\textwidth]{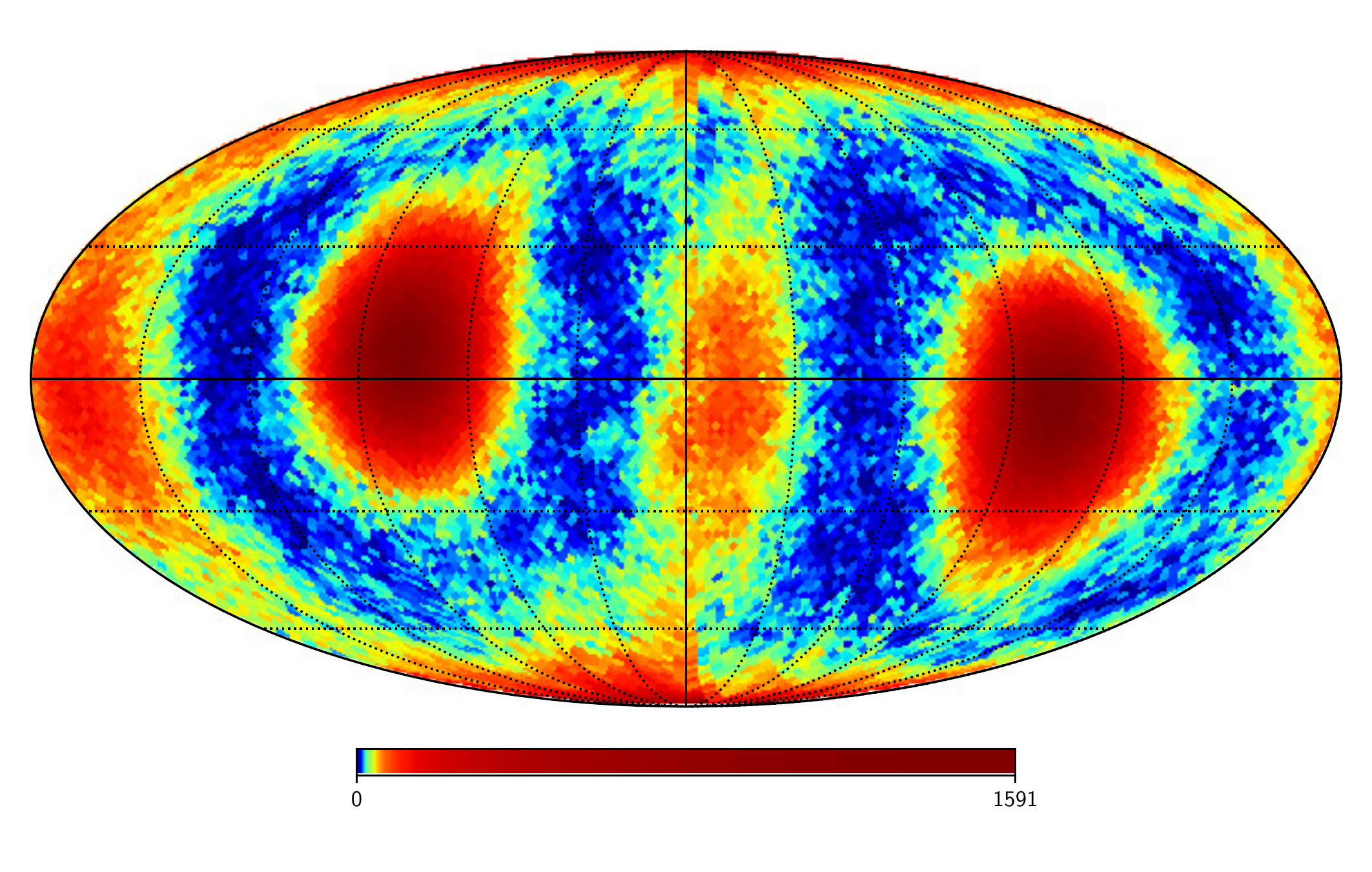}}
\caption{Equalised histograms of possible \emph{directions} of the quadrupole modulation for power law modulations of the form $g_{2M}(k) \propto (k/k_{\ast})^{q}$ ranging from $q=-2$ (upper left figure) to $q=2$ (centred lower figure) derived from Monte Carlo runs.}
\label{fig:dirhist}
\end{figure}

Fig.~\ref{fig:dirhist} shows the Mollweide projection of the posterior distributions for the other values of the power-law index $q$.

\begin{figure}[h!]
\centering
\subfigure{\includegraphics[width=0.45\textwidth]{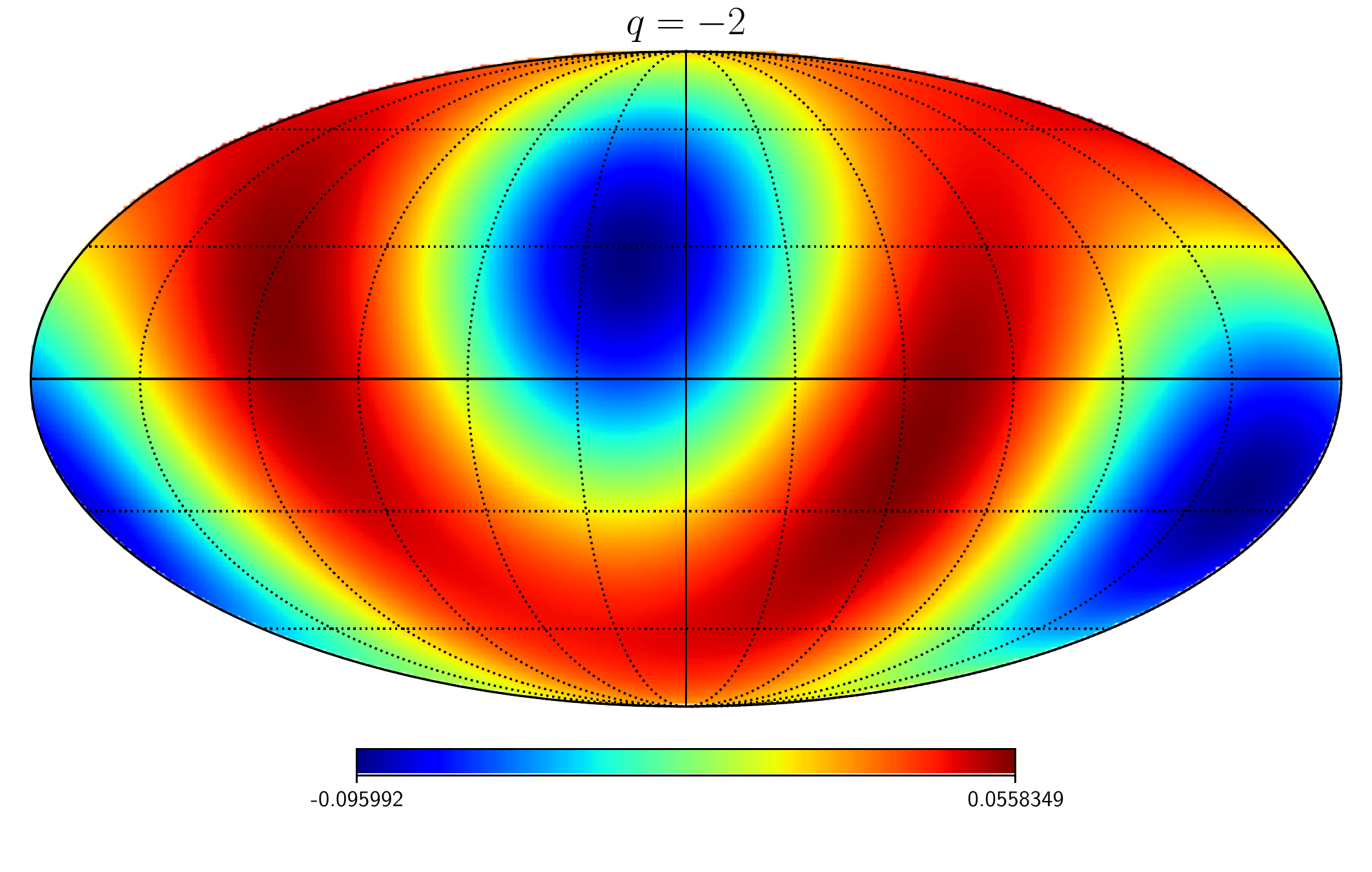}}
\subfigure{\includegraphics[width=0.45\textwidth]{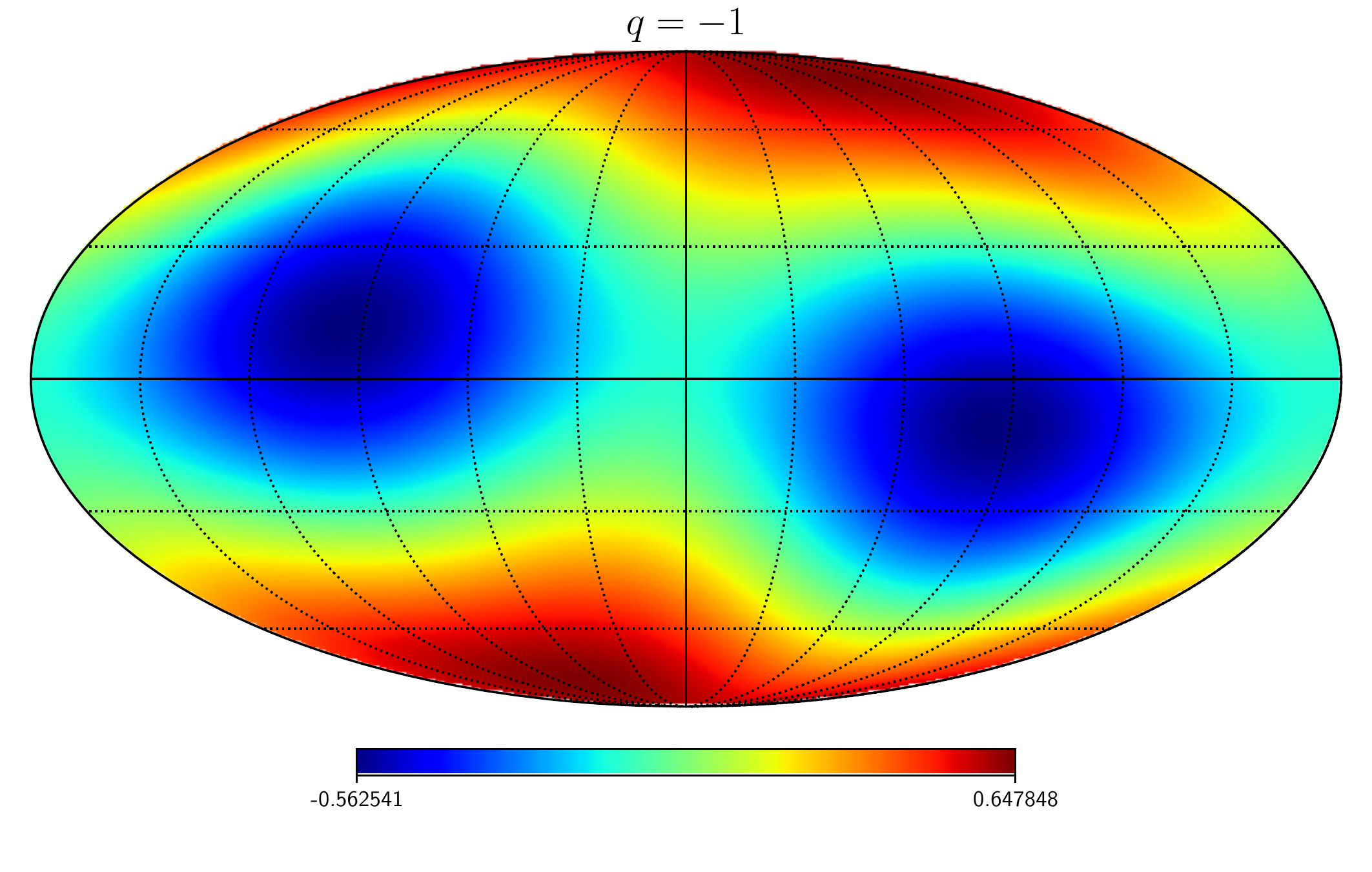}}
\subfigure{\includegraphics[width=0.45\textwidth]{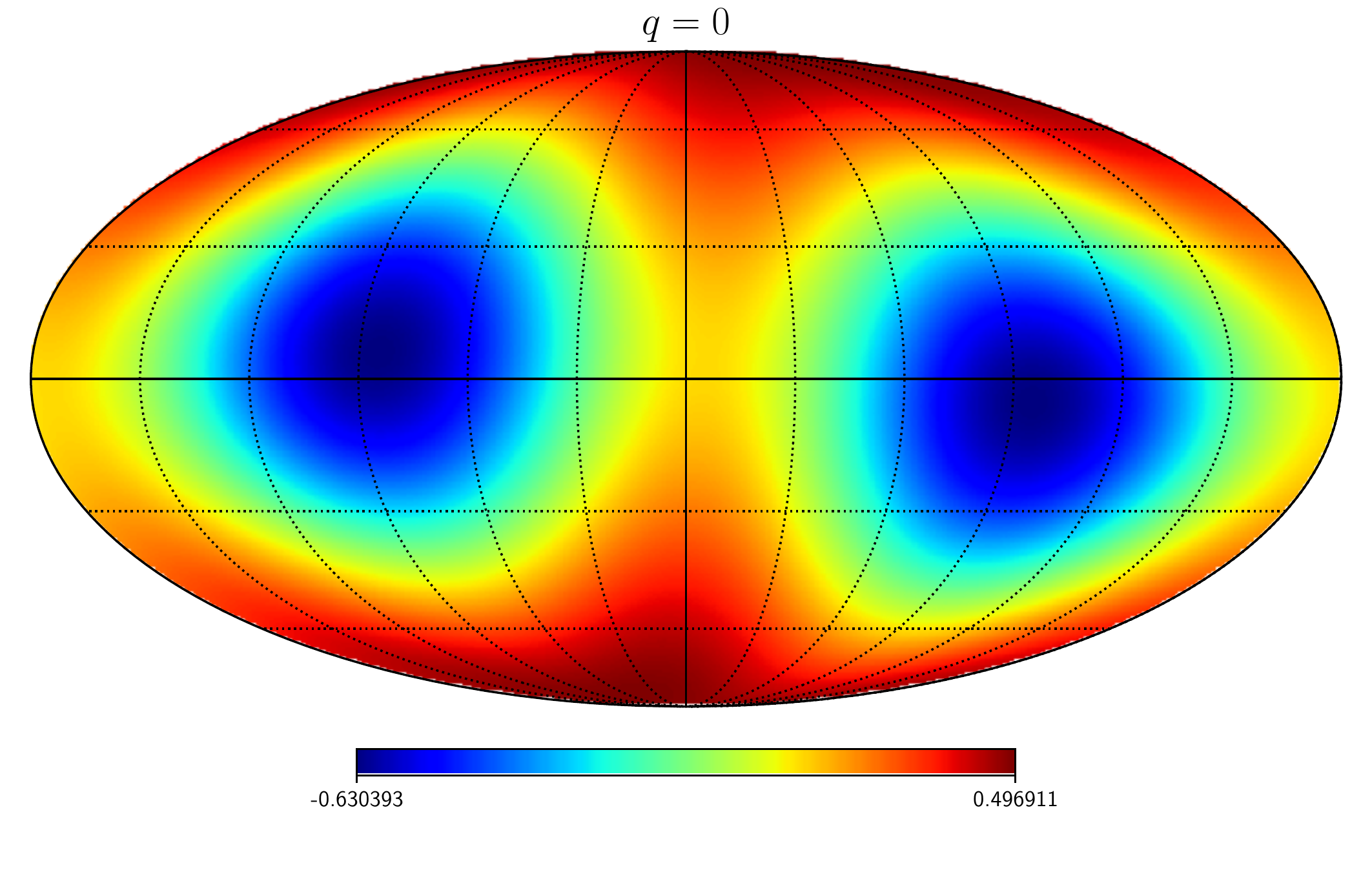}}
\subfigure{\includegraphics[width=0.45\textwidth]{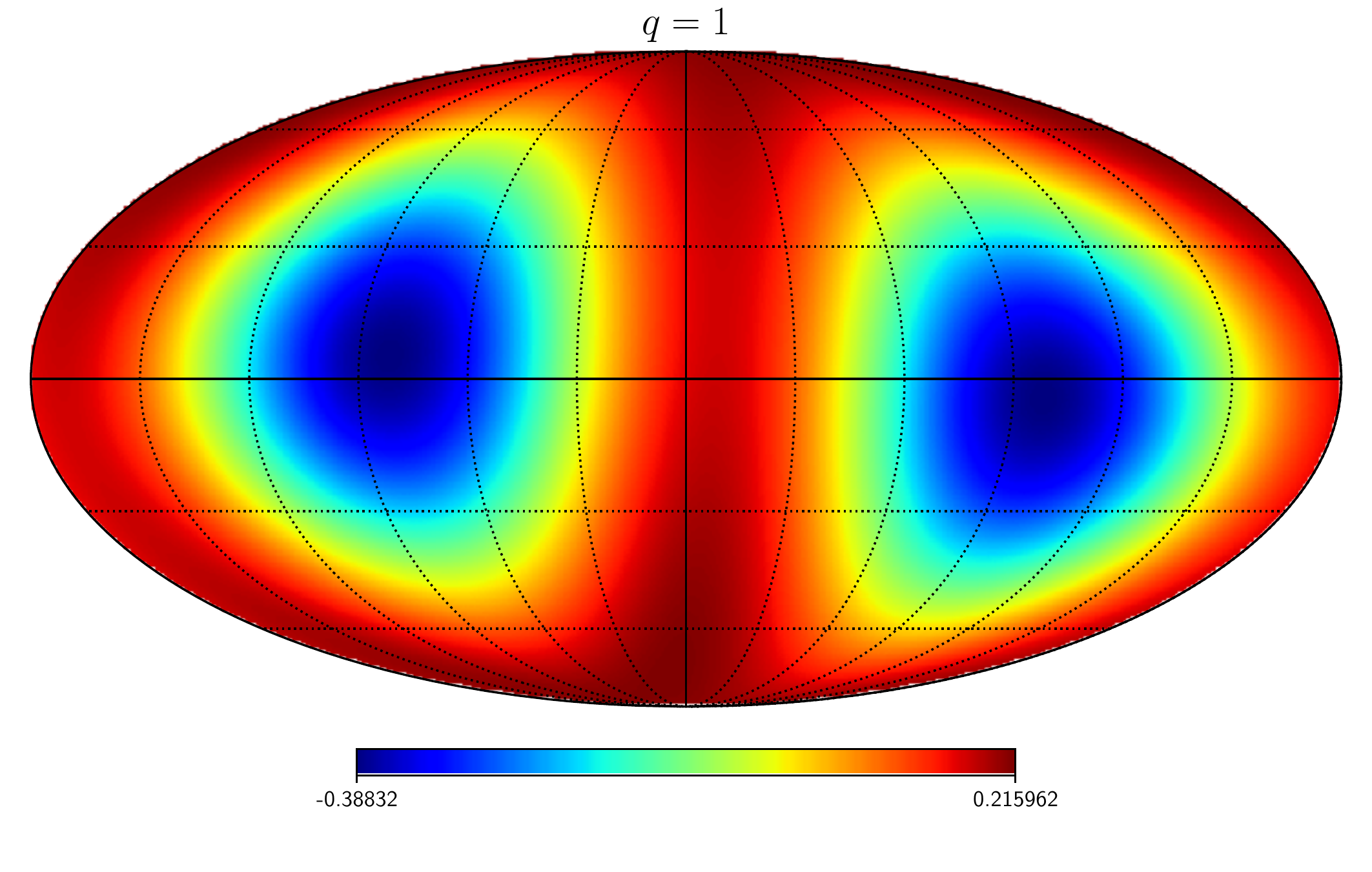}} \\
\subfigure{\includegraphics[width=0.45\textwidth]{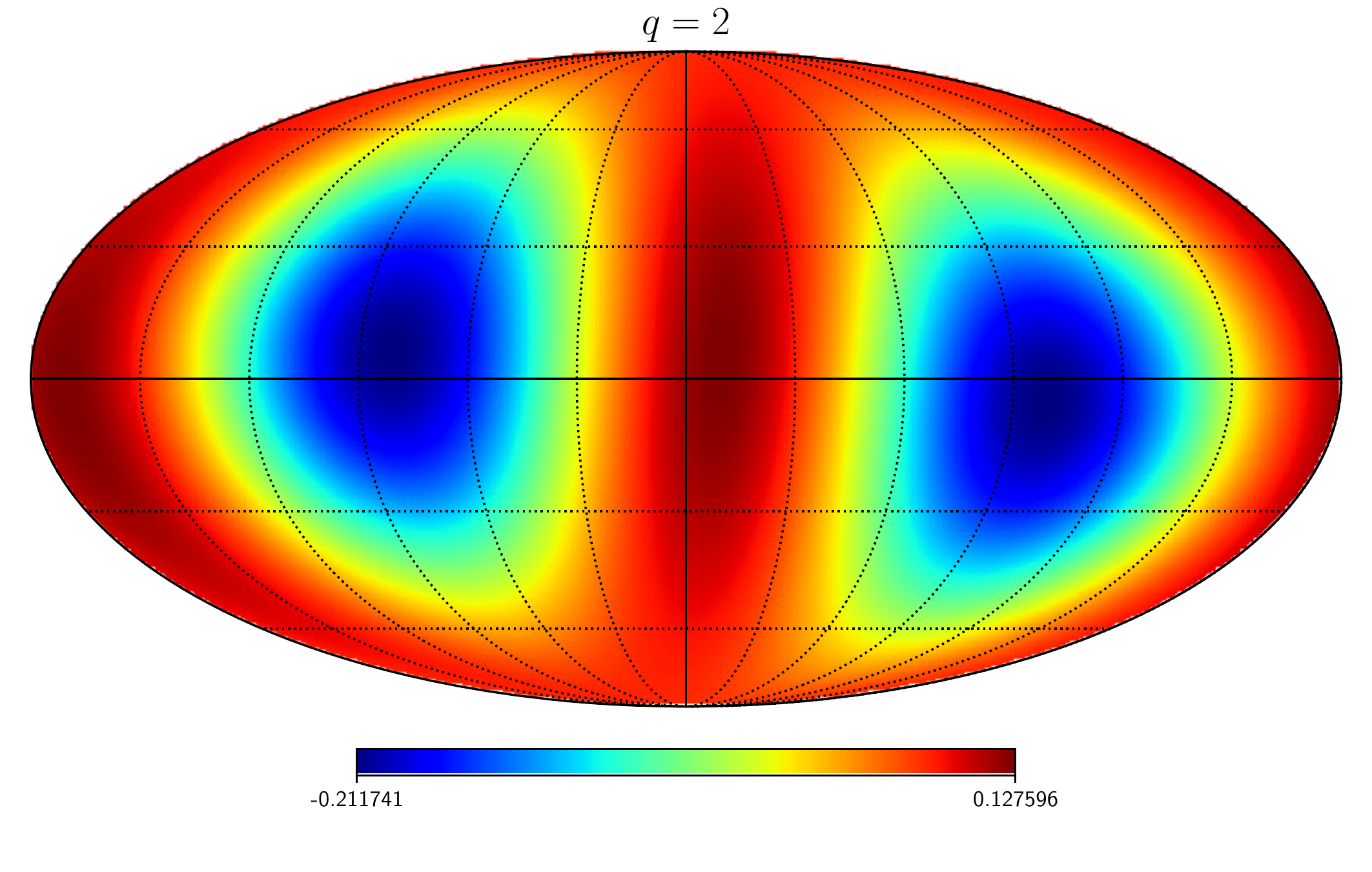}}
\caption{Mollweide projection of the best-fit power law quadrupole modulation $g_{2M}(k) \propto \mathcal{P}(k) (k/k_{\ast})^{q}$ ranging from $q=-2$ to $q=2$.}
\label{fig:powl}
\end{figure}

Eq.\eqref{eq:best-fit} can also be generalised to include a power-law quadrupole modulation:
\begin{align}
 		g_{2M}(k) = (4 \pi)^{-1/2} g_{2M \star} (k/k_{\ast})^{q} \mathcal{P}(k), \label{eq:genepowlaw}
 \end{align} 
with power $q$ and amplitude $g_{2M \star}$. Unlike the previously considered form \eqref{eq:newform}, this allows for the \emph{most} general quadrupole power law modulation which means that the directions of the two dipoles that a quadrupole can be factorised into need \emph{not} coincide. The maximum likelihood amplitudes and uncertainties of such modulations are given by the same expressions as those for the constant quadrupole modulation upon replacing the discretised version of $\mathcal{P}(k)$ (\emph{i.e.}, $\mathbf{p}$ in eqs.\eqref{eq:bfc} and \eqref{eq:bfce}) with a discretised version of $\mathcal{P}(k) (k/k_{\ast})^{q}$. The results are shown in Table~\ref{tab:powlawquad} and Figure~\ref{fig:powl}. 
Their statistical significance will be evaluated in Section~\ref{sec:statsig}.
\begin{table}[h!]
\begin{center}
\begin{tabular}{|l|r|r| l| r| r|}
\hline
$M$ & $ 10^2 \times \mathrm{Re} \, {g}_{2M \star}$ & $10^2 \times \mathrm{Im} \, {g}_{2M \star}$ & $M$ & $ 10^2 \times \mathrm{Re} \, {g}_{2M \star}$ & $10^2 \times \mathrm{Im} \, {g}_{2M \star}$ \\
\hline
\multicolumn{3}{|c|}{$q=-2$} & \multicolumn{3}{|c|}{$q=2$} \\
\hline
$0$ & $0.02\pm 0.12$ & & $0$ & $0.13\pm 0.12$ &\\
\hline
$1$ & $0.07\pm 0.09$ & $-0.02 \pm 0.09$ & $1$ & $0.00\pm 0.07$ & $-0.04 \pm 0.07$ \\
\hline
$2$ & $-0.06 \pm 0.09$ & $0.05 \pm 0.09$ & $2$ & $0.21 \pm 0.07$ & $0.07 \pm 0.07$ \\
\hline
\multicolumn{3}{|c|}{$q=-1$} & \multicolumn{3}{|c|}{$q=1$} \\
\hline
$0$ & $0.95\pm 0.39$ & & $0$ & $0.33\pm 0.21$ & \\
\hline
$1$ & $0.04\pm 0.25$ & $-0.30 \pm 0.25$ & $1$ & $0.02\pm 0.13$ & $-0.06 \pm 0.12$ \\
\hline
$2$ & $0.28 \pm 0.26$ & $-0.05 \pm 0.25$ & $2$ & $0.35 \pm 0.13$ & $0.11 \pm 0.12$ \\
\hline
\end{tabular}
\caption{Power-law quadrupole modulations of the form $g_{2M} \propto g_{2M \star} (k/k_{\ast})^{q} \mathcal{P}(k) $.}
\label{tab:powlawquad}
\end{center}
\end{table}

\begin{figure}[ht]
\centering
\subfigure{\includegraphics[width=0.48\textwidth]{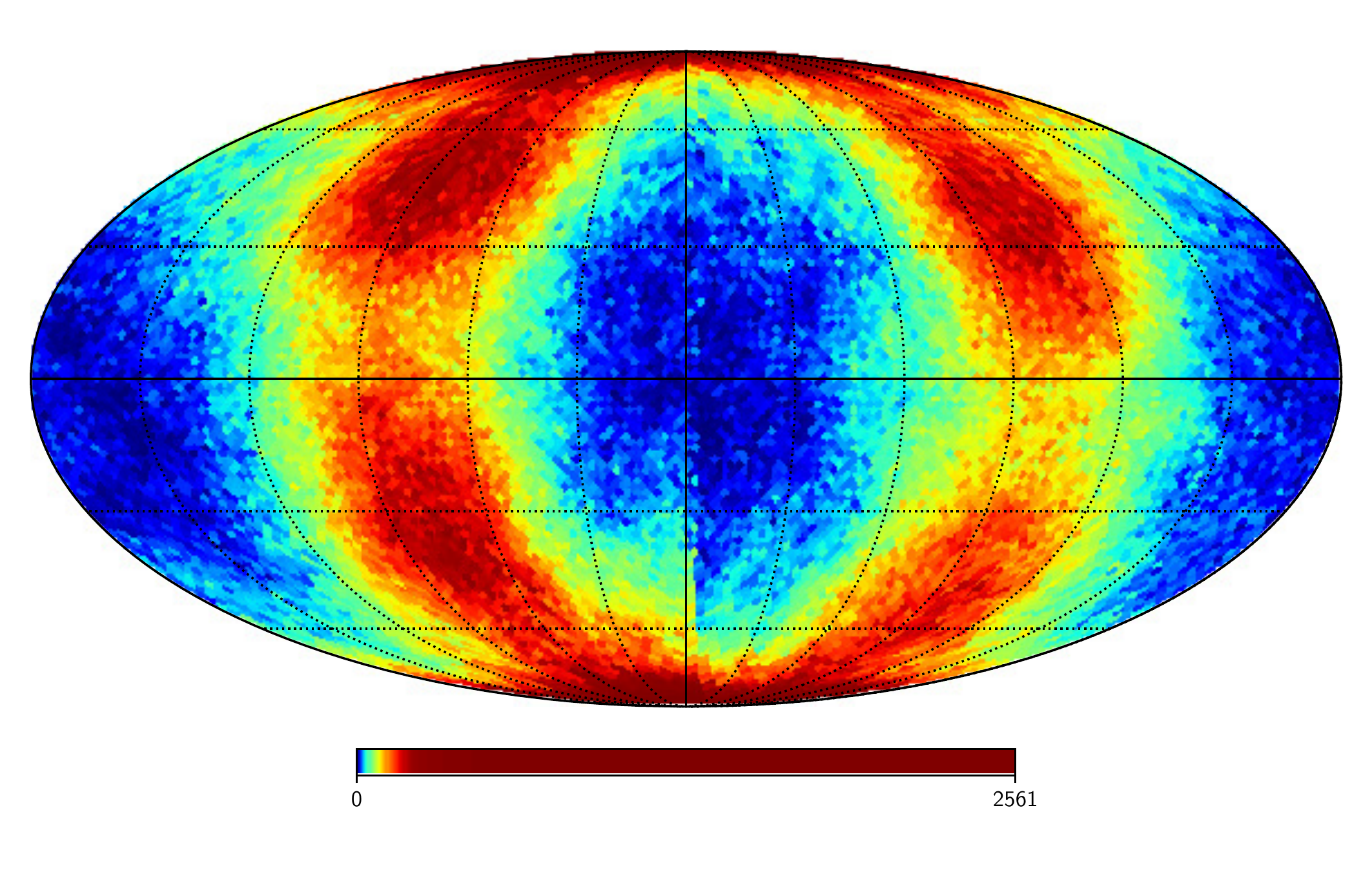}}
\subfigure{\includegraphics[width=0.48\textwidth]{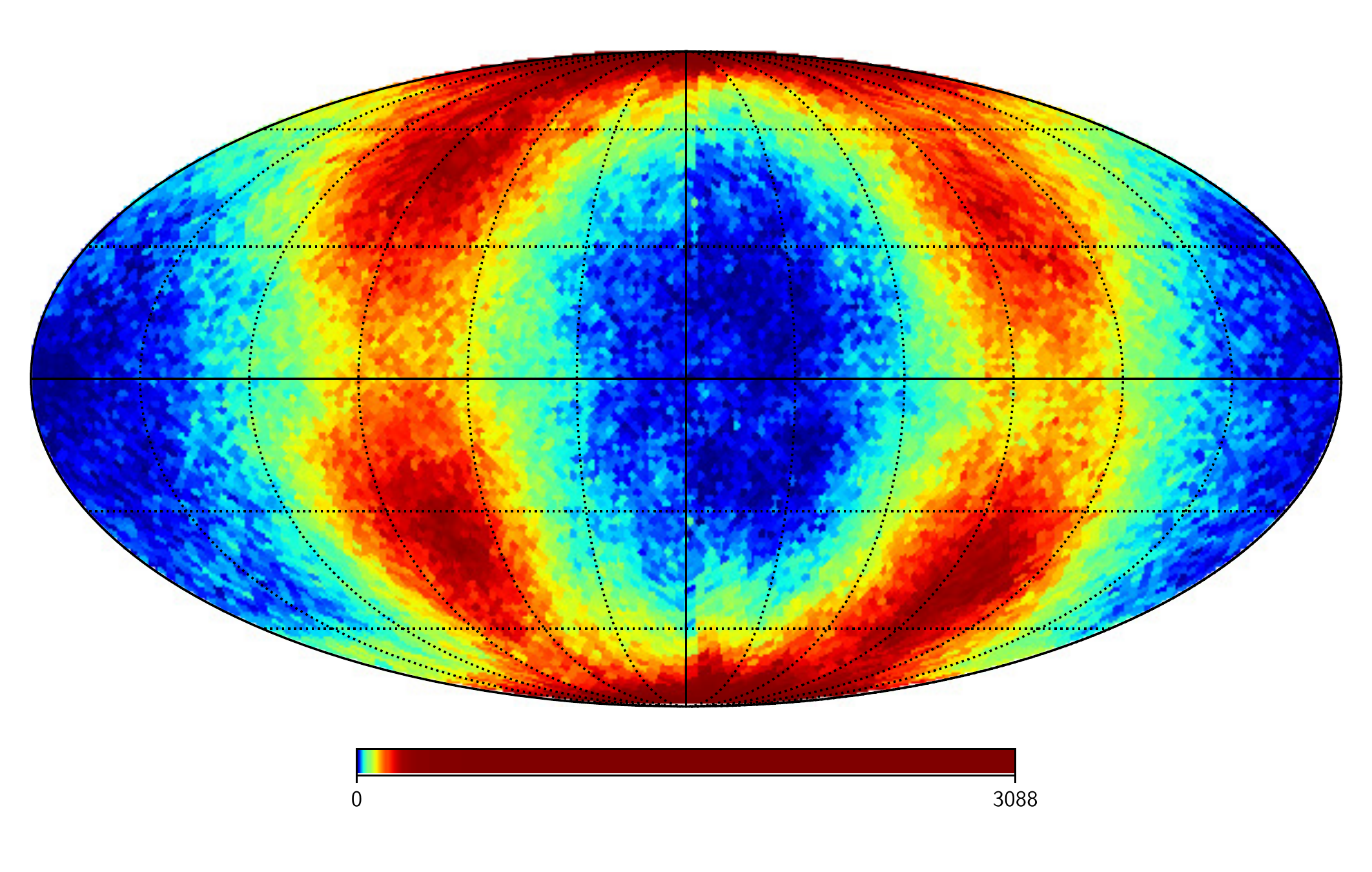}}
\caption{Mollweide projection of equalised histograms of possible directions $\hat{\mathbf{n}}_1$ (left) and $\hat{\mathbf{n}}_2$ (right) of the most general \emph{scale-independent} quadrupole modulation.}
\label{fig:dirhist2}
\end{figure}

The most general quadrupole is described by two directions, $\hat{\mathbf{n}}_1$ and $\hat{\mathbf{n}}_2$, and an amplitude $g_{A}$ and is of the form $g_{A}(\hat{\mathbf{k}}\cdot \mathbf{n}_1)(\hat{\mathbf{k}}\cdot \hat{\mathbf{n}}_2)$ where $\hat{\mathbf{k}} = (\theta,\phi)$. This  has previously been studied by ref.\cite{Ramazanov:2016gjl}. The relation between the quadrupole modulations $g_{2M}$ and the direction is found by decomposing this form into spherical harmonics
\begin{align}
		g_{2M} = g_{A} \int \mathrm{d}\Omega\, (\hat{\mathbf{k}} \cdot \hat{\mathbf{n}}_1) (\hat{\mathbf{k}} \cdot \hat{\mathbf{n}}_2) Y^{\ast}_{2M}(\theta,\phi). \label{eq:genquad}
\end{align}
The best-fit quadrupole modulations for the different powers $q$ are displayed in Fig.~\ref{fig:powl}, although they are not particularly informative as there are large uncertainties in the directions of $\hat{\mathbf{n}}_1$ and $\hat{\mathbf{n}}_2$ (as seen for the case of the constant quadrupole modulation $q=0$ in Fig.~\ref{fig:dirhist2}). 

\subsection{Non-parametric quadrupole modulations}
\begin{figure}[ht]
\centering
\subfigure{\includegraphics[width=0.48\textwidth]{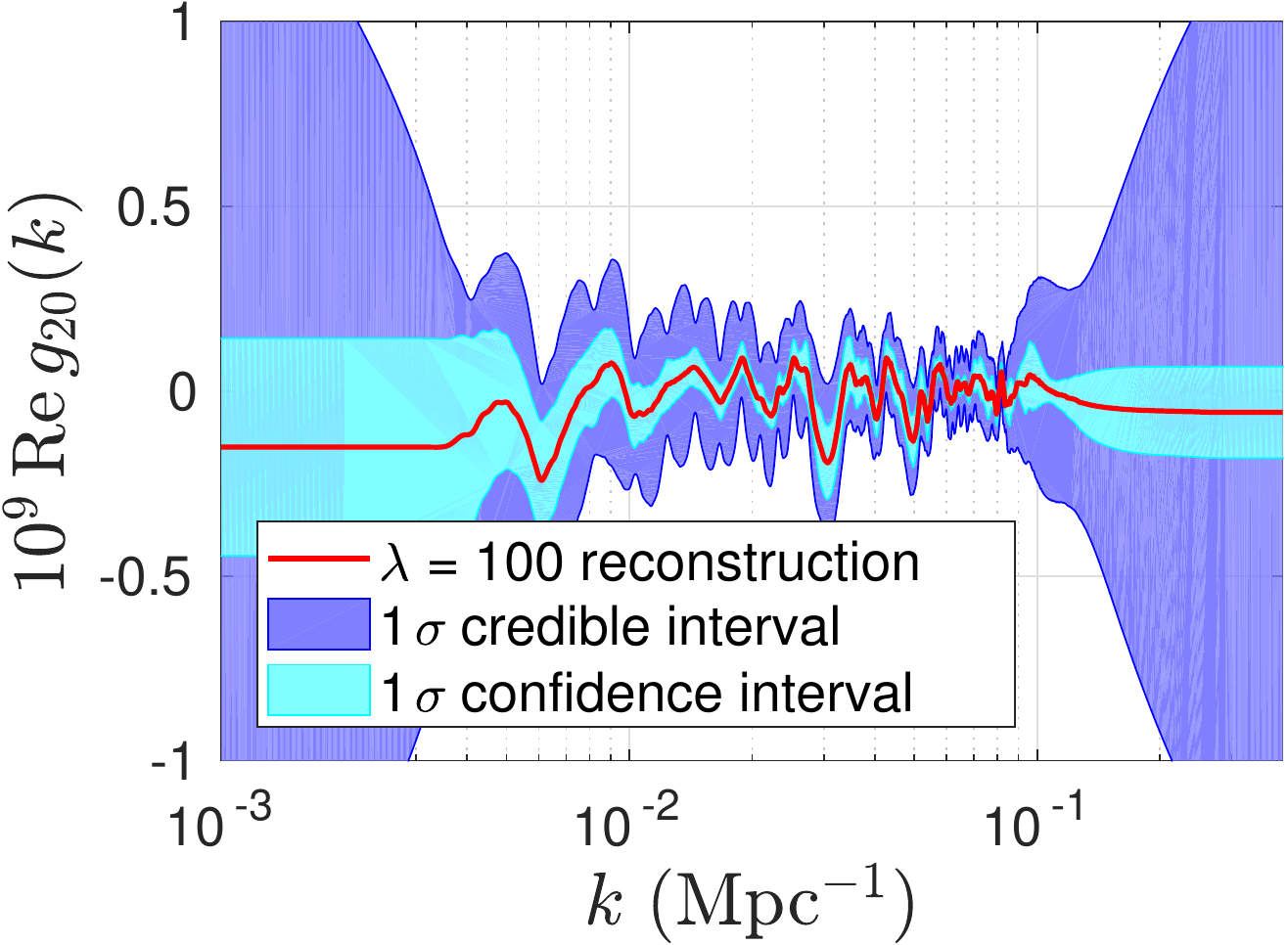}
\label{fig:m0}
} \\
\subfigure{\includegraphics[width=0.48\textwidth]{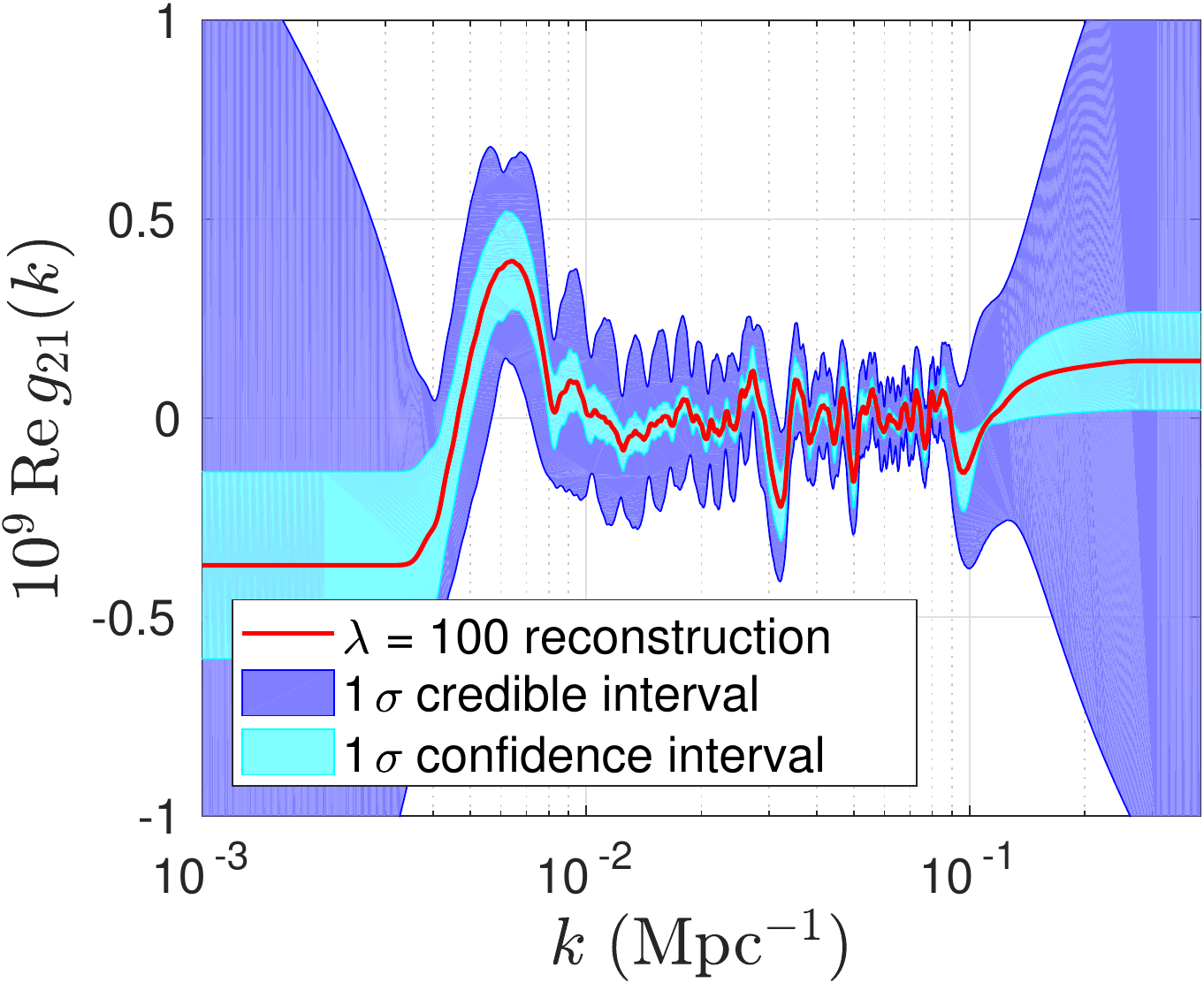}
\label{fig:m1}
}
\subfigure{\includegraphics[width=0.48\textwidth]{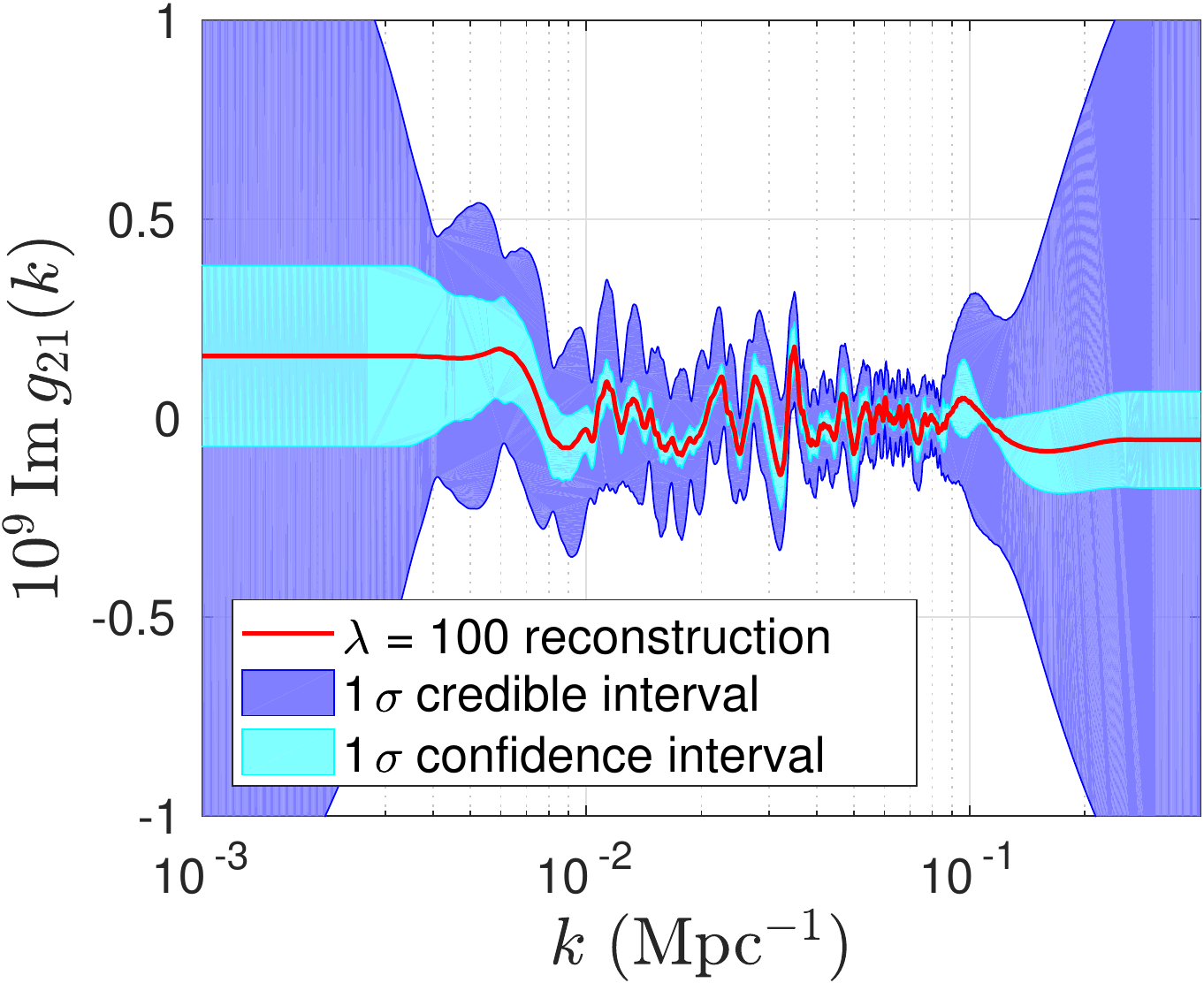}
\label{fig:m1im}
}
\subfigure{\includegraphics[width=0.48\textwidth]{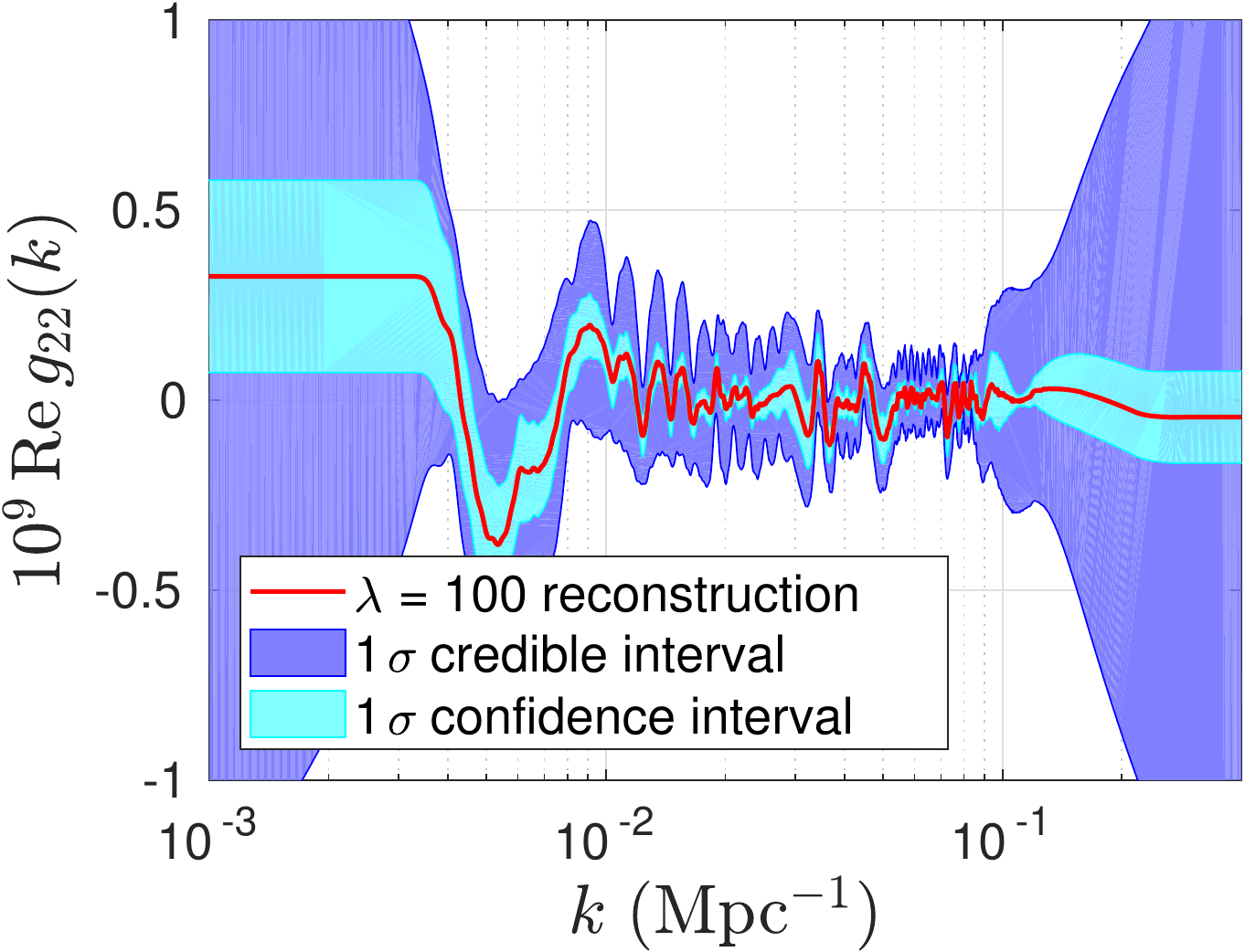}
\label{fig:m2}
}
\subfigure{\includegraphics[width=0.48\textwidth]{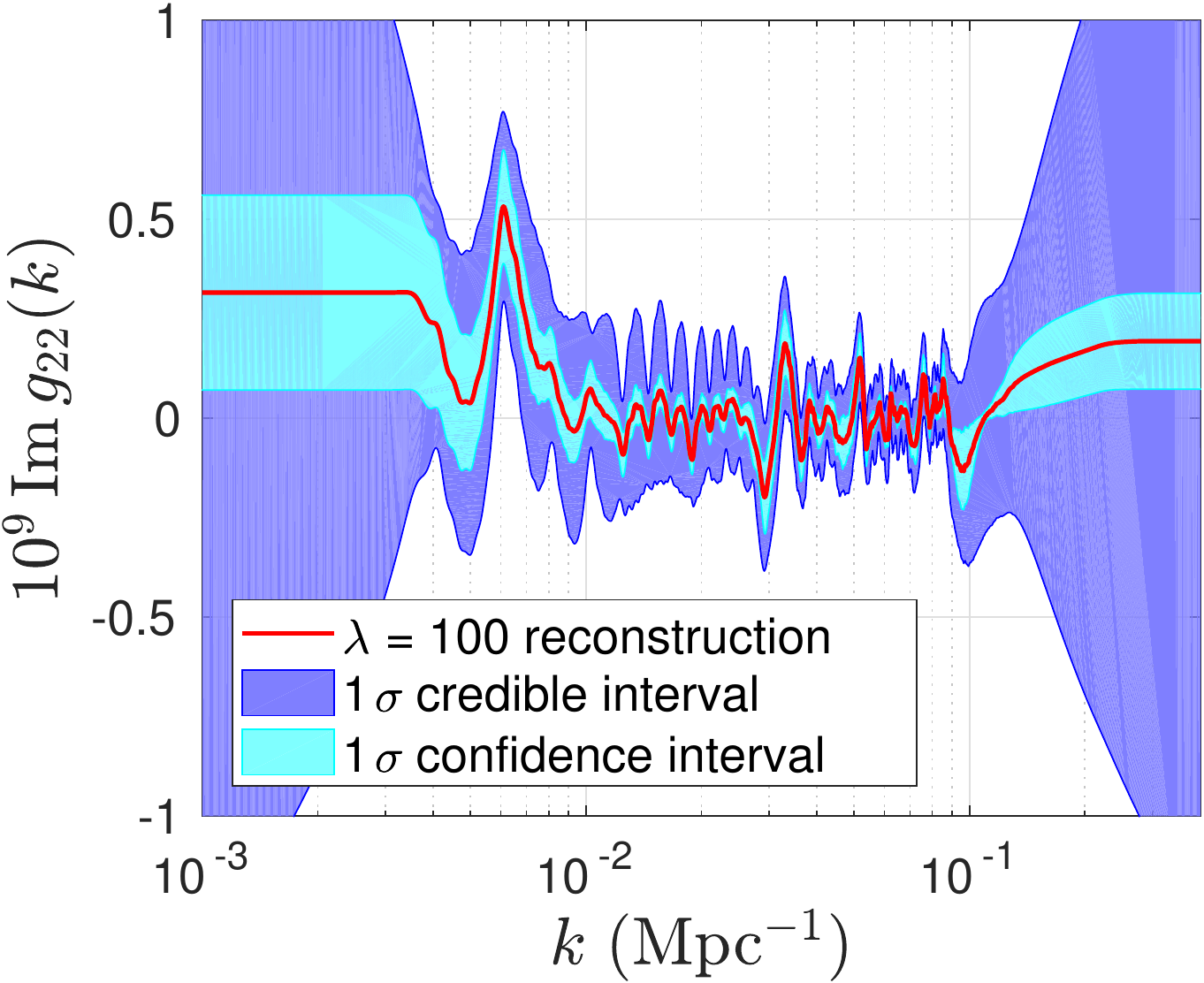}
\label{fig:m2im}
}
\caption{Reconstruction (full red line) of the quadrupole modulation from masked and binned ($\Delta \ell = 30$) PR2--2015 SMICA temperature data in the multipole range $30\leq \ell \leq 1200$. Purple bands and cyan bands indicate the $1\sigma$ credible intervals and $1\sigma$ confidence intervals, respectively. The regularisation parameter was set to $\lambda = 100$. Note the feature at $k \sim 6 \times 10^{-3} \, \mathrm{Mpc}^{-1}$ in $\mathrm{Im}\, g_{22}$.}
\label{fig:mainres}
\end{figure}

We now present new results concerning a possible scale-\emph{dependent} quadrupole modulation of the primordial power spectrum. The quadrupole modulation reconstructed  using eq.\eqref{eq:recon} from the masked PR2--2015 SMICA temperature map with uncertainties modelled by the masked FFP9 simulations is shown in Fig.~\ref{fig:mainres} for $\lambda = 100$ and in Fig.~\ref{fig:olambda5000} for $\lambda = 5000$. Plots for other choices of $\lambda$ are shown in Appendix~\ref{sec:olambda}.

These non-parametric reconstructions suggest the presence of a feature at $k \sim 0.006 \, \mathrm{Mpc}^{-1}$ in $\mathrm{Im} \, g_{22}$ with respect to both Bayesian and frequentist uncertainties. The dominant contributions to this feature are from the BipoSH coefficients $\mathrm{Im}\, A^{22}_{74~74}$ and $\mathrm{Im}\, A^{22}_{74~76}$ which are themselves bins of BipoSH coefficients in the multipole range $60 \leq \ell \leq 90$. We now proceed to investigate the direction of this possible feature and its relation to known regions of interest. 

One such is the hemispherical asymmetry, or dipole modulation of the CMB sky, first noted in ref.\cite{Eriksen:2003db}, which adds to the statistically isotropic CMB sky $T_{\mathrm{iso}}(\hat{\mathbf{n}})$ a dipole $A \hat{\mathbf{p}}\cdot \hat{\mathbf{n}}$ resulting in a statistically anisotropic CMB sky
\begin{align}
		T(\hat{\mathbf{n}}) = T_{\mathrm{iso}}(\hat{\mathbf{n}}) (1 + A \hat{\mathbf{p}}\cdot \hat{\mathbf{n}} )
\end{align}
where $\hat{\mathbf{p}}$ is the preferred direction and $A$ is the amplitude of the modulation. Proposed models \cite{Erickcek:2008sm,Dai:2013kfa} invoke a large amplitude super-horizon perturbation of an additional field during inflation, or any mechanism that directionally modulates the optical depth of reionisation, scalar spectral index or tensor amplitude. 
Different analyses \cite{Ade:2015hxq} have converged on there being a dipole modulation in the range $2 \leq \ell \leq 64$ at $\sim3 \sigma$ significance. We are sensitive to the latter half of this range. Higher multipole ranges are consistent with there being no dipole modulation at $2 \sigma$ level \cite{Ade:2015hxq}, although if it persists beyond $\ell \sim 64$, it is potentially observable in B mode polarisation \cite{Mukherjee:2015wra}. The best-fit amplitude and direction from a BipoSH analysis of the PR2--2015 SMICA map in the quoted range give $A= 0.069 \pm 0.022$ and direction $(l,b) = (228^{\circ},-18^{\circ})\pm 30^{\circ}$. Another region of interest is the well-known CMB dipole, which is interpreted as being a Doppler boost due to our motion relative to the `CMB rest frame' in the direction $(264^{\circ},48^{\circ})$ \cite{Ade:2015hxq}. The Doppler boost itself induces a statistical anisotropy due to the aberration of light \cite{Challinor:2002zh} and this too has a BipoSH description \cite{Mukherjee:2013zbi,Aluri:2015tja}.

\begin{figure}[ht]
\centering
\includegraphics[width=0.48\textwidth¤]{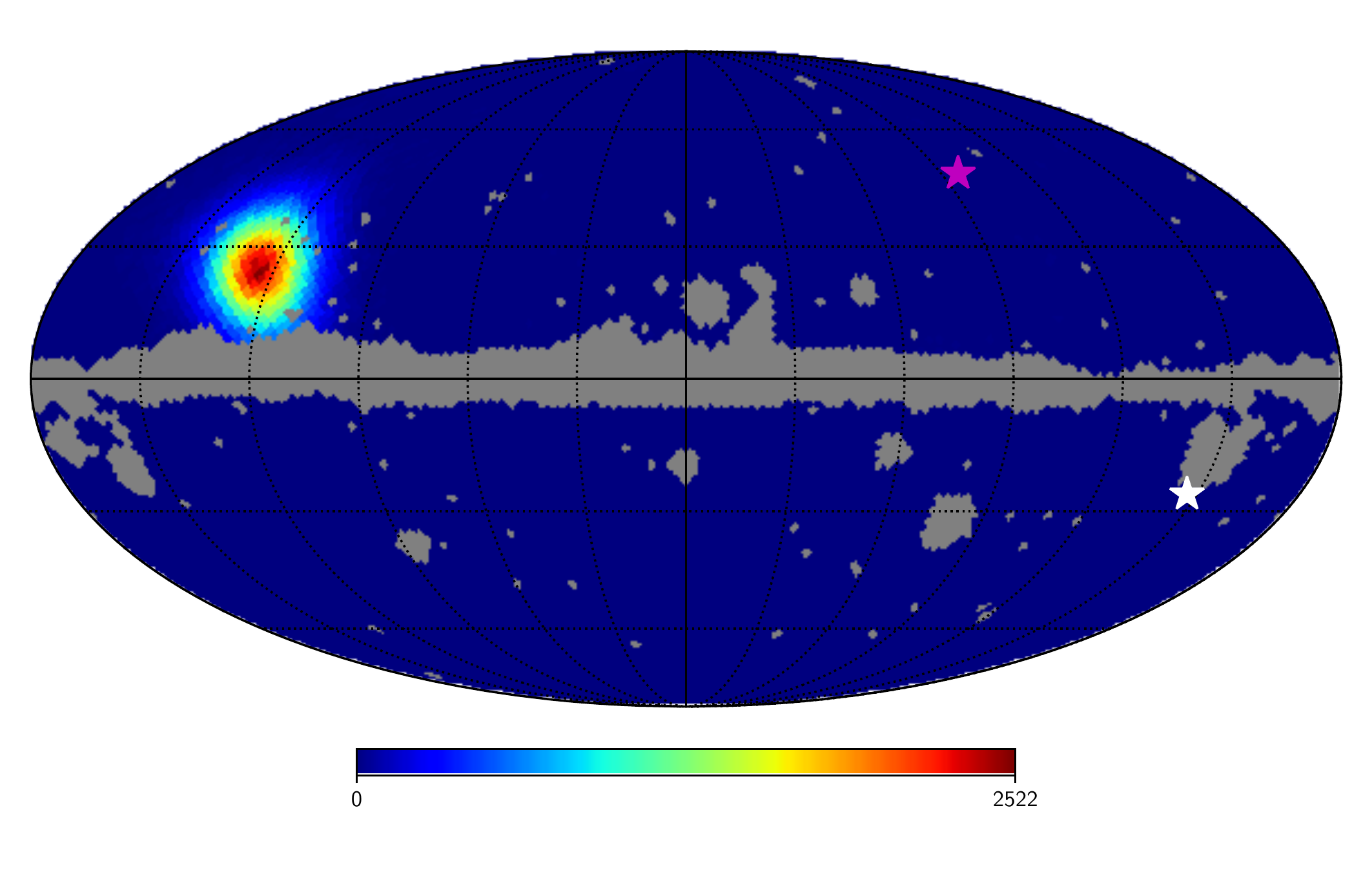}
\includegraphics[width=0.48\textwidth]{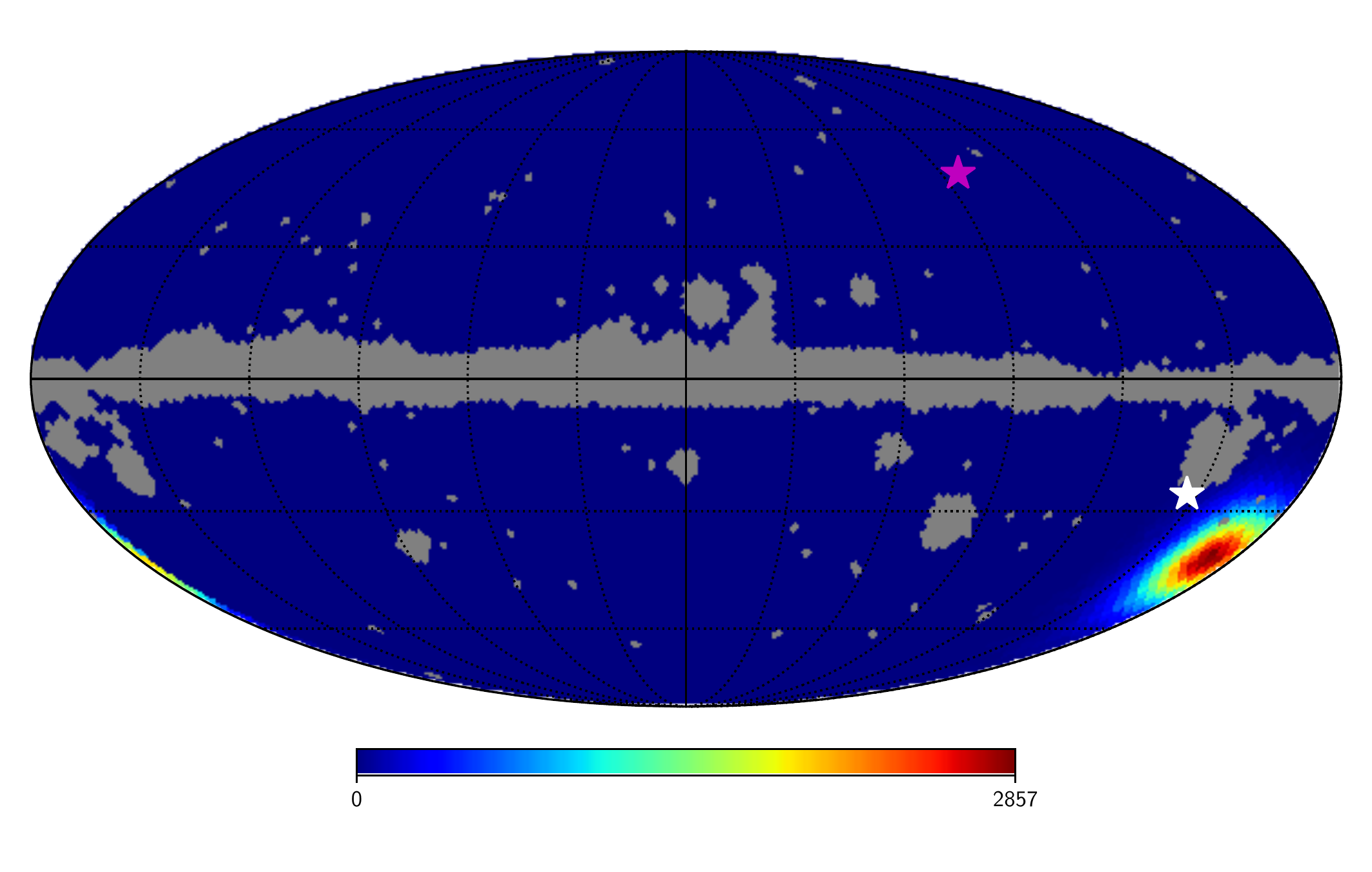}
\caption{Histograms of the posterior distribution of the direction $\hat{\mathbf{n}}$ for the case of a hot ($10^{9} g_{A} = 0.75\pm 0.22$) constant quadrupole modulation (left panel) and a cold ($10^{9} g_{A} = -0.82 \pm 0.21$) modulation (right panel) in the wave number range $0.005 \leq k/\mathrm{Mpc}^{-1}\leq 0.008$. The SMICA mask (in grey) is superimposed for comparison. The magenta and white stars indicate the directions of the CMB dipole and the hemispherical asymmetry, respectively. The best-fit directions are $({128^{\circ}}^{+14}_{-14},{25^{\circ}}^{+11}_{-9})$ for positive amplitude and $({191^{\circ}}^{+15}_{-14},{-41^{\circ}}^{+10}_{-11})$ for negative amplitude. }
\label{fig:featdir}
\end{figure}

When fitting the quadrupole modulation of eq.\eqref{eq:genquad} to the data over a limited wave number range, $0.005 \leq k/\mathrm{Mpc}^{-1} \leq 0.008$, assuming just one direction $\hat{\mathbf{n}}_1 = \hat{\mathbf{n}}_2$ and a constant amplitude $g_{A}$, we find that the posterior probability distribution of the direction is bimodal.
Histograms of the posterior distribution of the direction for both cases are shown in Fig.~\ref{fig:featdir}. Details of the directions are provided in Table~\ref{tab:modes} where angular distances to the CMB dipole and the hemispherical asymmetry (as determined in ref.\cite{Ade:2015hxq}) are calculated.

\begin{table}[ht]
\begin{center}
\begin{tabular}{|l|c|c|c|}
\hline
\multicolumn{2}{|l|}{For $k=0.005$-$0.008 \, \mathrm{Mpc}^{-1}$: } & \multicolumn{2}{|c|}{Angular distances to:} \\
\hline
Amp. $10^{9} g_A$ & Direction $(l,b)$ & CMB dipole $(264^{\circ},48^{\circ})$ & Hemisph. asym. $(213^{\circ},-26^{\circ})$ \\
\hline
$0.76\pm 0.22$  & $({128^{\circ}}^{+14}_{-14},{25^{\circ}}^{+11}_{-9})$ & $ 97^{\circ}$ & $97^{\circ}$ \\
\hline
$-0.82 \pm 0.21$ & $({191^{\circ}}^{+15}_{-14},{-41^{\circ}}^{+10}_{-11})$ & $110^{\circ}$ & $24^{\circ}$ \\
 \hline
\end{tabular}
\caption{Details of the two modes of the posterior distribution of the quadrupole modulation direction shown in Fig.~\ref{fig:featdir}. The first mode is for the case of a hot modulation. Angular distances to the directions of the CMB dipole and the hemispherical asymmetry are also indicated.}
\label{tab:modes}
\end{center}
\end{table}

In the case of a `hot' quadrupole modulation ($g_A>0$), the direction of the quadrupole modulation is, within the $\sim 10^{\circ}$ uncertainty, roughly perpendicular to the best-fit direction of hemispherical asymmetry. Within this uncertainty, it is also perpendicular to the best-fit direction of the CMB dipole. For the case of a `cold' quadrupole modulation ($g_A<0$) the coincidences are less pronounced, however the direction is approximately aligned with that of the hemispherical asymmetry within the uncertainty. 
It is quite unexpected that the direction of the quadrupole modulation should be related to that of either the CMB dipole or the hemispherical asymmetry.

We will relegate the discussion of features and their statistical significance to the next section where all five components will be compressed to a single amplitude $g_{2}(k)$.

\begin{figure}[ht]
\centering
\subfigure{\includegraphics[width=0.48\textwidth]{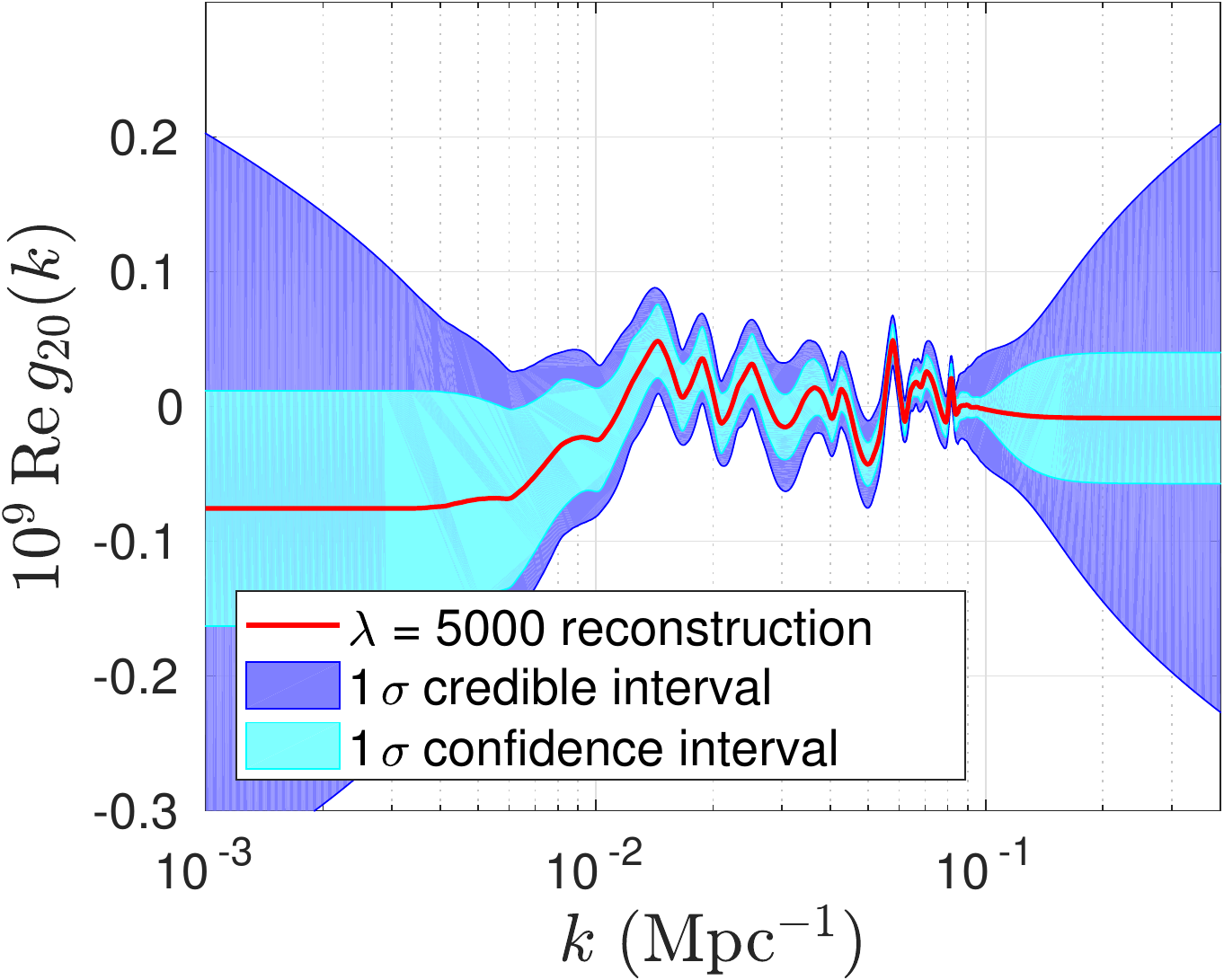}
\label{fig:m05000}
} \\
\subfigure{\includegraphics[width=0.48\textwidth]{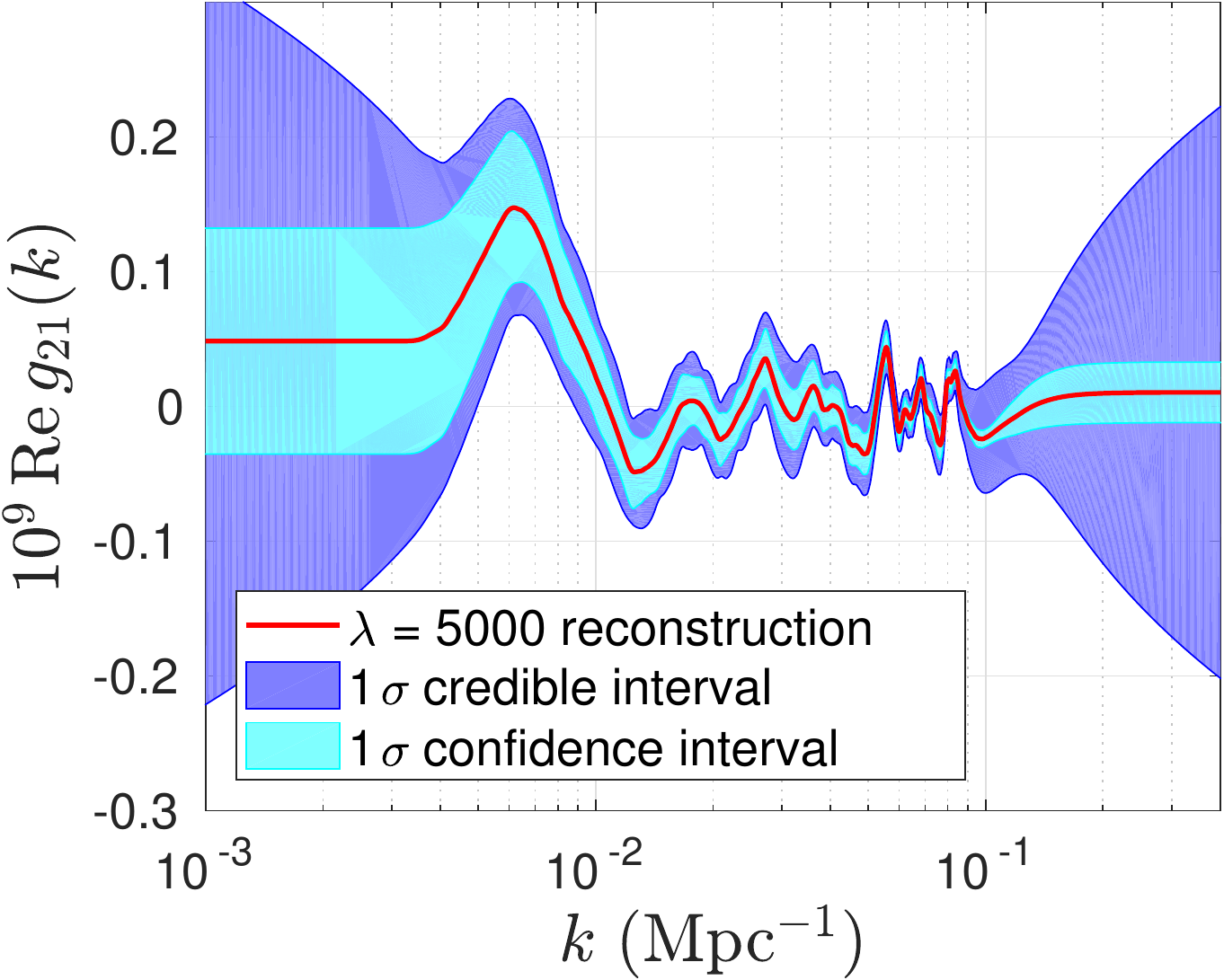}
\label{fig:m15000}
}
\subfigure{\includegraphics[width=0.48\textwidth]{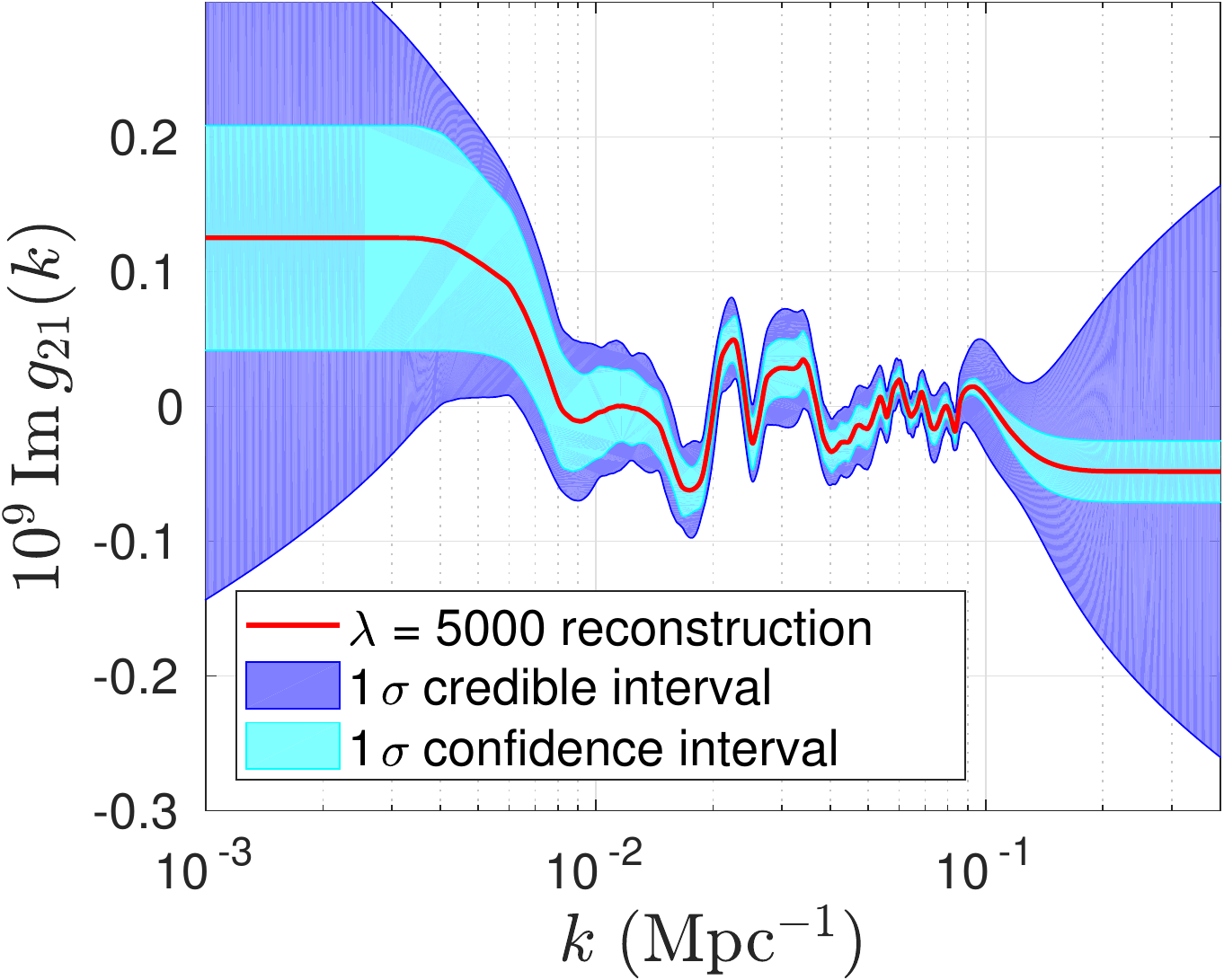}
\label{fig:m1im5000}
}
\subfigure{\includegraphics[width=0.48\textwidth]{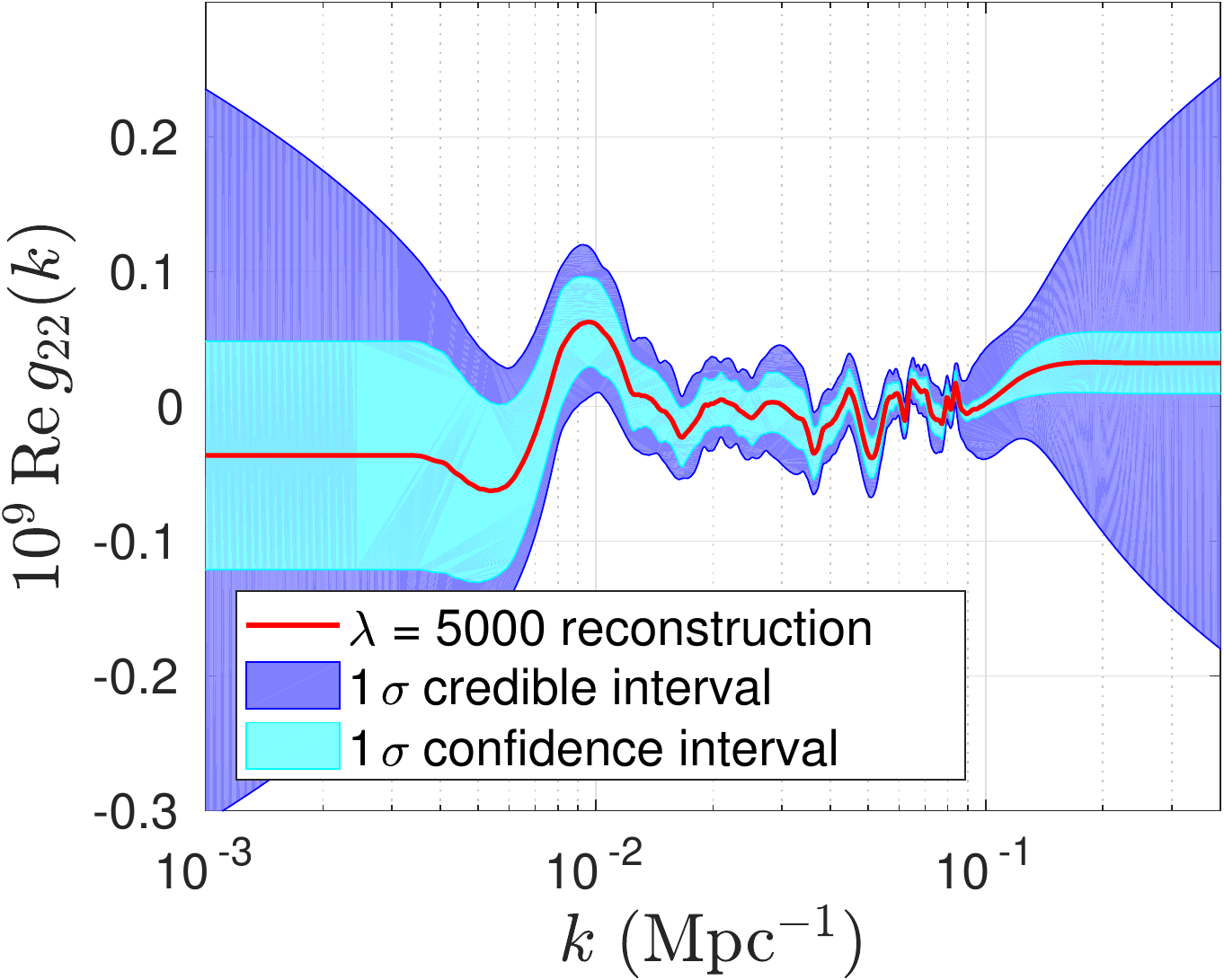}
\label{fig:m25000}
}
\subfigure{\includegraphics[width=0.48\textwidth]{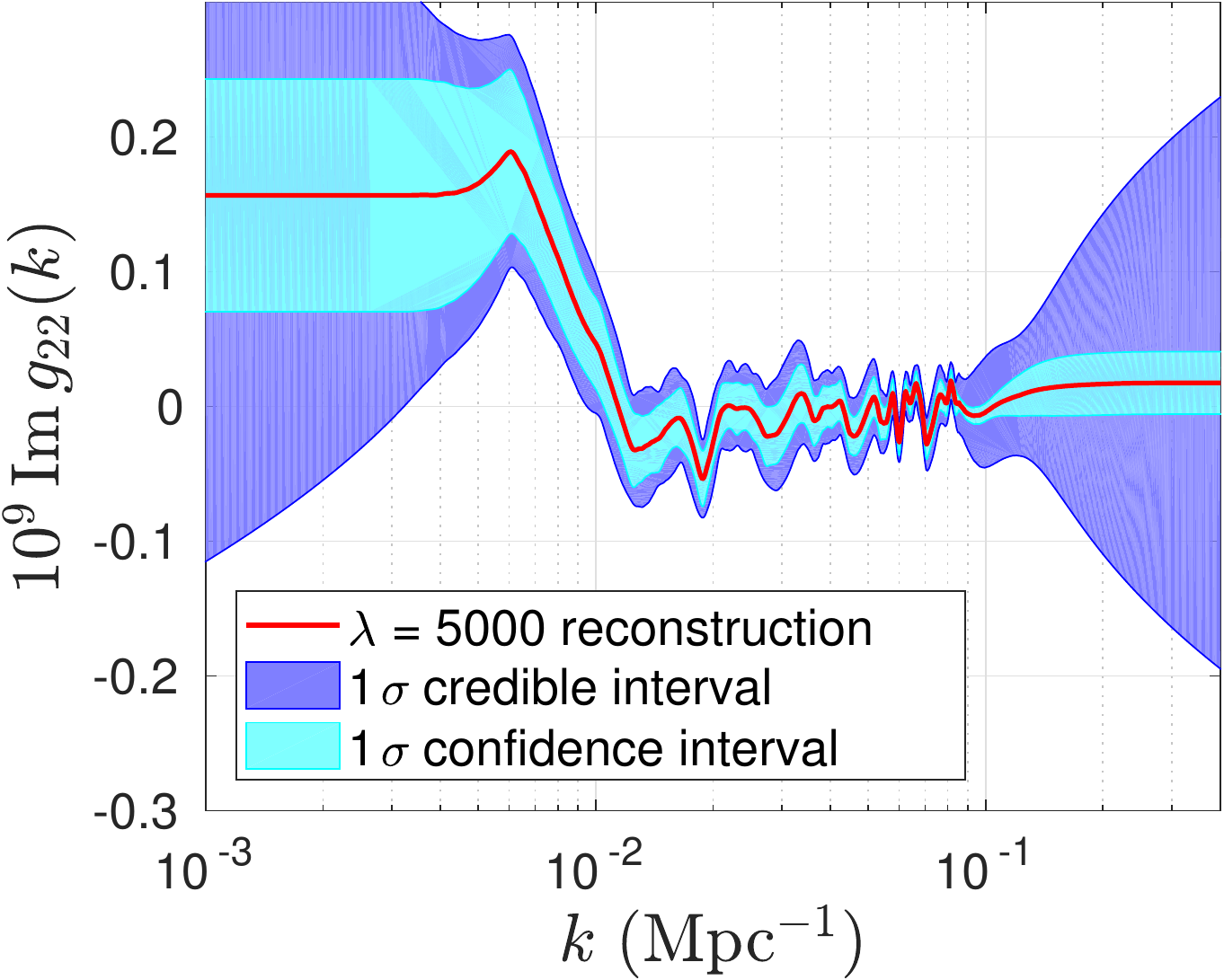}
\label{fig:m2im5000}
}
\caption{Reconstruction (full red line) with credible intervals (purple band) and confidence intervals (cyan band) for $\lambda = 5000$.}
\label{fig:olambda5000}
\end{figure}

\subsection{The diagonal approximation}

When forming an estimator for $g_{2M}$ it is common for computational
reasons to neglect in the Fisher matrix the off-diagonal elements of
$\mathbf{C}^{-1}$, the inverse of the covariance $C_{\ell m, \ell^\prime
m^\prime}\equiv \langle b_{\ell m} b^{\ast}_{\ell^\prime m^\prime}
\rangle$ of the masked spherical harmonic coefficients
\cite{Kim:2013gka,Ade:2015hxq}. To enable comparison with these results we
make here the equivalent approximation of truncating the BipoSH covariance
matrix $\mathbf{\Sigma}$ to its diagonal alone. In this case we find
\emph{no} evidence of a constant quadrupole modulation nor one scaling with wave number as a power law. However when fitting a quadrupole modulation to the anomalous wave number range $0.005$-$0.008 \, \mathrm{Mpc}^{-1}$ in the diagonal approximation, the coincidences with the hemispherical asymmetry and CMB dipole are even more pronounced. The details are given in Appendix~\ref{sec:diagap}.

\section{Statistical significance of a quadrupole modulation} \label{sec:statsig}
\subsection{Power law quadrupole modulations}

To test the statistical significance of power-law quadrupole modulation
\begin{equation}
		g_{2M}(k) = (4 \pi)^{-1/2} g_{2M \star} (k/k_{\ast})^{q} \mathcal{P}(k),
\end{equation}
we calculate the overall amplitude
\begin{equation}
		g_{2} = \sqrt{\frac{1}{5} \sum_{M} |g_{2M \star}|^2} \label{eq:g2}
\end{equation}
for the components $\hat{g}_{2M \star}$ that best fit the data and compare it with the estimates of $g_2$ calculated from $10^4$ realisations of noise modelled by the covariance matrix $\mathbf{\Sigma}$ derived from the FFP9 simulations. Each realisation is a set of BipoSH coefficients $\mathbf{A}^{2M}$ from which $g_{2M \star}$ can be deduced using eq.\eqref{eq:bfc}. These are then used to compute $g_{2}$. The p-value is equal to the fraction of realisations with $g_2$ greater than or equal to the value $\hat{g}_{2}$ obtained from data and is shown in Table~\ref{tab:g2}. The power-law cases $-1 \leq q \leq 2$ are unusual at the $\sim 2\sigma$ significance level, with $q=0$ the most significant at $2.16\sigma$. 
\begin{table}[ht]
\begin{center}
\begin{tabular}{|l|r|r|r|r|r|}
\hline
$q$ & -2 & -1 & 0 & 1 & 2 \\
\hline
p-value & 0.91 & 0.08 & 0.03 & 0.05 & 0.06 \\
\hline
\end{tabular}
\caption{P-value test of $g_2$ \eqref{eq:g2} calculated for power law quadrupole modulations from data and compared with $10^4$ noise realisations.}
\label{tab:g2}
\end{center}
\end{table}

\subsection{Non-parametric quadrupole modulations}

\begin{figure}[ht]
\centering
\subfigure{\includegraphics[width=0.48\textwidth]{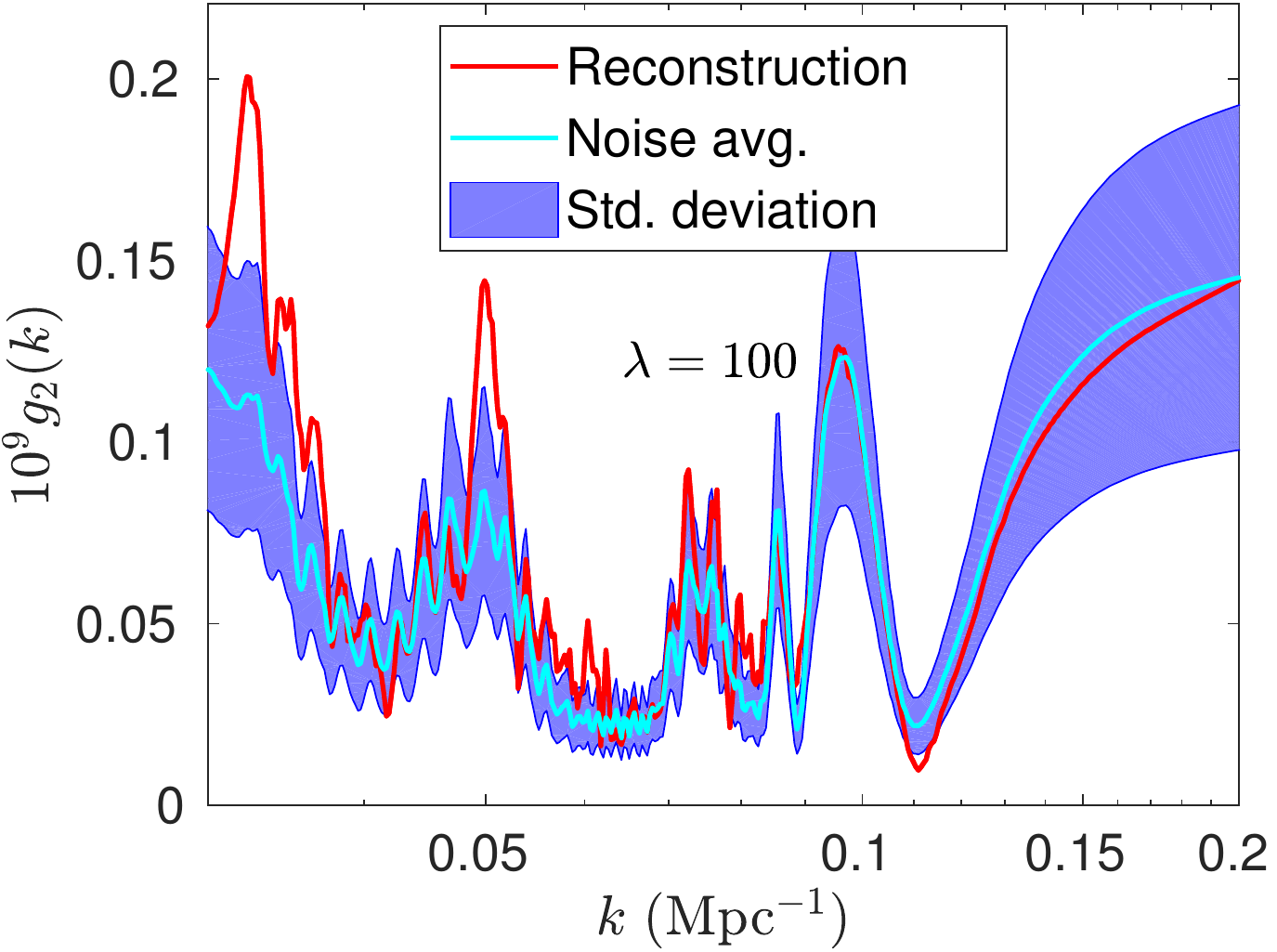}}
\subfigure{\includegraphics[width=0.48\textwidth]{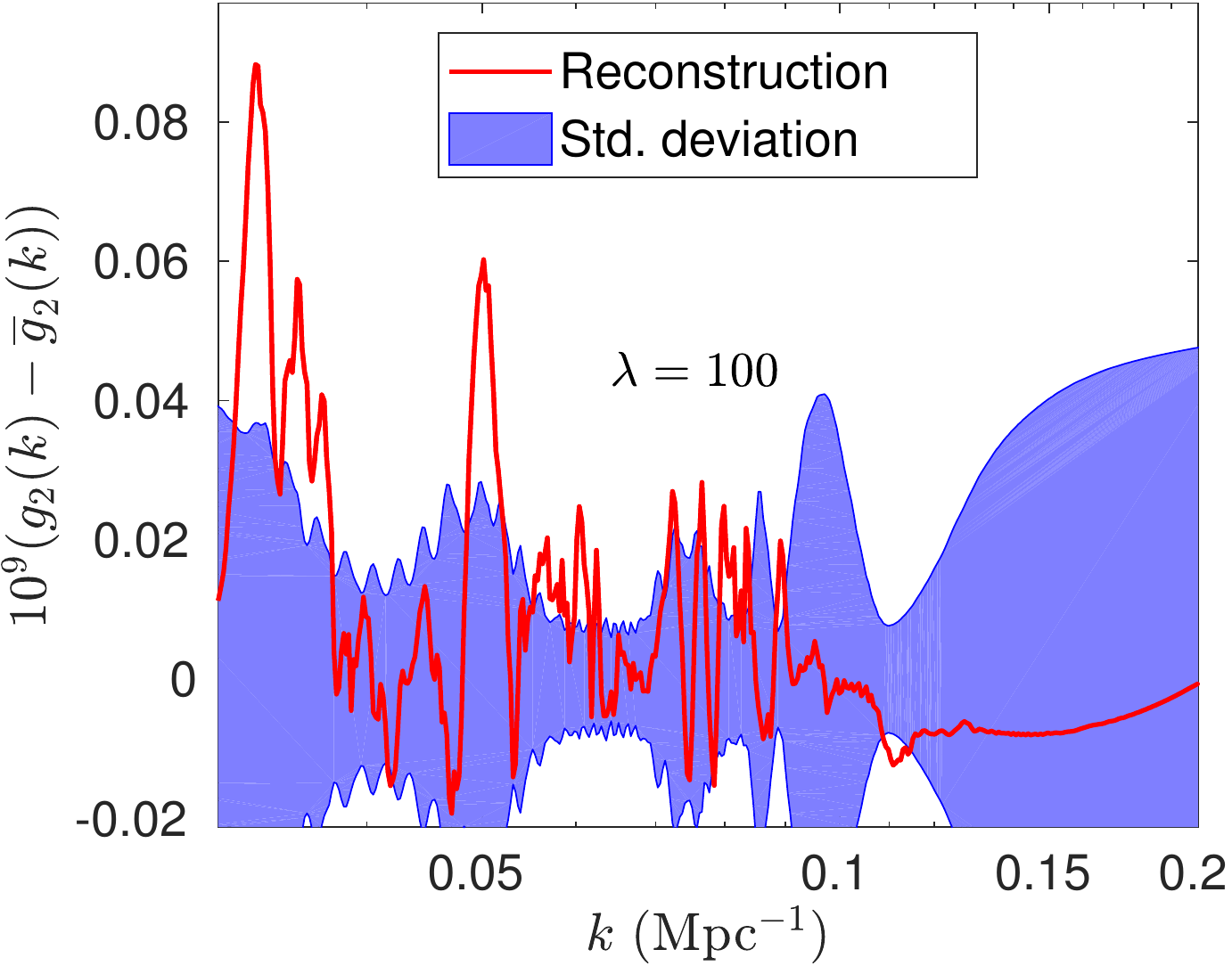}}
\caption{The left panel shows the amplitude $g_{2}(k)$ for the reconstruction from data (red line) and its mean (cyan line) and standard deviation (purple band) of $g_2(k)$ for the noise realisations. In the right panel, this mean $\overline{g}_{2}(k)$ has been subtracted from the $g_{2}(k)$ reconstructed from data.}
\label{fig:g2rec}
\end{figure}

\begin{figure}[ht]
 \centering
 \subfigure{\includegraphics[width=0.48\textwidth]{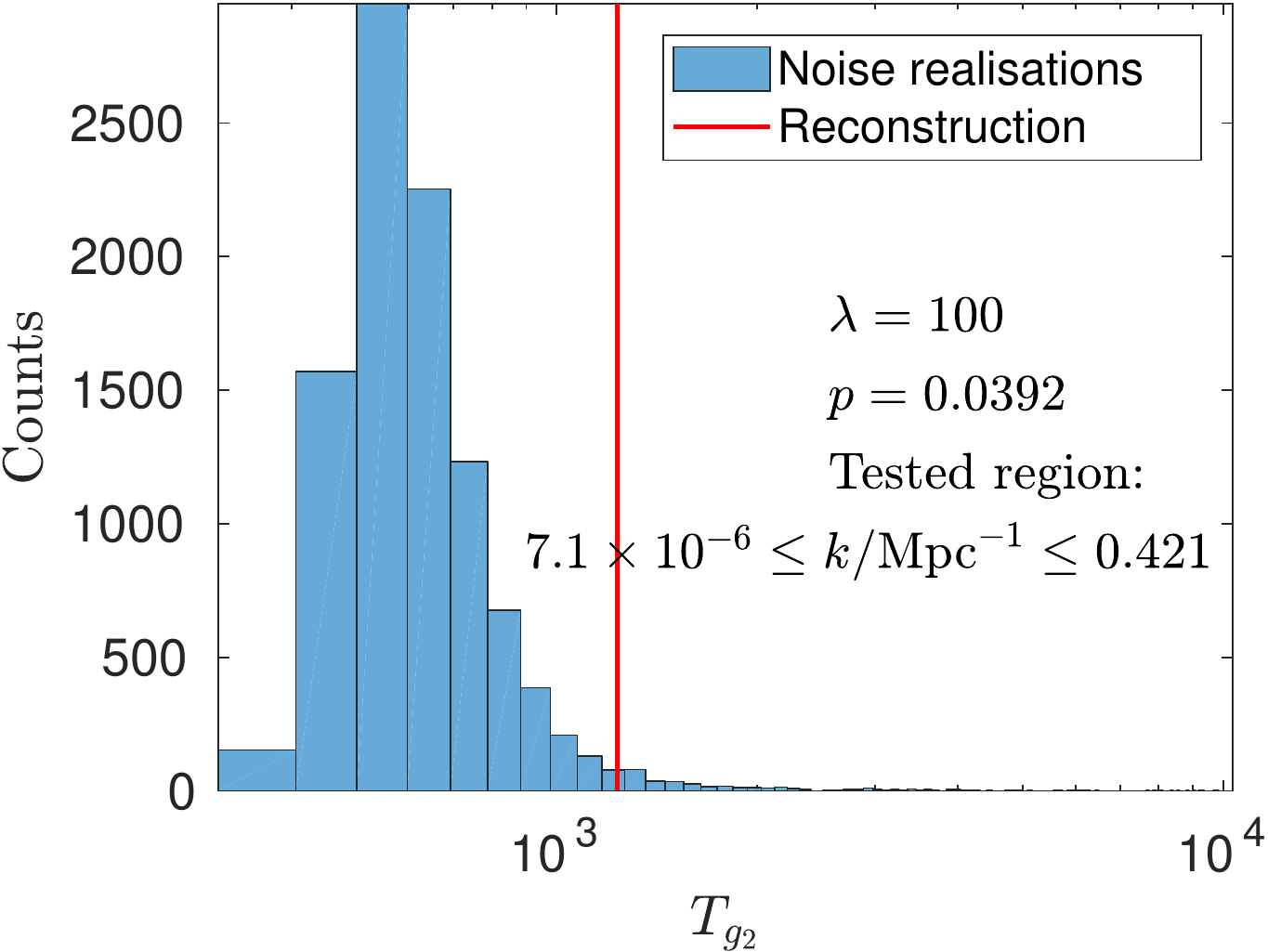}}
 \subfigure{\includegraphics[width=0.48\textwidth]{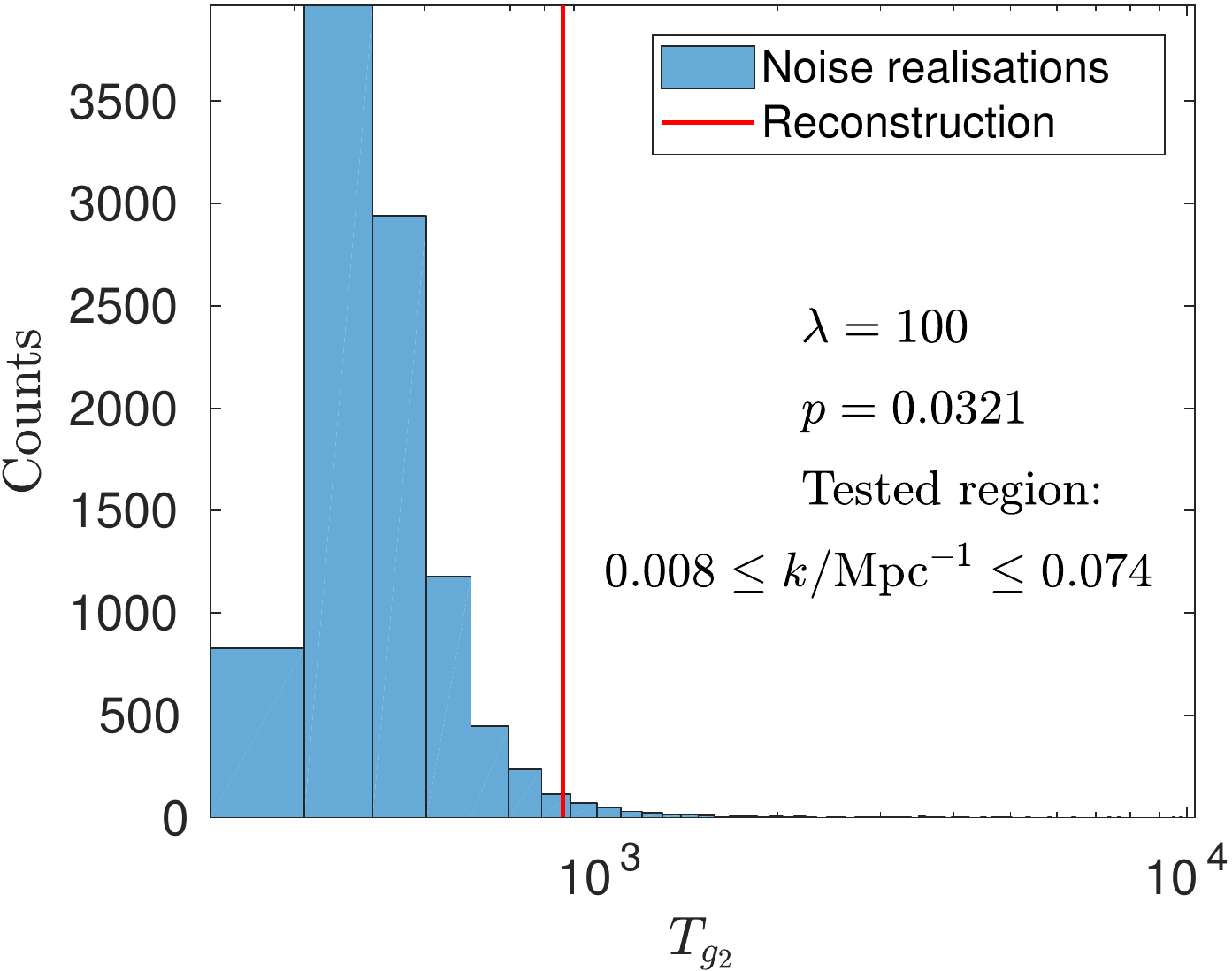}}
 \caption{P-value test using $T_{g_2}$. The distribution of $T_{g_2}$ for noise realisations and the value of $T_{g_2}$ for the reconstruction from data (red line). The right panel shows the same test, but for a more restricted range of wave numbers.}
 \label{fig:g2p}
 \end{figure}

We repeat the same procedure for the reconstructions $g_{2M}(k)$ from data, calculating
\begin{align}
		g_{2}(k) = \sqrt{\frac{1}{5} \sum_{M} |g_{2M}(k)|^2}
\end{align}
for the ($\lambda = 100$) reconstruction from data and for the ($\lambda = 100$) reconstructions from $10^4$ realisations of noise only. The results are shown in Fig.~\ref{fig:g2rec}, where on the right the mean of the noise realisations, $\overline{g}_{2}(k) = N^{-1} \sum_{j=N} g_{2,j}(k)$ has been subtracted from $g_{2}(k)$. Using the noise realisations, a covariance matrix for $g_{2}(k)$ was constructed:
\begin{align}
		\Sigma_{g_2\, k k'} = \frac{1}{N-1} \sum_{j=1}^{N} (g_{2,j}(k)-\overline{g}_2(k)) (g_{2,j}(k')-\overline{g}_2(k')),
\end{align}
the square root of whose diagonal values have been used as standard deviations in the plots of Fig.~\ref{fig:g2rec}. The amplitude $g_{2}(k)$ suggests the presence of features at 0.006, 0.019, 0.033, 0.061, 0.0826, 0.086, and 0.096 $\mathrm{Mpc}^{-1}$.
However, a test of the \emph{global} significance should be made. To do so, the constructed covariance matrix was inverted and used in a $\chi^2$ statistic:
\begin{align}
T_{g_2}(g_{2}(k)) =  (\mathbf{g}_{2}-\overline{\mathbf{g}}_{2})^{T} \mathbf{\Sigma}^{-1}_{g_2} (\mathbf{g}_{2}-\overline{\mathbf{g}}_{2}) \label{eq:thetg2}
\end{align}
where the bold notation has been used to suppress the wave number indices. Strictly speaking, the covariance matrix is not invertible, so a pseudoinverse must be used instead. This statistic was calculated for the noise realisations as well as the reconstruction from data and a p-value was thus derived. The results are shown in Fig.~\ref{fig:g2p} where the right panel shows the results of the p-value test for a region of wave numbers limited to $0.008 \leq k/\mathrm{Mpc}^{-1} \leq 0.074$.
The p-values translate to $2.06\sigma$ and $2.14\sigma$ significance levels, respectively. These are quite marginal, however, we shall now argue that another test statistic should be adopted that is more sensitive and under better theoretical control. Though $T_{g_2}$ is based on the amplitude $g_2$ which is rotationally invariant, it is not the case that all components $g_{2M}$ are equally uncertain since the submatrices of the block matrix $\mathbf{\Sigma}$ corresponding to different components $M$, are not equal to one another, and so equal statistical weight should not be given to all components. Since the statistic is based on the amplitude, it is insensitive to the direction. If a certain direction is systematically singled out across $k$, but with a small amplitude, that will count no more than randomly oriented quadrupole modulations across $k$. Also, we do not have good theoretical control over the test statistic $T_{g_2}$. It can be seen in both panels of Fig.~\ref{fig:g2p} that the possible values of $T_{g_2}$ for the noise realisations, although peaked around 500, still appear at values up to 10,000.

We shall now present a test statistic which addresses these issues. We construct a $\chi^2$ statistic using the frequentist covariance matrix
\begin{align}
T(g_{2M}) = \mathbf{g}^T {\mathbf{\Sigma}_{\mathrm{F}}}^{-1} \mathbf{g}. \label{eq:t1}
\end{align}
Strictly speaking, the inverse of the frequentist covariance matrix does not exist, but its pseudoinverse does. We have proven in Appendix~\ref{sec:tchidist} that (using the pseudoinverse) $T(g_{2M})$ is distributed according to a $\chi^2$ distribution with degrees of freedom equal to the rank of the pseudoinverse covariance matrix. This should equal the rank of the original covariance matrix of the data, which is $78$ per component of $g_{2M}$ or 390 for all five (three real and two imaginary) components considered together. 
Therefore when using all the data the test statistic $T$ should have a $\chi^2$ distribution with 390 degrees of freedom.

\begin{figure}[ht]
\centering
\includegraphics[width=0.5\textwidth]{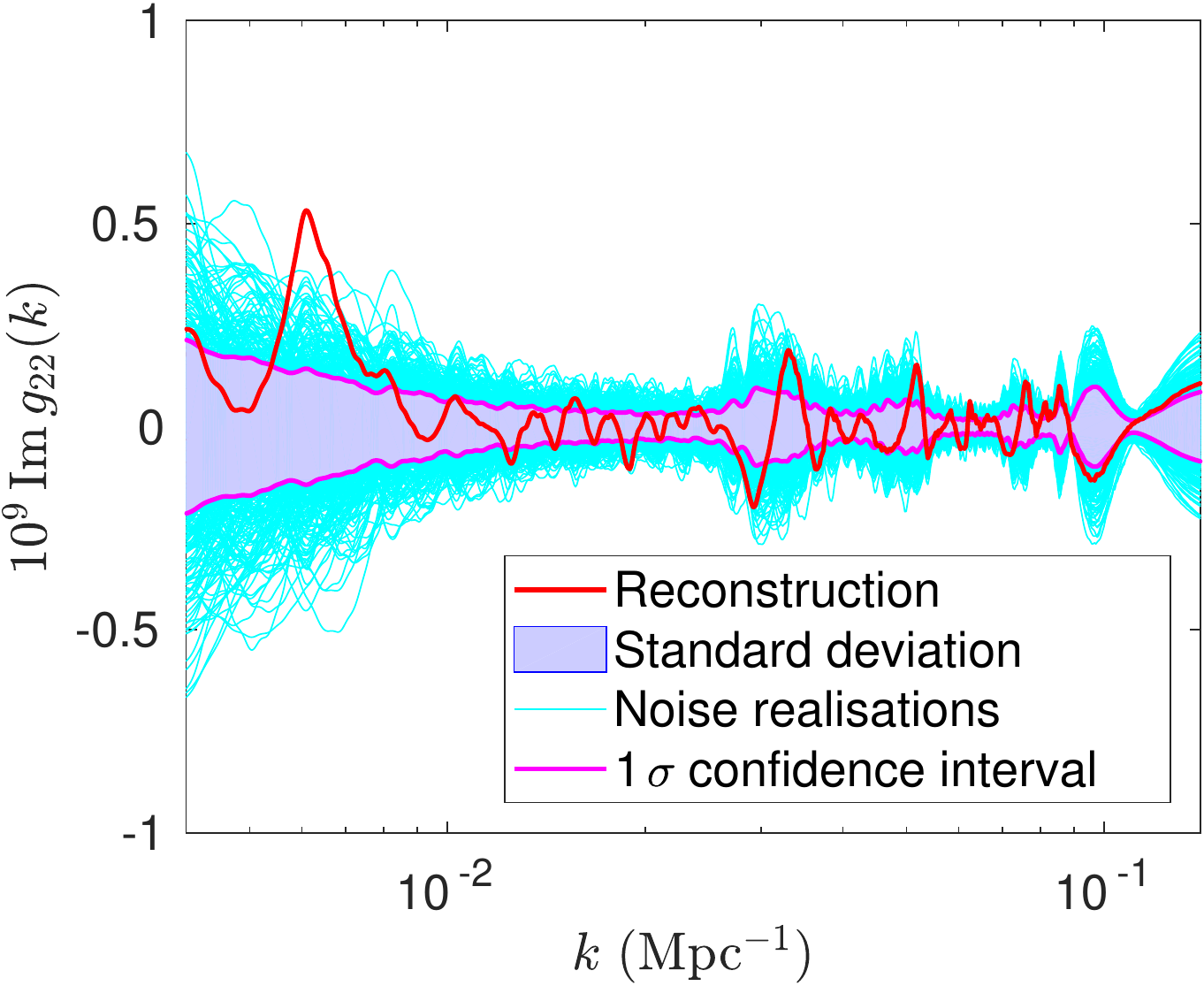}
\caption{Noise realisations (thin cyan lines), their standard deviation (purple band) as well as the reconstruction from data (red line). The $1\sigma$ confidence interval (magenta lines) derived from the frequentist covariance matrix matches the standard deviation exactly.} 
\label{fig:twodisfigs}
\end{figure}

\begin{figure}[ht]
\centering
\subfigure{\includegraphics[width=0.48\textwidth]{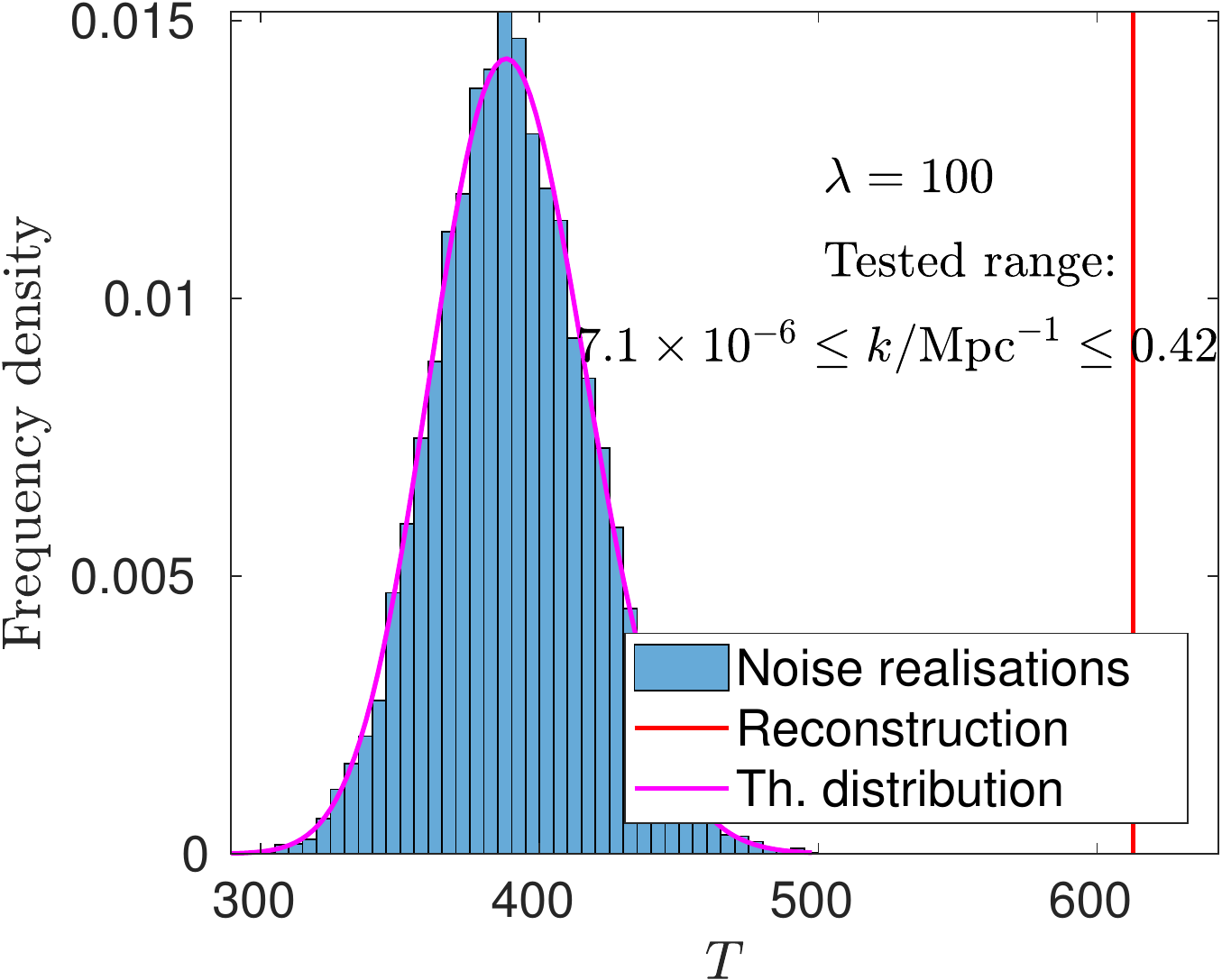}}
\subfigure{\includegraphics[width=0.48\textwidth]{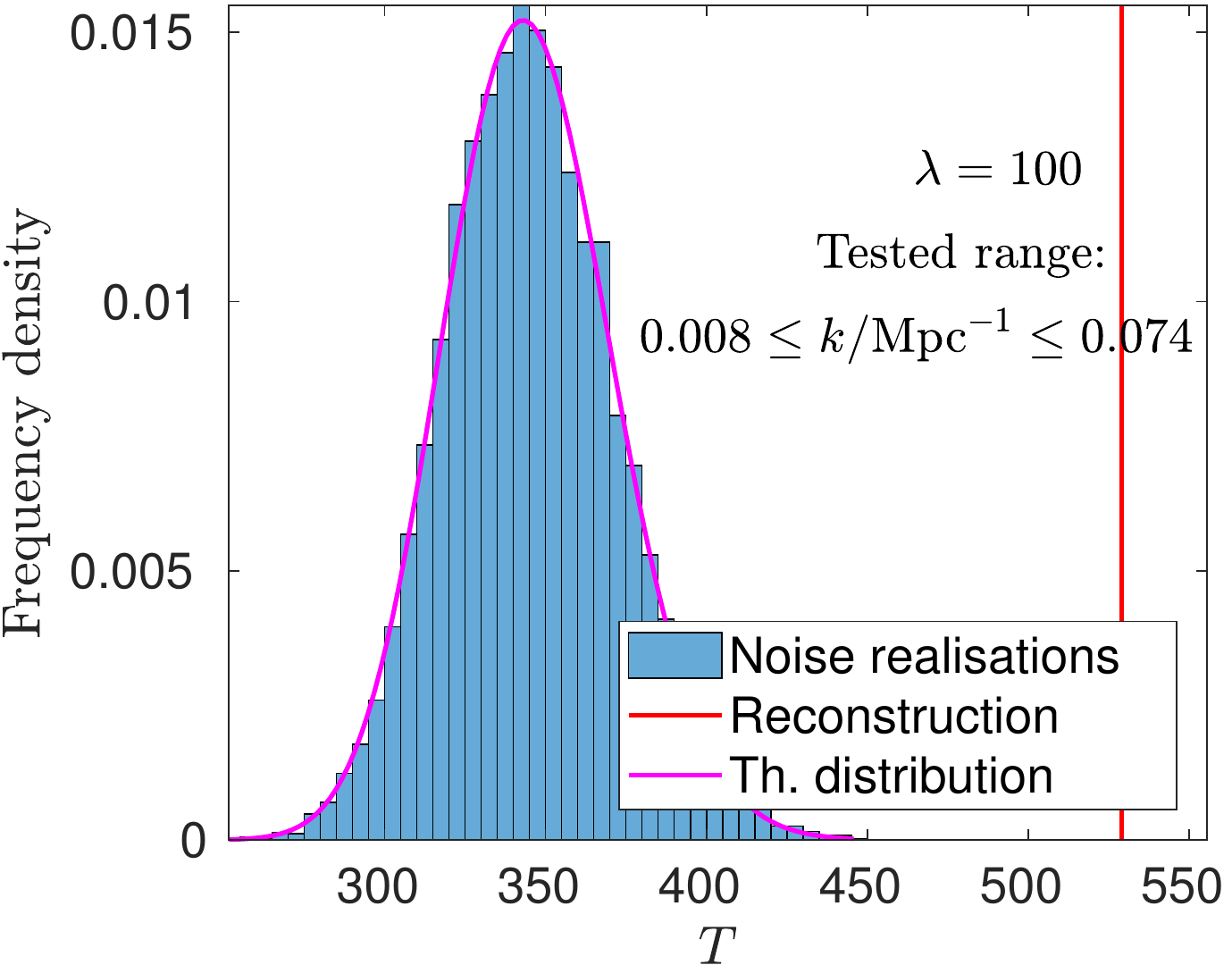}}
\caption{P-value tests of $T$ for the full range of wave numbers (left panel) and a more limited range (right panel). The value of $T$ for the reconstruction from data is indicated with a red line.}
\label{fig:pvaltests}
\end{figure}

First we illustrate in the left panel of Fig.~\ref{fig:twodisfigs} that the frequentist covariance matrix corresponds to the standard deviation of reconstructed spectra of noise realisations. The standard deviation of the noise realisations of $\mathrm{Im}\,g_{22}$ defining the purple band is exactly bounded by the magenta line which is the square root of the diagonal entries of $\mathbf{\Sigma}_\mathrm{F}$ which give the $1\sigma$ confidence intervals. 
No noise realisations were found to have as extreme a value of $T$. In fact, the theoretical distribution is able to tell us the answer: $p = F_{390}(613) \sim 3.7 \times 10^{-12} \sim 6.95\sigma$ where $F_{390}$ is the cumulative distribution function of $\chi_{390}^2$. By restricting the wave number range to $0.008 \leq k/\mathrm{Mpc}^{-1} \leq 0.074$, we find that the reconstruction from data is made more likely. The results of the p-value tests are shown in Fig.~\ref{fig:pvaltests}. For the restricted range $0.008 \leq k/\mathrm{Mpc}^{-1} \leq 0.074$, the p-value is $F_{390}(529) \sim 3.2\times 10^{-6}$, corresponding to $4.66\sigma$ significance. This illustrates clearly that even after cutting the data, the reconstructed spectrum is very unlikely to be a realisation of noise only. 


When noise realisations are tested against the non-parametric reconstructions using a covariance matrix truncated to its diagonal entries, they are only marginally significant using the test statistic $T_{g_2}$ ($2.5\sigma$) but \emph{still} rare ($4.6\sigma$) when using the directionally sensitive test statistic $T$. For the restricted wave number range $0.008 \leq k/\mathrm{Mpc}^{-1} \leq 0.074$ the significance drops to $2.2\sigma$ and $2.7\sigma$ for $T_{g_2}$ and $T$, respectively.


\section{Summary and Conclusions}\label{sec:sum}

We have described a method for reconstructing a direction- and scale-dependent modulation of the primordial power spectrum, using Tikhonov regularisation and the bipolar spherical harmonic representation. The BipoSH coefficients capture the statistical anisotropy of a CMB map and Tikhonov regularisation can then be used to infer the direction-dependent primordial power spectrum responsible. The uncertainties were estimated using the Planck FFP9 simulations and a simplified Gaussian likelihood was constructed. The method was then used on the PR2--2015 SMICA temperature map for the multipole range $30 \leq \ell \leq 1200$, binned with $\Delta \ell = 30$. The full covariance matrix of the binned BipoSH coefficients was used.

We find a $2.16\sigma$ indication of a constant quadrupole modulation, but less significance, $2 \sigma$, for the power law case. The best-fit directions are uncertain, but lie near the Galactic plane $(l,b) = ({84^{\circ}}^{+13}_{-15},{7^{\circ}}^{+13}_{-12})$ for negative amplitudes, and orthogonal for positive amplitudes: $b={1.92^{\circ}}^{+21}_{-21}$ --- suggesting contamination by Galactic foregrounds in the SMICA map. We find no evidence of either a constant or quadrupolar modulation only when we truncate the covariance matrix of the BipoSH coefficients to the diagonal.

However our \emph{non}-parametric reconstruction suggests spectral features at wave number $k/\mathrm{Mpc}^{-1} \sim 0.006, 0.019, 0.033, 0.061, 0.0826, 0.086~\mathrm{and}~0.096$. The most prominent feature is at $0.006 \, \mathrm{Mpc}^{-1}$, the dominant contribution to which comes from multipoles $60 \leq \ell \leq 90$. When a constant quadrupole modulation is fitted to data in the wave number range $0.005 \leq k/\mathrm{Mpc}^{-1} \leq 0.008$, the preferred direction of this modulation is related to the directions of the CMB dipole and hemispherical asymmetry. We find that a hot quadrupole modulation is perpendicular to both the CMB dipole and hemispherical asymmetry, while a  cold quadrupole modulation is, within the uncertainties in direction, \emph{aligned} with the hemispherical asymmetry. This is an unexpected result that merits further investigation.

In order to assess the global significance of the features, we construct two test statistics to evaluate reconstructions from data against reconstructions of noise only. With a test statistic sensitive only to the amplitude, we find the reconstruction from data to be unusual with $2.06\sigma$ significance for the full wave number range, and $2.14\sigma$ when we only consider intermediate wave numbers $0.008 \leq k/\mathrm{Mpc}^{-1} \leq 0.074$. With the test statistic we believe to be more appropriate  which is sensitive to a preferred direction as well, the significance increases to $6.95 \sigma$ and $4.66 \sigma$, respectively, for the full and intermediate range of wave numbers.

Our method goes beyond previous work in allowing for non-parametric scale-dependence in a possible quadrupolar modulation and can be easily generalised to include higher order modulations as well. The binning in multipoles and restriction to $\ell\geq 30$ can be avoided by constructing a non-Gaussian likelihood which makes full use of the Planck data and simulations. Other data sets such as polarisation, large-scale structure and future 21-cm line observations can also be incorporated.
An immediate suggestion for future work would be to consider different Planck maps, and also consider different masks.
 
\acknowledgments{
We thank Jeppe Tr{\o}st Nielsen for helpful suggestions and Sunny Vagnozzi, Sebastian von Hausegger, Hao Liu, Pavel Naselsky, Benjamin Wandelt, Guilhem Lavaux, Roya Mohayaee, Chris Pethick and Andrew Jackson for discussions. AD and SS were supported by the Danish National Research Foundation. SM acknowledges support from the Simons Foundation, the Labex ILP (ANR-10-LABX-63) part of the Idex SUPER, the Agence Nationale de la Recherche as part of the programme Investissements d'avenir (ANR-11-IDEX-0004-02), and hospitality at the Niels Bohr Institute where this work was initiated. SM also acknowledges the NERSC and Planck collaboration for providing the FFP simulations and auxiliary files.

\bibliographystyle{JHEP}
\bibliography{refs}

\appendix

\section{The general quadrupole} 
\label{sec:genquadr}

The resolution of
\begin{equation}
	g_{\ast} (\mathbf{\hat{k}} \cdot \mathbf{\hat{n}})^2 \label{eq:ng}
\end{equation}
\emph{not} being the most general quadrupole modulation is simple:
it is in fact
\begin{equation}
g_{\ast} (\mathbf{\hat{k}} \cdot \mathbf{\hat{n}_1})(\mathbf{\hat{k}}\cdot \mathbf{\hat{n}_2}) .\label{eq:gen}
\end{equation}
Clearly \eqref{eq:ng} is a special case of \eqref{eq:gen} when $\hat{\mathbf{n}}_1 = \hat{\mathbf{n}}_2$. Moreover, \eqref{eq:gen} is the tensor product of two dipoles and therefore contains a quadrupole. Let us determine $g_{2M}$ given the directions $\hat{\mathbf{n}}_1$ and $\hat{\mathbf{n}}_2$. Let these be $(\theta_1,\phi_1)$ and $(\theta_2,\phi_2)$ while $\hat{\mathbf{k}}$ points in the direction $(\theta,\phi)$.
Then
\begin{align}
	g_{20} &= g_{\ast} \int \mathrm{d}\Omega \,  (\mathbf{\hat{k}} \cdot \mathbf{\hat{n}_1})(\mathbf{\hat{k}}\cdot \mathbf{\hat{n}_2}) Y_{20}^{\ast}(\theta,\phi)
	\nonumber\\ &= g_{\ast} \frac{1}{\sqrt{4 \pi}} \int \mathrm{d}\Omega \, (\sin(\theta) \sin(\theta_1) \cos(\phi-\phi_1) + \cos(\theta) \cos(\theta_1)) \nonumber \\ &\quad\quad\quad\quad (\sin(\theta) \sin(\theta_2) \cos(\phi-\phi_2) + \cos(\theta) \cos(\theta_2)) Y_{20}^{\ast}(\theta,\phi) \\
	&= g_{\ast} \frac{2}{3} \sqrt{\frac{\pi}{5}} (2 \cos(\theta_1) \cos(\theta_2) - \cos(\phi_1-\phi_2) \sin(\theta_1) \sin(\theta_2))
\end{align}
\begin{align}
	g_{21} &= g_{\ast} \int \mathrm{d}\Omega \,  (\mathbf{\hat{k}} \cdot \mathbf{\hat{n}_1})(\mathbf{\hat{k}}\cdot \mathbf{\hat{n}_2}) Y_{21}^{\ast}(\theta,\phi) \\ &= g_{\ast} \frac{1}{\sqrt{4 \pi}} \int \mathrm{d}\Omega \, (\sin(\theta) \sin(\theta_1) \cos(\phi-\phi_1) + \cos(\theta) \cos(\theta_1)) \nonumber \\ &\quad\quad\quad\quad (\sin(\theta) \sin(\theta_2) \cos(\phi-\phi_2) + \cos(\theta) \cos(\theta_2)) Y_{21}^{\ast}(\theta,\phi) \\ &= - g_{\ast} \sqrt{\frac{2 \pi}{15}} (\exp(-i\phi_1) \cos(\theta_2) \sin(\theta_1) + \exp(-i\phi_2) \cos(\theta_1) \sin(\theta_2) ) 
\end{align}
and
\begin{align}
	g_{22} &= g_{\ast} \int \mathrm{d}\Omega \,  (\mathbf{\hat{k}} \cdot \mathbf{\hat{n}_1})(\mathbf{\hat{k}}\cdot \mathbf{\hat{n}_2}) Y_{22}^{\ast}(\theta,\phi) \\ &= g_{\ast} \frac{1}{\sqrt{4 \pi}} \int \mathrm{d}\Omega \, (\sin(\theta) \sin(\theta_1) \cos(\phi-\phi_1) + \cos(\theta) \cos(\theta_1)) \nonumber \\ &\quad\quad\quad\quad (\sin(\theta) \sin(\theta_2) \cos(\phi-\phi_2) + \cos(\theta) \cos(\theta_2)) Y_{22}^{\ast}(\theta,\phi) \\ &= g_{\ast} \sqrt{\frac{2 \pi}{15}} \exp(-i (\phi_1 + \phi_2)) \sin(\theta_1)\sin(\theta_2).
\end{align}
\section{Relating quadrupole modulations to BipoSH coefficients} 
\label{sec:relate}

We investigate in detail the relation between a given quadrupole modulation $g_{L M}(k)$ and the BipoSH coefficient $A^{LM}_{\ell \ell'}$. In linear theory, there is a simple relation between the temperature perturbation $\Delta T/T_0$ in the direction $\mathbf{\hat{n}}$ at the position $\mathbf{x}$ at time $t$ and the curvature perturbation $\mathcal{R}(\mathbf{k})$, \emph{viz.}
\begin{align}
\frac{\Delta T}{T_0}(\mathbf{x},\mathbf{\hat{n}},t) &= \int \frac{\mathrm{d}^3 k}{\left(2 \pi\right)^3} \exp(i \mathbf{k} \cdot \mathbf{x}) \Theta (\mathbf{k},\mathbf{\hat{n}},t) \mathcal{R}(\mathbf{k}),
\end{align}
where $\Theta(\mathbf{k},\mathbf{n},t)$ is the transfer function. Assuming that the cosmological evolution itself does not induce any anisotropy, the transfer function can depend only on $\mathbf{\hat{k}} \cdot \mathbf{\hat{n}}$ which takes values between -1 and 1. It can therefore be expressed in terms of Legendre polynomials $P_\ell(\mathbf{\hat{k}}\cdot \mathbf{\hat{n}})$ which are complete on the interval $[-1,1]$, with coefficients $\Theta_\ell$:
\begin{align}
\Theta(\mathbf{k},\mathbf{\hat{n}}, t) &= \sum_{\ell} (-i)^{\ell} (2\ell + 1) \Theta_{\ell}(k, t) P_{\ell} (\mathbf{\hat{k}} \cdot \mathbf{\hat{n}})\mathrm{.}
\end{align}
The Legendre polynomials may be expressed as spherical harmonics:
\begin{align}
P_{\ell} (\mathbf{\hat{k}} \cdot \mathbf{\hat{n}}) = \frac{4\pi}{2\ell + 1} \sum_{m = -\ell}^{\ell} Y_{\ell m}(\mathbf{\hat{k}}) Y^{\ast}_{\ell m}(\mathbf{\hat{n}}),
\end{align}
so that:
\begin{align}
\Theta (\mathbf{k}, \mathbf{\hat{n}}, t) &= 4\pi \sum_{\ell m} (-i)^{\ell} \Theta_{\ell}(k, t) Y_{\ell m}(\mathbf{\hat{k}}) Y^\ast_{\ell m}(\mathbf{\hat{n}})\mathrm{.}\label{eq:thet}
\end{align}
The central object is the angular correlation function
\begin{align}
C (\mathbf{\hat{n}}, \mathbf{\hat{n}}') &= \Big\langle \frac{\Delta T}{T_0} (\mathbf{x}, \mathbf{\hat{n}}, t) \frac{\Delta T}{T_0} (\mathbf{x}, \mathbf{\hat{n}}', t) \Big\rangle,
\end{align}
which can be re-expressed using (\ref{eq:thet}):
\begin{align}
C (\mathbf{\hat{n}}, \mathbf{\hat{n}}') &= \sum_{\ell m} \sum_{\ell' m'} (4\pi)^{2} (-i)^{\ell + \ell'} Y_{\ell m}(\mathbf{\hat{n}}) Y_{\ell' m'}(\mathbf{\hat{n}}') 
\nonumber \\
&\quad \int \frac{\mathrm{d}^3 k}{(2\pi)^3} \int \frac{\mathrm{d}^3 k'}{(2\pi)^3} \mathrm{e}^{i (\mathbf{k} + \mathbf{k'}) \cdot \mathbf{x}} \Theta_{\ell}(k,t) \Theta_{\ell'}(k',t) Y^{\ast}_{\ell m}(\mathbf{\hat{k}}) Y_{\ell' m'}^{\ast}(\mathbf{\hat{k}}') \langle \mathcal{R}(\mathbf{k}) \mathcal{R}(\mathbf{k}') \rangle.
\end{align}
Now, we know that
\begin{align}
\langle \mathcal{R}(\mathbf{k}) \mathcal{R}( \mathbf{k}') \rangle &= (2\pi)^3 \delta^{(3)}(\mathbf{k}+\mathbf{k}') P(\mathbf{k})\\ &= (2\pi)^3 \delta^{(3)}(\mathbf{k} + \mathbf{k}') \frac{2\pi^2}{k^3} \mathcal{P}(\mathbf{k}) \\  &= (2\pi)^3 \delta^{(3)}(\mathbf{k} + \mathbf{k}') \frac{2\pi^2}{k^3} \sqrt{4\pi} \sum_{LM} g_{LM} (k) Y_{LM} (\mathbf{\hat{k}}),
\end{align}
so, replacing $\langle\mathcal{R}(\mathbf{k})\mathcal{R} (\mathbf{k}')\rangle$, we get
\begin{align}
C (\mathbf{\hat{n}}, \mathbf{\hat{n}}') &= \sqrt{4\pi} \sum_{\ell m} \sum_{\ell' m'} (4\pi)^{2} (-i)^{\ell + \ell'} Y_{\ell m} (\mathbf{\hat{n}}) Y_{\ell' m'}(\mathbf{\hat{n}}') \int \frac{\mathrm{d}^3 k}{(2\pi)^3} \int \frac{\mathrm{d}^3 k'}{(2\pi)^3} \mathrm{e}^{i (\mathbf{k} + \mathbf{k'}) \cdot \mathbf{x}} \nonumber \\
&\quad \Theta_{\ell} (k, t) \Theta_{\ell'} (k', t) Y^{\ast}_{\ell m}(\mathbf{\hat{k}}) Y_{\ell' m'}^{\ast}(\mathbf{\hat{k}'}) (2\pi)^3 \delta^{(3)}(\mathbf{k} + \mathbf{k}') \frac{2\pi^2}{k^3} \sum_{LM} g_{LM}(k) Y_{LM} (\mathbf{\hat{k}})\mathrm{.}
\end{align}
The 3-dimensional integral over $\mathbf{k}'$ is trivial due to the $\delta$ function and the 3-dimensional integral over $\mathbf{k}$ is split into a radial part and an angular part:
\begin{align}
C (\mathbf{\hat{n}},\mathbf{\hat{n}}') &= \sqrt{4\pi} \sum_{LM} \sum_{\ell m} \sum_{\ell' m'} (4\pi)^2 (-i)^{\ell + \ell'} \frac{1}{(2 \pi)^3} Y_{\ell m}(\mathbf{\hat{n}}) Y_{\ell' m'}(\mathbf{\hat{n}}') \nonumber \\ & \quad \int_{0}^{\infty} \mathrm{d}k \, k^2 \frac{2\pi^2}{k^3} g_{LM} (k) \Theta_{\ell} (k, t) \Theta_{\ell'} (k, t) \int \mathrm{d} \mathbf{\hat{k}}\, Y_{LM} (\mathbf{\hat{k}}) Y^{\ast}_{\ell m} (\mathbf{\hat{k}}) Y_{\ell' m'}^{\ast}(-\mathbf{\hat{k}}).
\end{align}
Using $Y_{\ell' m'}(-\mathbf{\hat{k}}) = (-1)^{\ell'} Y_{\ell' m'}(\mathbf{\hat{k}}) = (-i)^{2\ell'} Y_{\ell' m'}(\mathbf{\hat{k}})$ we get
\begin{align}
C (\mathbf{n}, \mathbf{n}') &= \sqrt{4\pi} \sum_{LM} \sum_{\ell m} \sum_{\ell' m'} (4\pi)^2 (-i)^{\ell + \ell'} \frac{1}{(2\pi)^3} Y_{\ell m}(\mathbf{\hat{n}}) Y_{\ell' m'}(\mathbf{\hat{n}}') \nonumber \\ & \quad \int_{0}^{\infty} \mathrm{d}k \, k^2 \frac{2 \pi^2}{k^3} g_{LM} (k) \Theta_{\ell} (k, t) \Theta_{\ell'} (k, t) \int \mathrm{d} \mathbf{\hat{k}} \, Y_{LM} (\mathbf{\hat{k}}) Y^{\ast}_{\ell m}(\mathbf{\hat{k}}) (-i)^{-2\ell' } Y_{\ell' m'}^{\ast}(\mathbf{\hat{k}}),
\end{align}
where the numerical factors may be combined, and factors shuffled, to write:
\begin{align}
C (\mathbf{\hat{n}}, \mathbf{\hat{n}'}) &= \sqrt{4\pi} \sum_{LM} \sum_{\ell m} \sum_{\ell' m'} (4\pi) (-i)^{\ell - \ell'} Y_{\ell m}(\mathbf{\hat{n}}) Y_{\ell' m'}(\mathbf{\hat{n}}') \nonumber \\ & \quad  \int_{0}^{\infty} \frac{\mathrm{d}k}{k} g_{LM}(k) \Theta_{\ell} (k, t) \Theta_{\ell'} (k, t) \int \mathrm{d} \mathbf{\hat{k}} \, Y_{LM}(\mathbf{\hat{k}}) Y^{\ast}_{\ell m}(\mathbf{\hat{k}}) Y_{\ell' m'}^{\ast}(\mathbf{\hat{k}}).
\end{align}
Now
\begin{align}
\int \mathrm{d} \mathbf{\hat{k}} \, Y_{LM} (\mathbf{\hat{k}} ) Y^{\ast}_{\ell m}(\mathbf{\hat{k}}) Y_{\ell' m'}^{\ast}(\mathbf{\hat{k}}) = (-1)^{m'} \sqrt{\frac{(2\ell +1) (2\ell' +1 )}{4\pi (2L + 1)}} C^{L 0}_{\ell 0 \ell' 0} C^{L M}_{\ell m \ell' -m'},
\end{align}
and the factors of $\sqrt{4 \pi}$ cancel, yielding
\begin{align}
C (\mathbf{\hat{n}}, \mathbf{\hat{n}'}) &= \sum_{LM} \sum_{\ell \ell'} \sum_{m m'} Y_{\ell m} (\mathbf{\hat{n}}) Y_{\ell' m'}(\mathbf{\hat{n}}') C^{LM}_{\ell m \ell' -m'} (-1)^{m'} C^{L 0}_{\ell 0 \ell' 0} \sqrt{\frac{(2\ell +1)(2\ell' +1)}{(2L + 1)}} \nonumber \\ &\quad \times \left[ 4\pi (-i)^{\ell - \ell'} \int_{0}^{\infty} \mathrm{d}\log k \, g_{LM}(k) \Theta_{\ell}(k, t) \Theta_{\ell'}(k, t)\right]\mathrm{.}
\end{align}
The square brackets isolate the BipoSH coefficients from the (WMAP normalised) basis functions and the expression matches eq.\eqref{eq:relate} after identifying $\Theta$ with $\Delta$.

\section{Other choices of \boldmath $\lambda$} 
\label{sec:olambda}

We present here the reconstructions for other values of the regularisation parameter $\lambda$. The results for $\lambda = 500$ are shown in Fig.~\ref{fig:olambda500}.
Note the feature at $k \sim 0.006 \, \mathrm{Mpc}^{-1}$ which persists when $\lambda$ is increased to 1000 as shown in Fig.~\ref{fig:olambda1000}. The reconstructions for $\lambda = 10000$ are shown in Fig.~\ref{fig:olambda10000}. The quadrupole modulation $g_{2M}$ at $k \sim 0.006 \, \mathrm{Mpc}^{-1}$ is tabulated in Table~\ref{tab:qatk}.

\begin{table}[ht]
\begin{center}
\begin{tabular}{|l|r|r|}
\hline
$M$ & $ 10^9 \, \mathrm{Re} \, {g}_{2M}$ & $10^9 \, \mathrm{Im} \, {g}_{2M}$ \\
\hline
$0$ & $-0.23\pm 0.16$ &  \\
\hline
$1$ & $0.38\pm 0.13$ & $0.17 \pm 0.13$  \\
\hline
$2$ & $-0.20\pm 0.14$ & $0.50 \pm 0.14$ \\
\hline
\end{tabular}
\caption{The quadrupole modulation at $k \sim 0.006 \,\mathrm{Mpc}^{-1}$ with frequentist errors, from the reconstruction with $\lambda=100$.}
\label{tab:qatk}
\end{center}
\end{table}

\begin{figure}[ht]
\centering
\subfigure
{\includegraphics[width=0.48\textwidth]{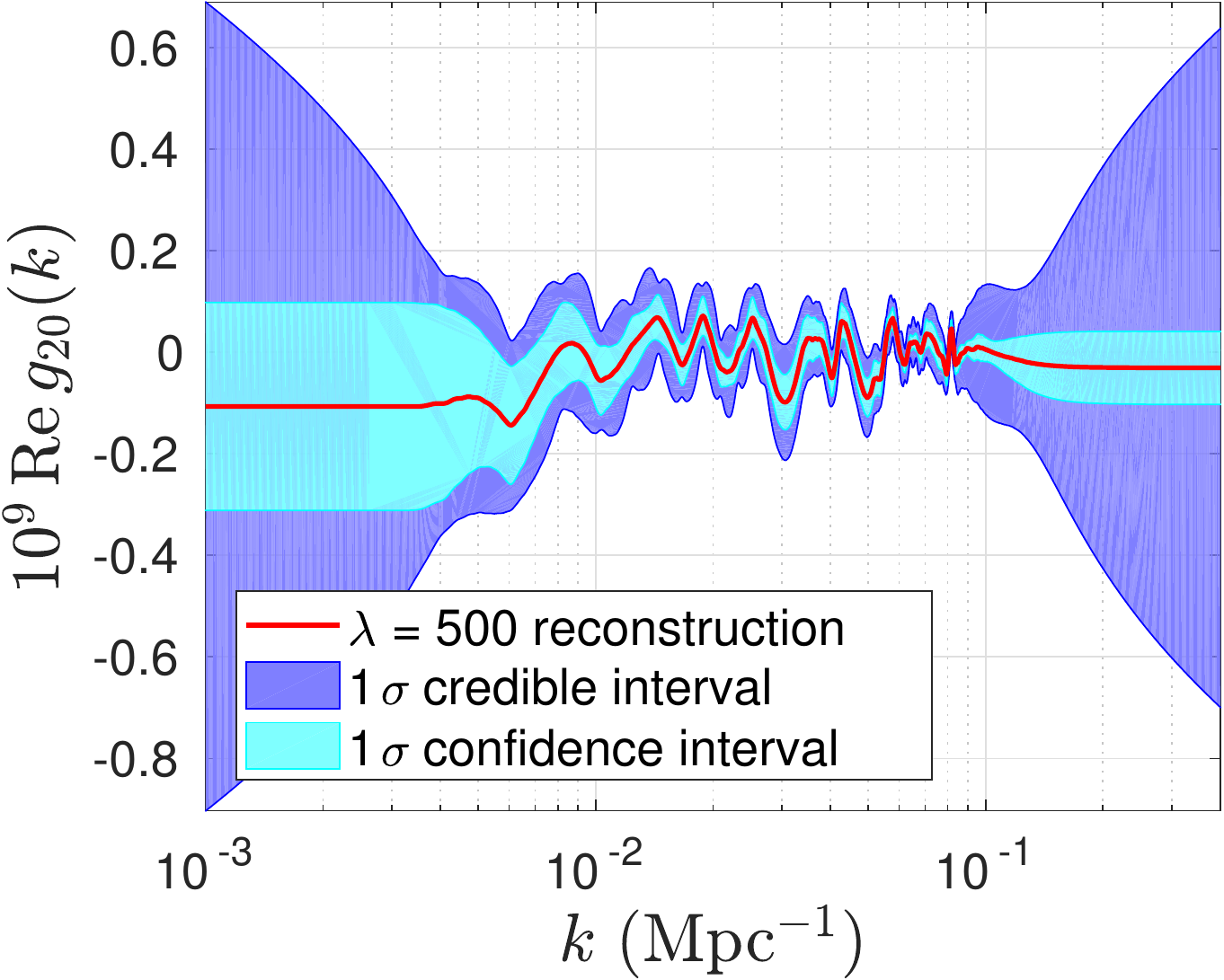}
\label{fig:m0500}
} \\
\subfigure
{\includegraphics[width=0.48\textwidth]{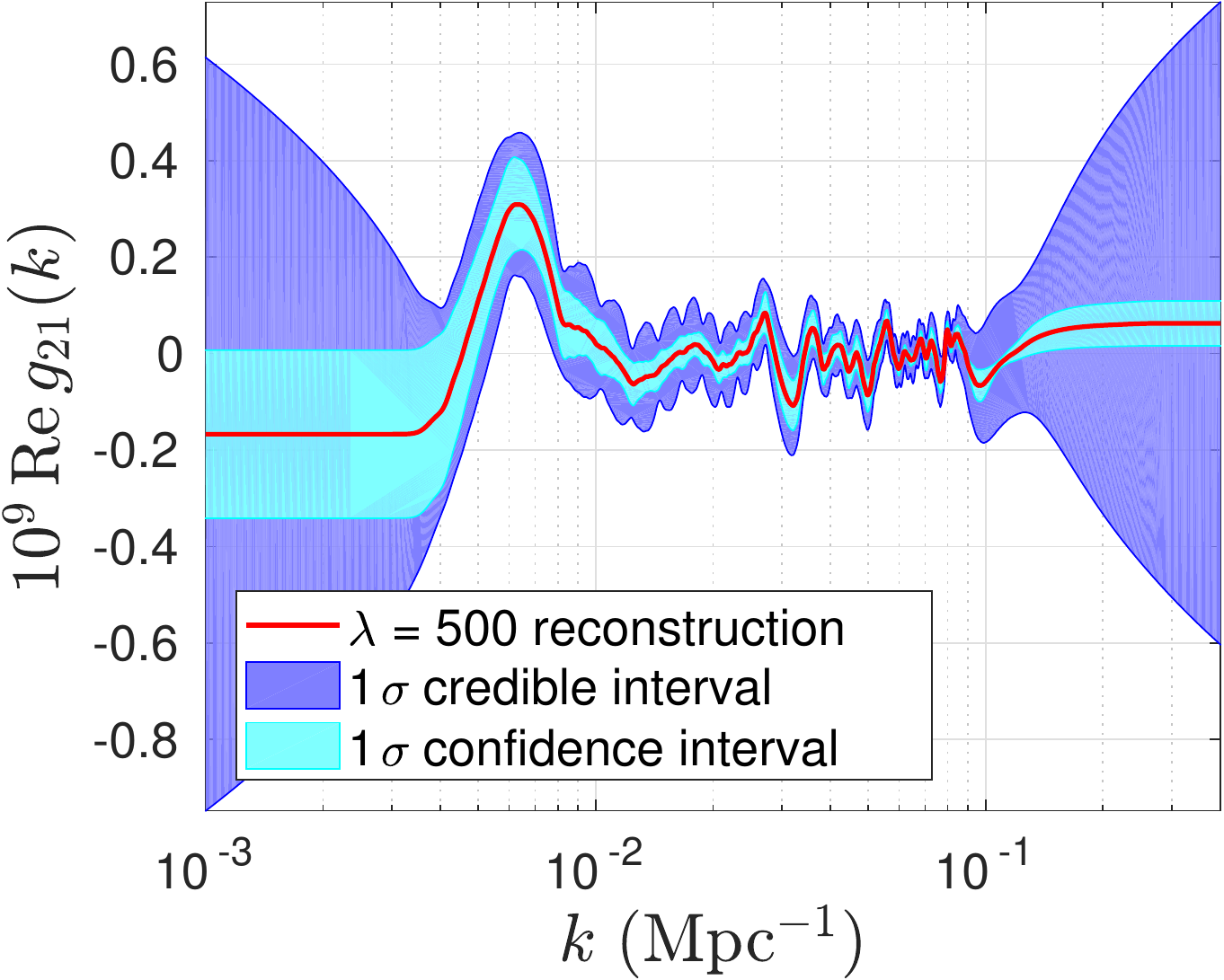}
\label{fig:m1500}
}
\subfigure
{\includegraphics[width=0.48\textwidth]{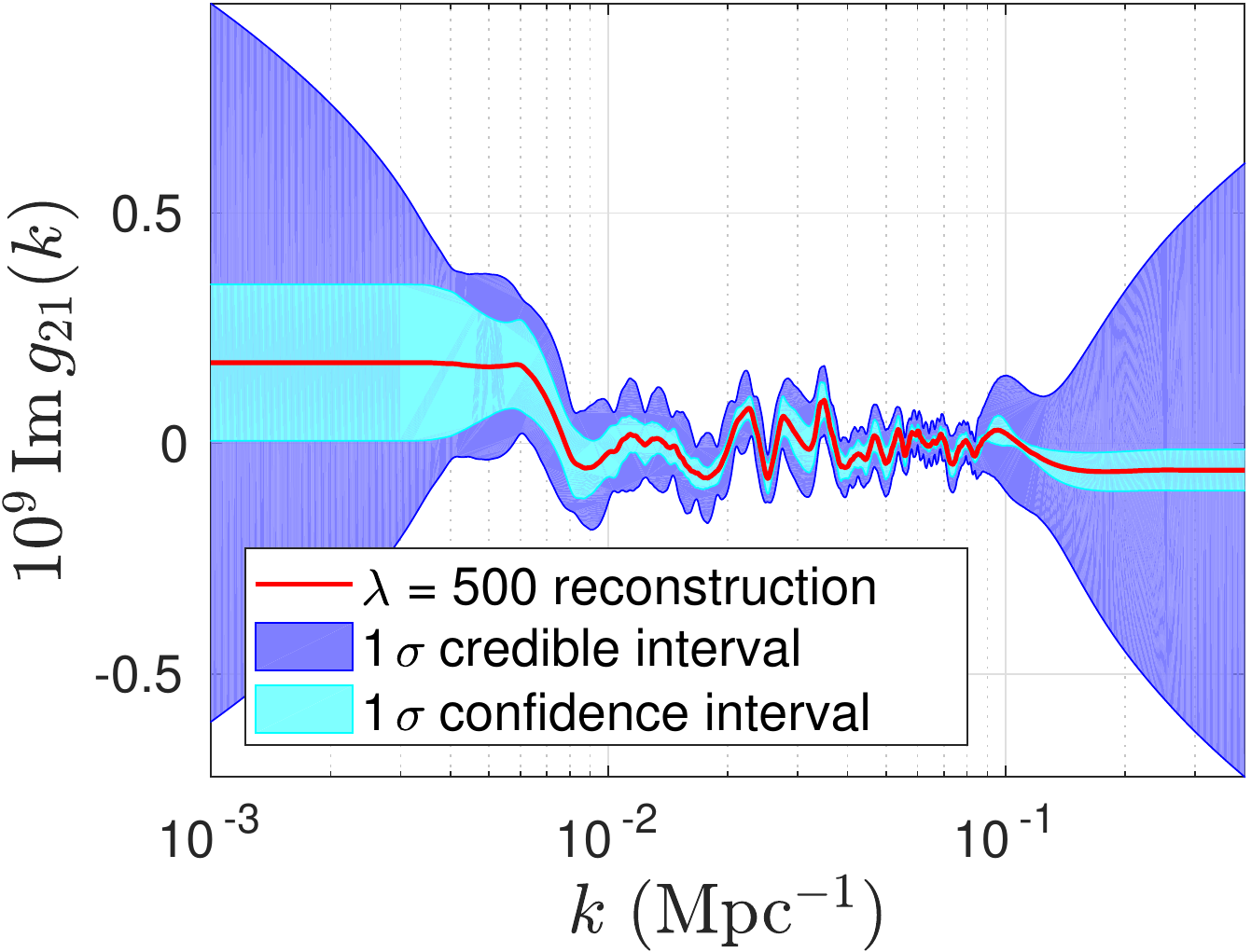}
\label{fig:m1im500}
}
\subfigure
{\includegraphics[width=0.48\textwidth]{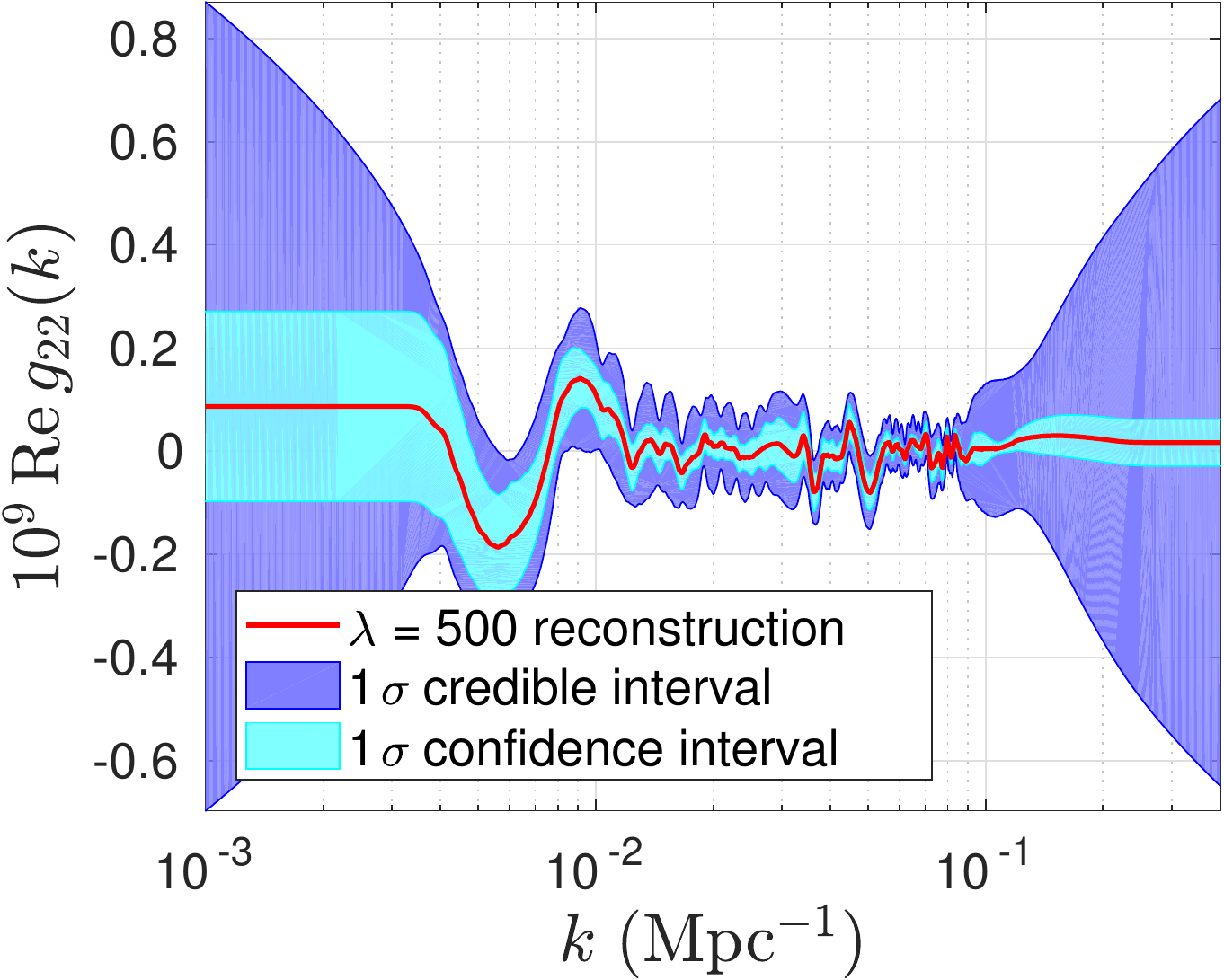}
\label{fig:m2500}
}
\subfigure
{\includegraphics[width=0.48\textwidth]{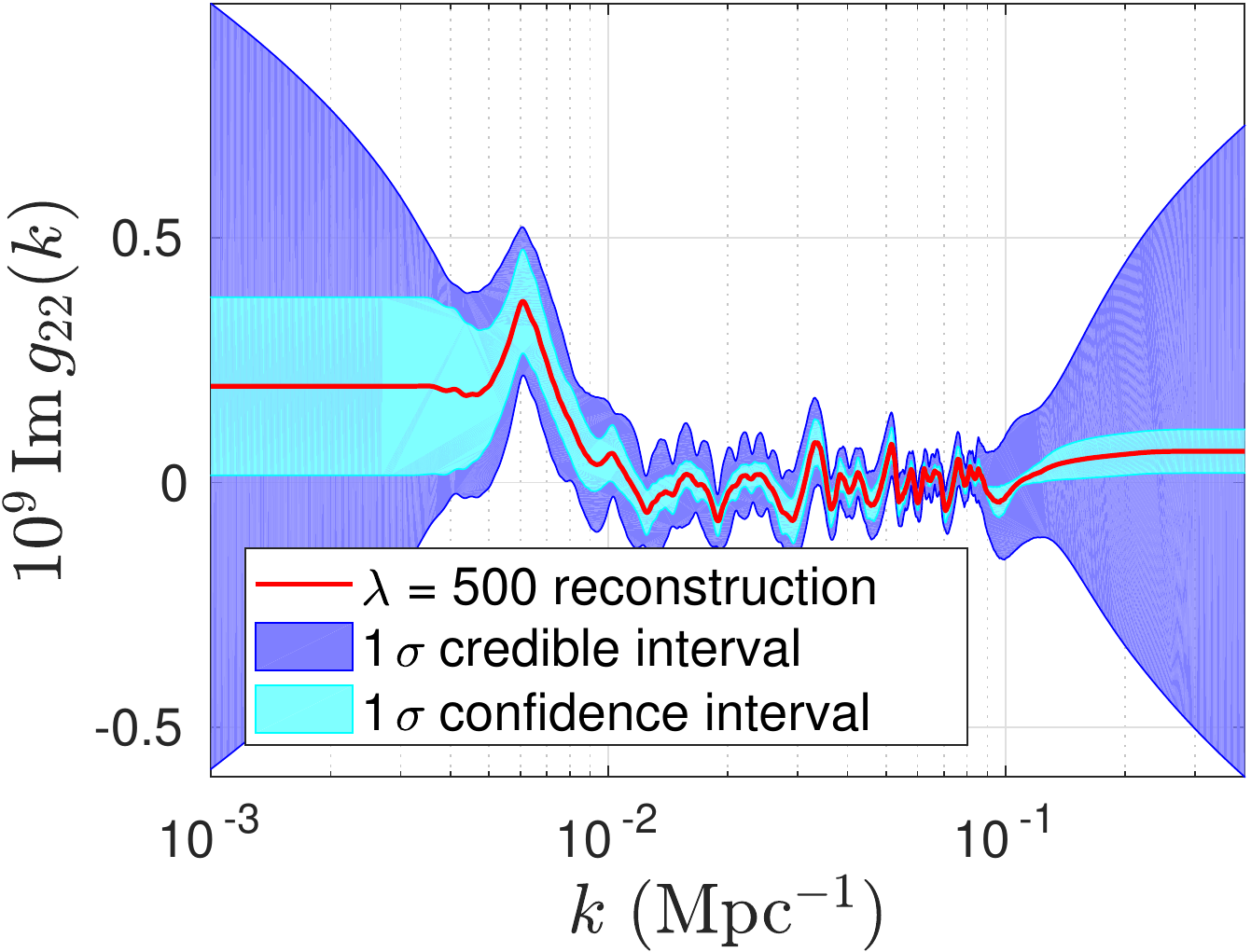}
\label{fig:m2im500}
}
\caption{Reconstructed quadrupole modulation (full red line) with $1 \sigma$ credible intervals (purple) and $1 \sigma$ confidence intervals (cyan) for $\lambda = 500$. Note the feature in 
$\mathrm{Im}\, g_{22}$ at $k \sim 0.006 \, \mathrm{Mpc}^{-1}$.}
\label{fig:olambda500}
\end{figure}

\begin{figure}[ht]
\centering
\subfigure
{\includegraphics[width=0.48\textwidth]{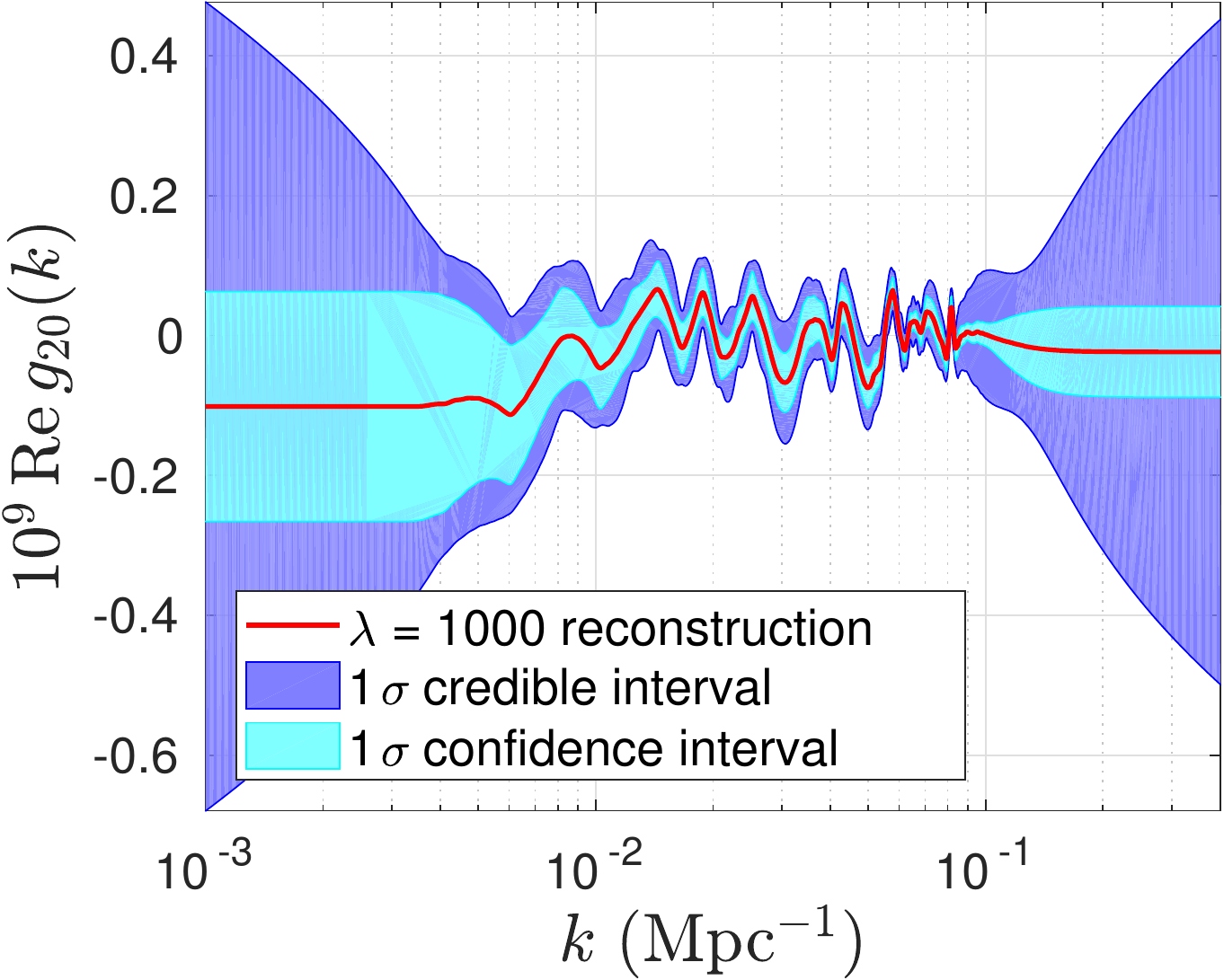}
\label{fig:m01000}
} \\
\subfigure
{\includegraphics[width=0.48\textwidth]{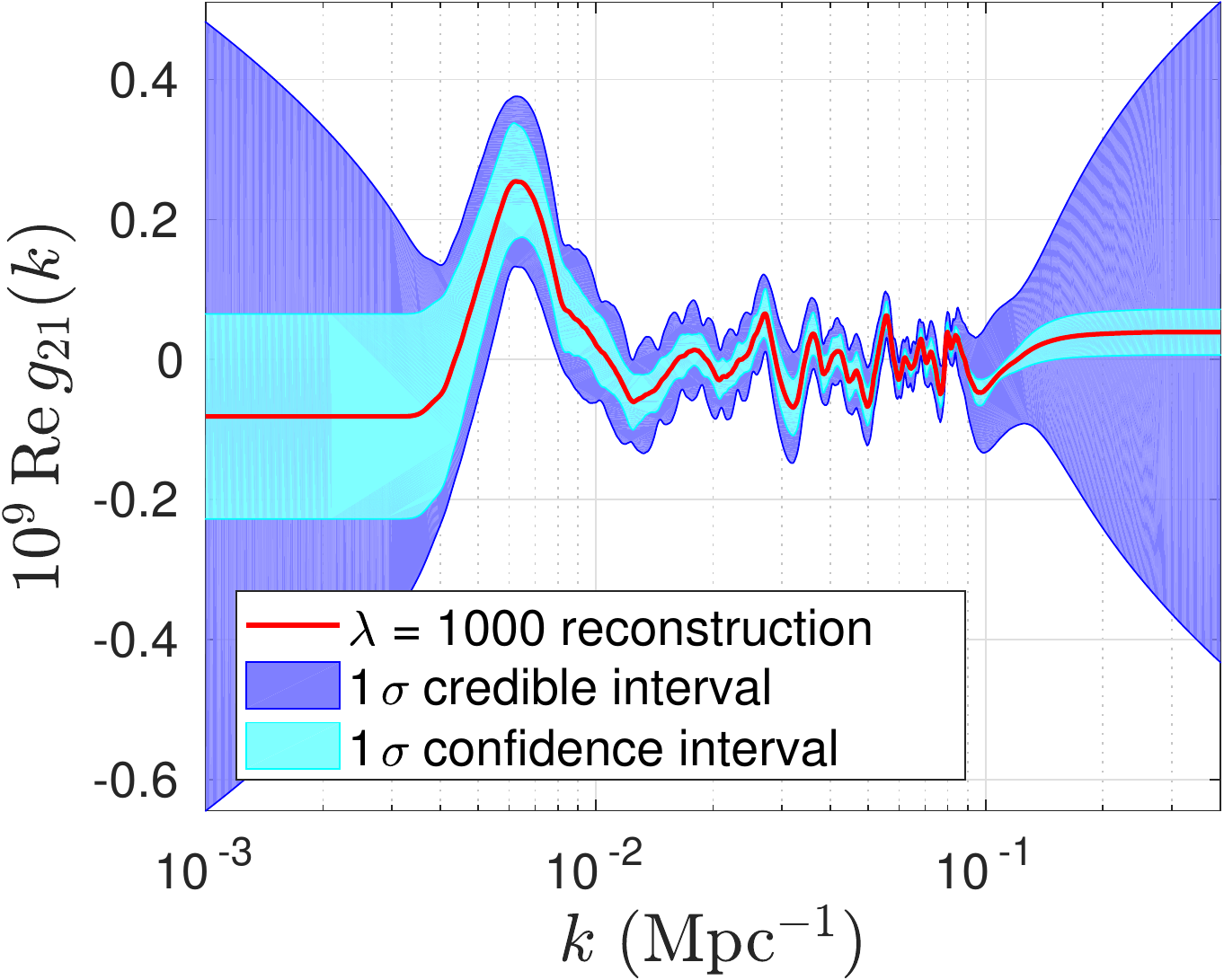}
\label{fig:m11000}
}
\subfigure
{\includegraphics[width=0.48\textwidth]{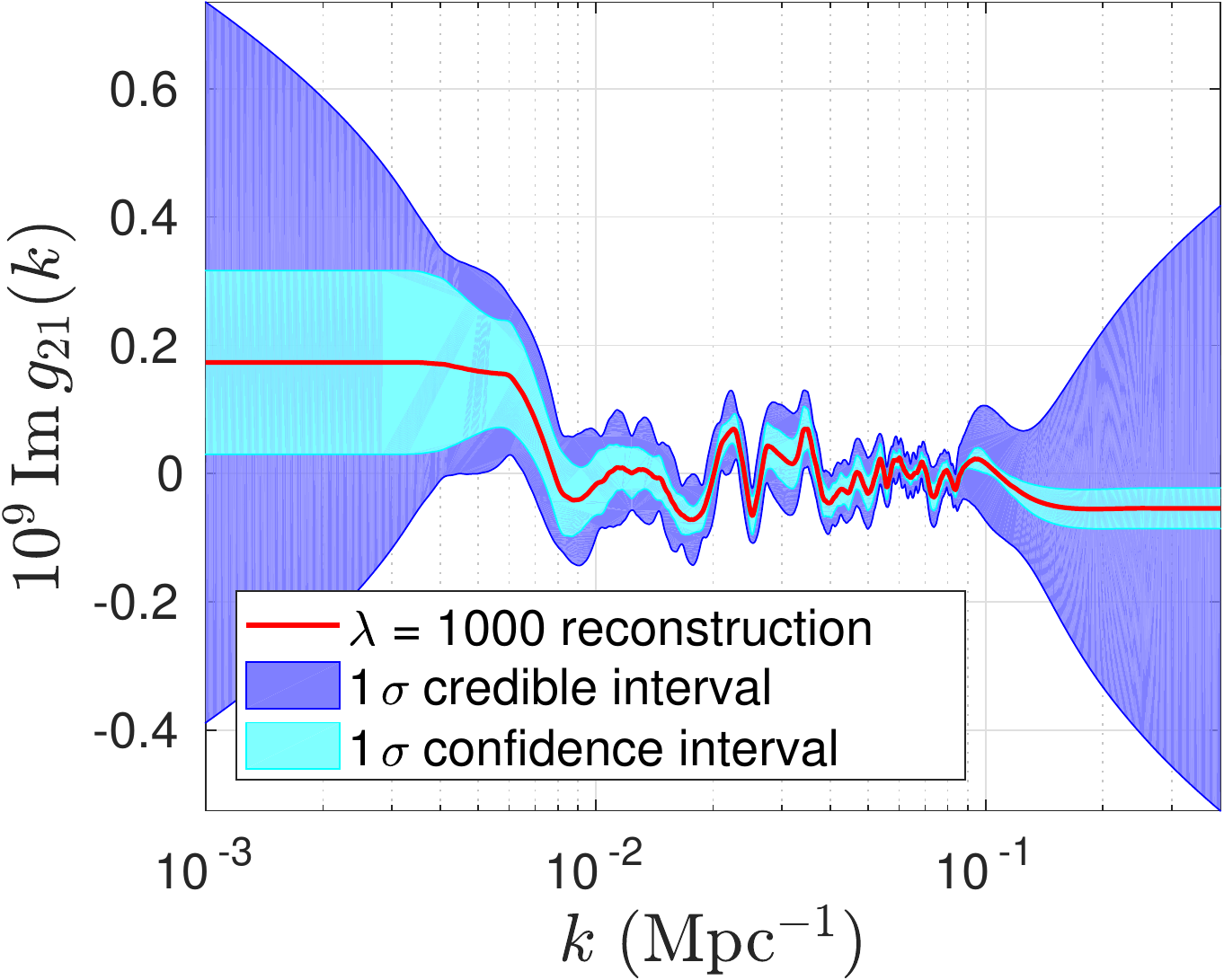}
\label{fig:m1im1000}
}
\subfigure
{\includegraphics[width=0.48\textwidth]{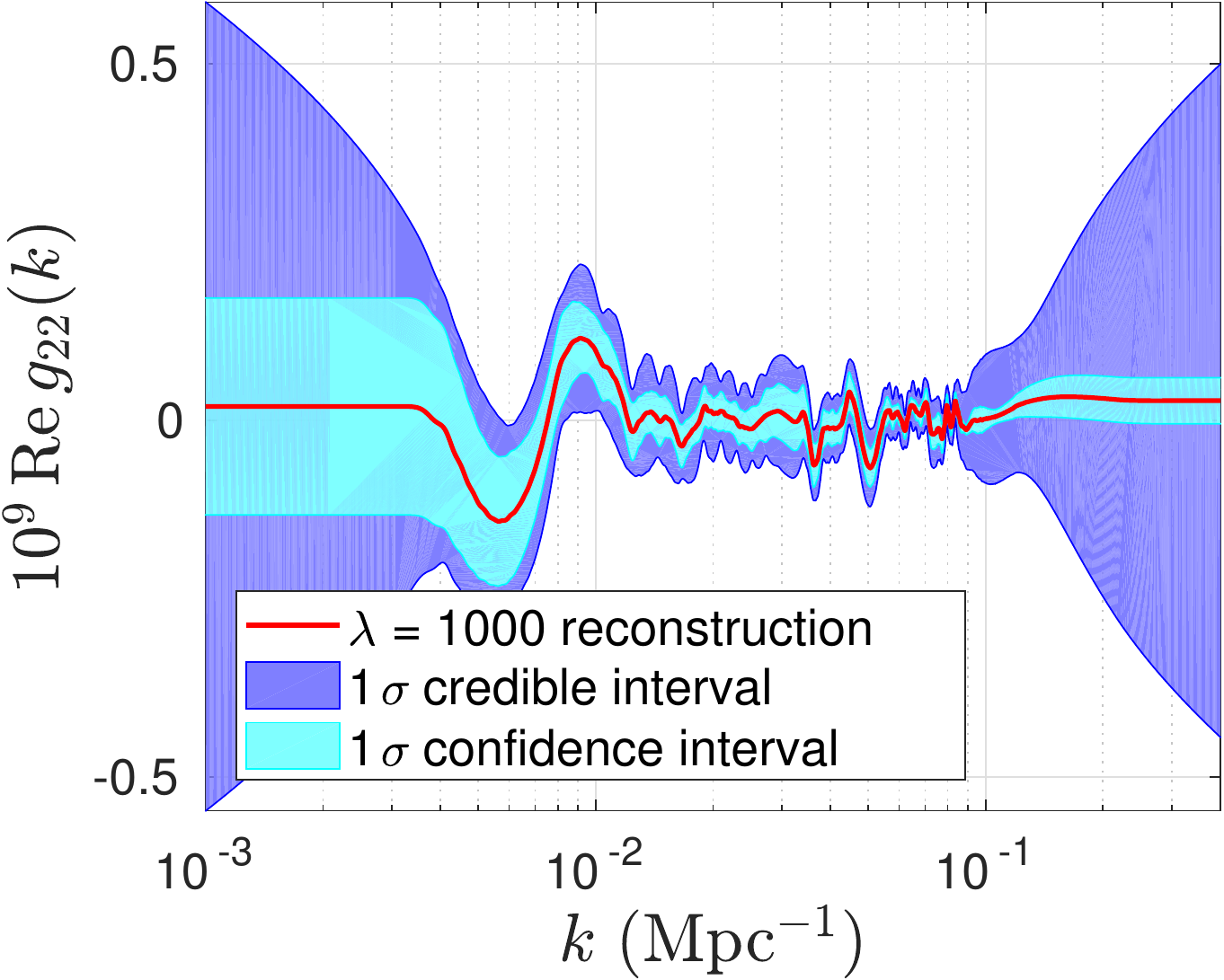}
\label{fig:m21000}
}
\subfigure
{\includegraphics[width=0.48\textwidth]{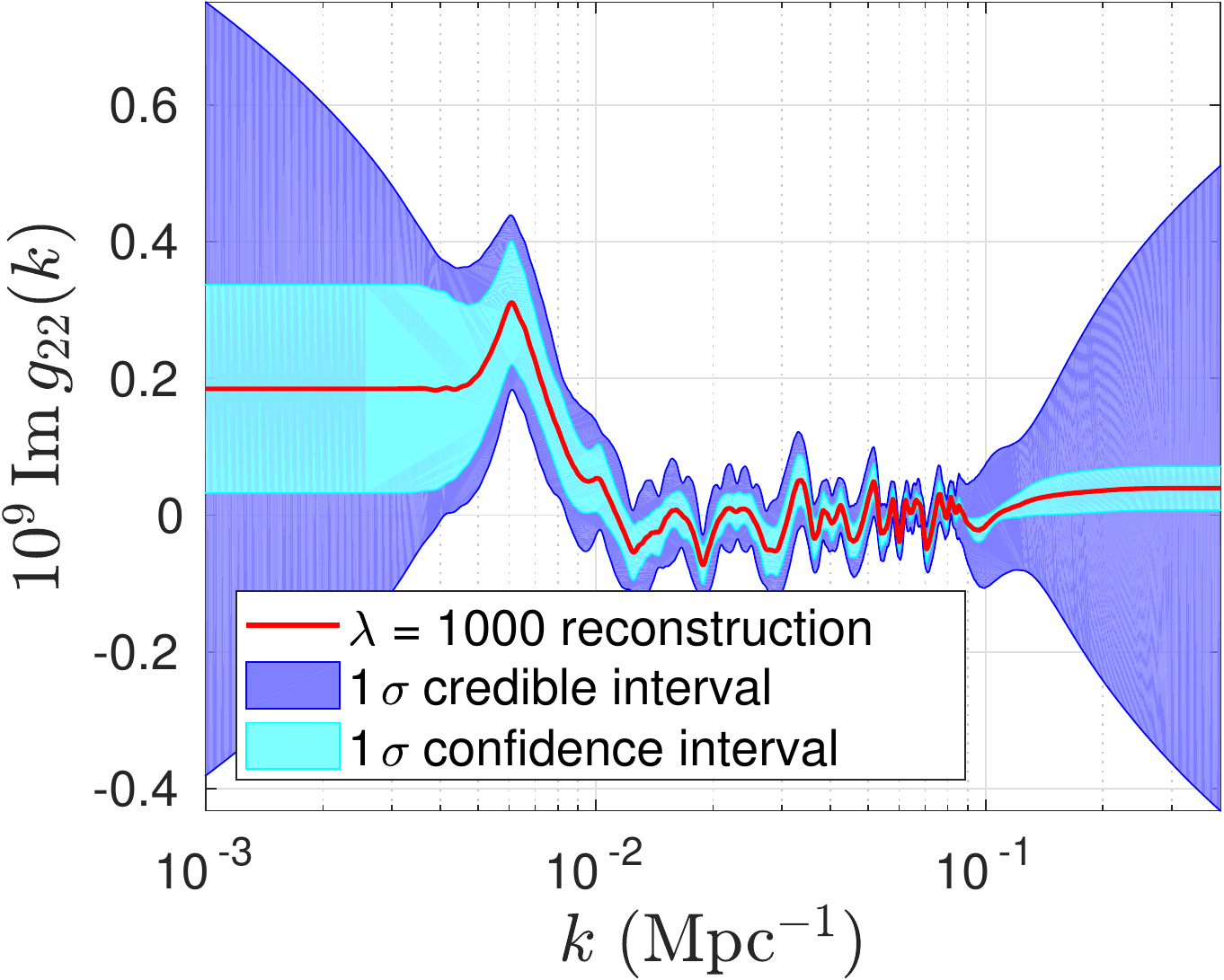}
\label{fig:m2im1000}
}
\caption{Reconstruction for $\lambda = 1000$.}
\label{fig:olambda1000}
\end{figure}

\begin{figure}[ht]
\centering
\subfigure
{\includegraphics[width=0.48\textwidth]{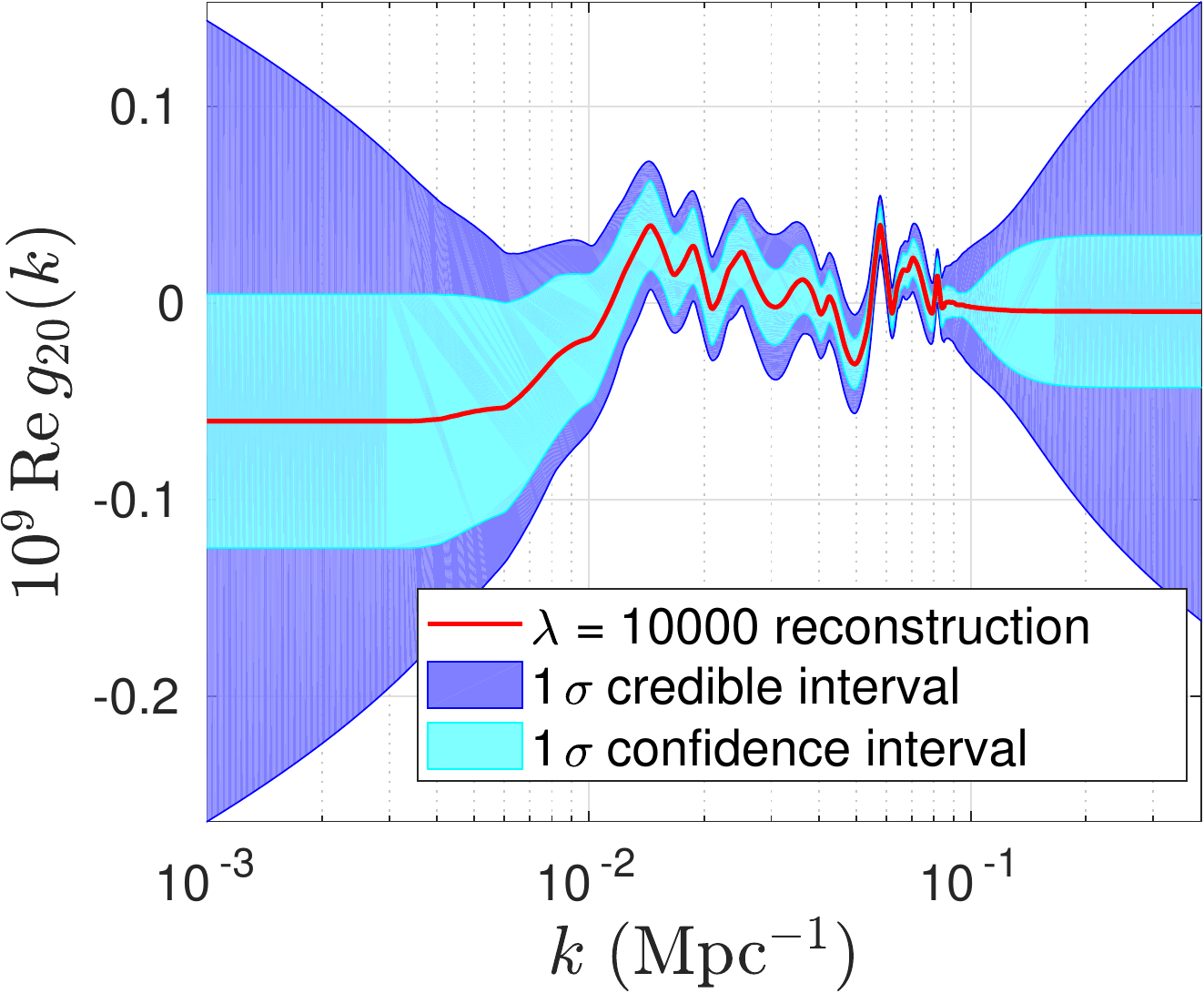}
\label{fig:m010000}
} \\
\subfigure
{\includegraphics[width=0.48\textwidth]{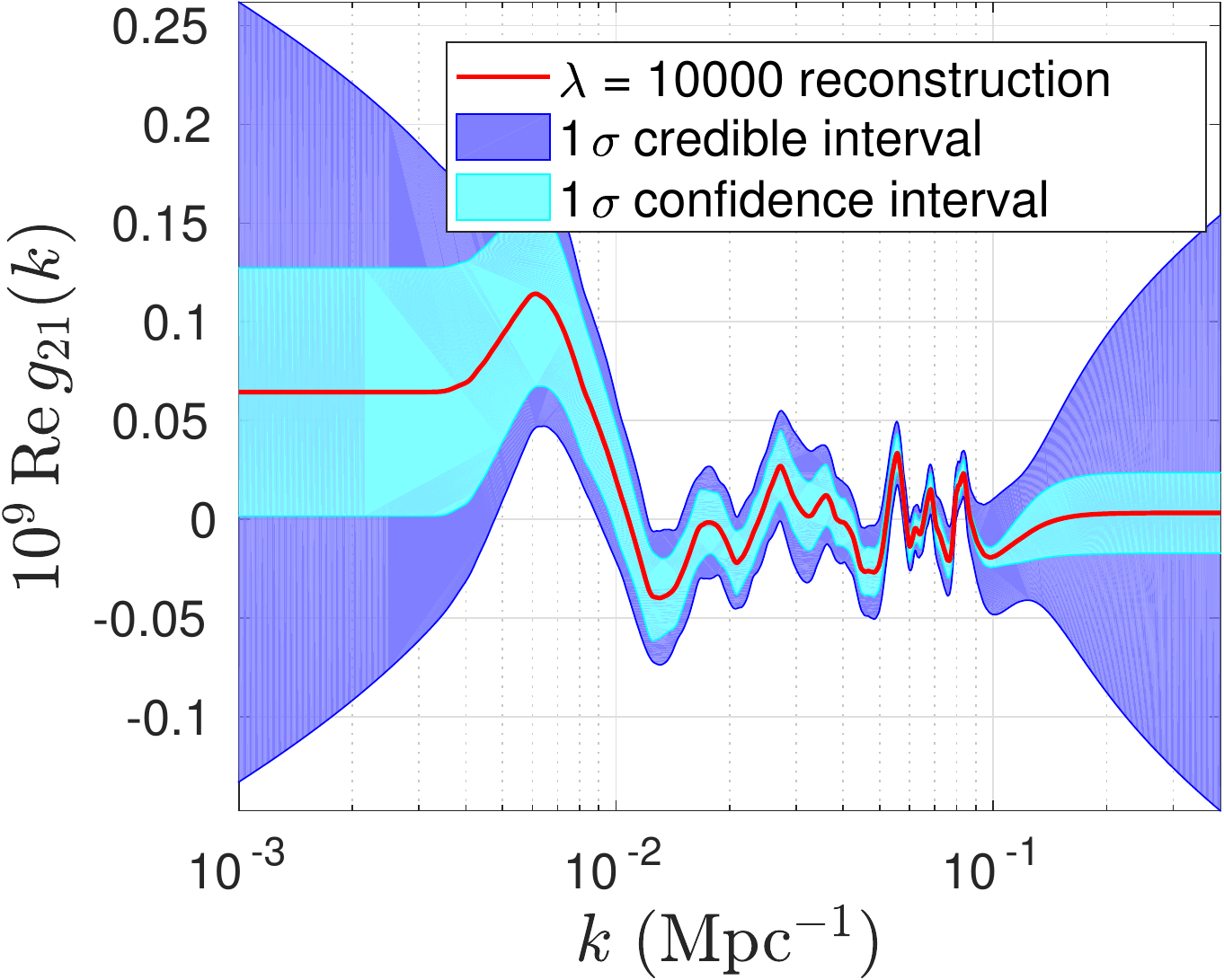}
\label{fig:m110000}
}
\subfigure
{\includegraphics[width=0.48\textwidth]{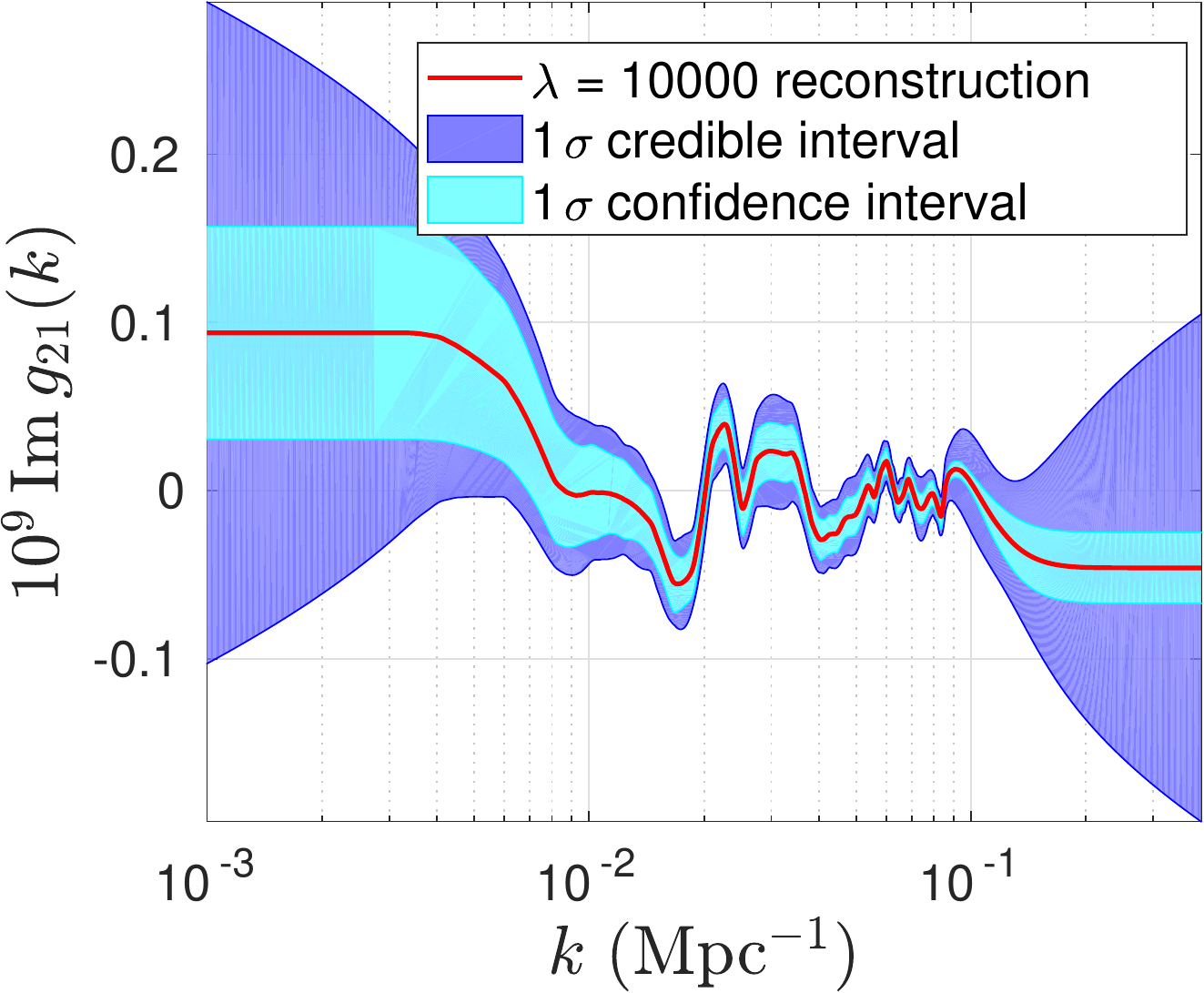}
\label{fig:m1im10000}
}
\subfigure{\includegraphics[width=0.48\textwidth]{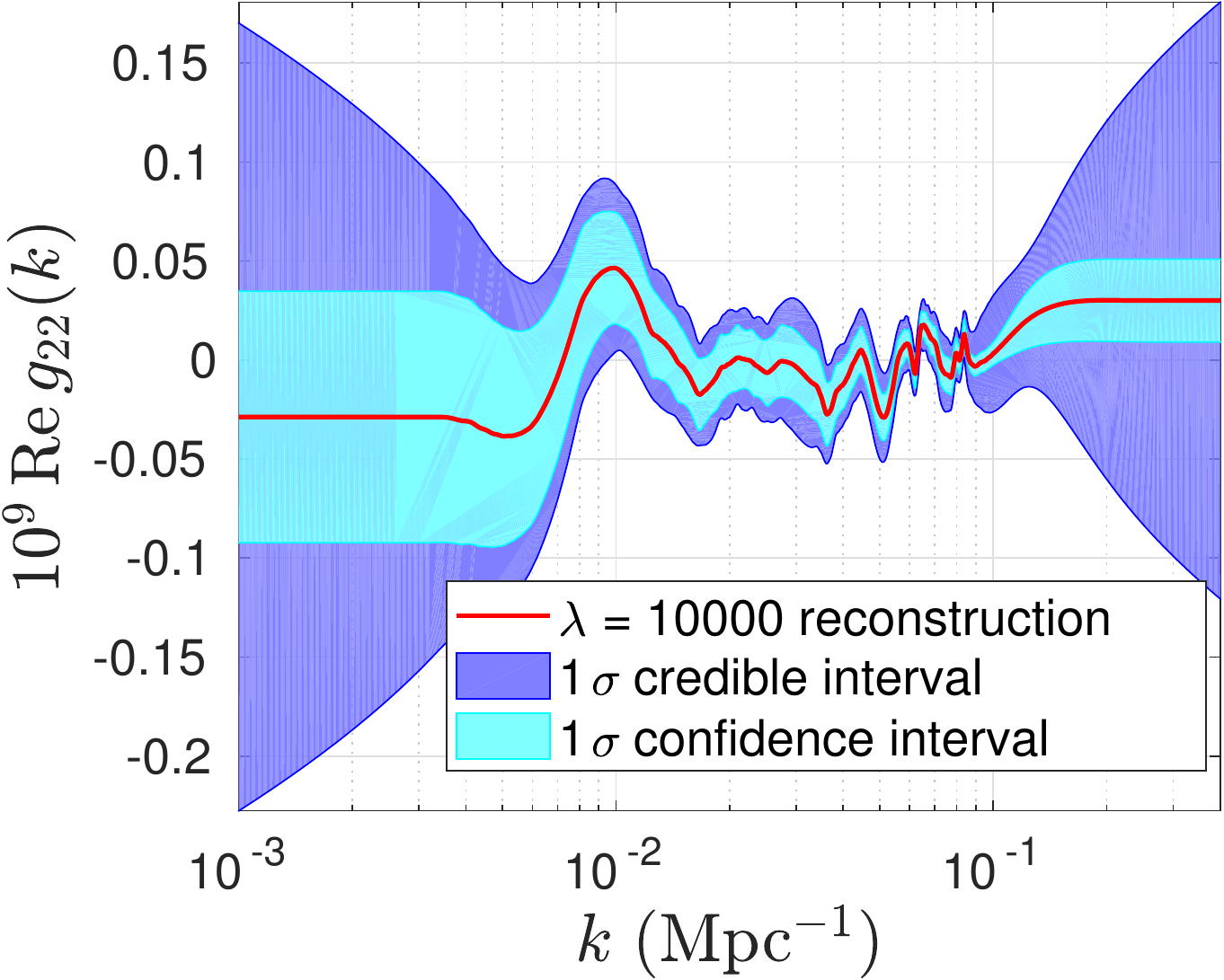}
\label{fig:m210000}
}
\subfigure
{\includegraphics[width=0.48\textwidth]{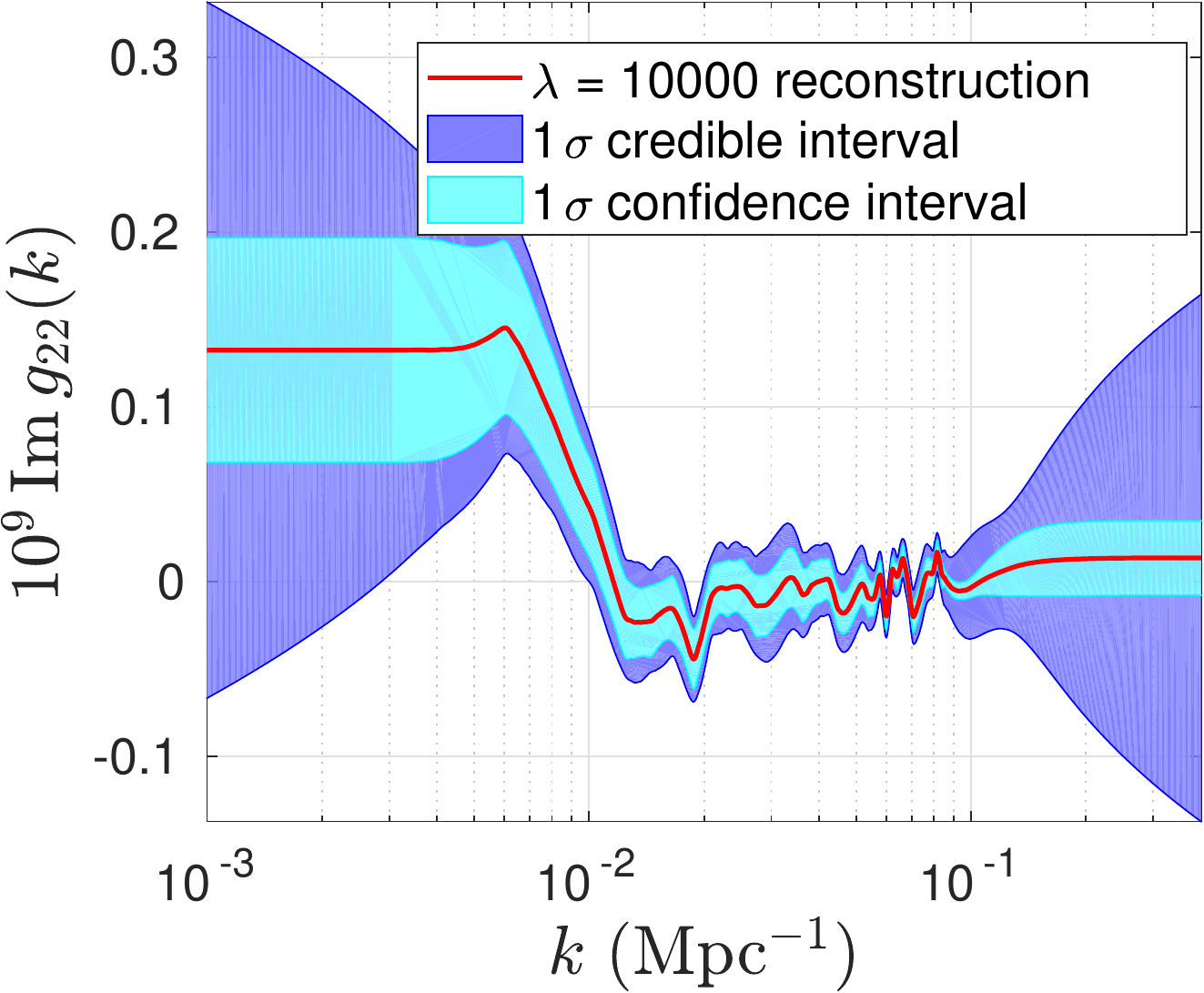}
\label{fig:m2im10000}
}
\caption{Reconstruction for $\lambda = 10000$.}
\label{fig:olambda10000}
\end{figure}

\section{The distribution of \boldmath $T$ } 
\label{sec:tchidist}

It is shown below that the test statistic in eq.\eqref{eq:t1}:
\begin{align}
T(g_{2 M},\lambda) &= \mathbf{g}^T \mathbf{\Sigma}_\mathrm{F}^{+} (\mathbf{g}),
\end{align}
follows a $\chi^2$ distribution with degrees of freedom equal to the rank of the pseudoinverse of the covariance matrix.

Let the $N \times N$ covariance matrix $\mathbf{\Sigma}_\mathrm{F}$ be diagonalised as
\begin{equation}
\mathbf{\Sigma}_{\mathrm{F}} = \mathbf{U} \mathbf{D} \mathbf{U}^{T},
\label{eq:diagd}
\end{equation}
where $\mathbf{D}$ is a diagonal matrix.
The pseudoinverse of the covariance matrix then reads
\begin{equation}
	\mathbf{\Sigma}_\mathrm{F}^{+} = \mathbf{U}  \mathbf{\tilde{D}} \mathbf{U}^{T}
\end{equation}
where $\mathbf{\tilde{D}}$ is the matrix that results from replacing the non-zero diagonal elements by their reciprocal. Further decompose $\mathbf{\tilde{D}}$ into $\mathbf{\tilde{D}} = \mathbf{\tilde{D}}^{1/2} (\mathbf{\tilde{D}}^{1/2})^{T}$ such that
\begin{equation}
	\mathbf{\Sigma}_\mathrm{F}^{+} = \mathbf{U}   \mathbf{\tilde{D}}^{1/2} (\mathbf{\tilde{D}}^{1/2})^{T} \mathbf{U}^{T},
\end{equation}
where $\mathbf{\tilde{D}}^{1/2}$ is a matrix with the diagonal elements equal to the square root of those of $\mathbf{D}$ and with the zero columns removed. There are $N - \mathrm{rank}(\mathbf{\Sigma}^{+}_\mathrm{F}) $ such columns. Transform to new variables
\begin{equation}
	\mathbf{y} = (\mathbf{\tilde{D}}^{1/2})^{T} \mathbf{U}^{T} \mathbf{g},
\end{equation}
which is a variable of dimension $\mathrm{rank}(\mathbf{\Sigma}^{+}_\mathrm{F})$ so the test statistic becomes 
\begin{equation}
	T(y_{2M},\lambda) = \mathbf{y}^{T} \mathbf{y}\textrm{.}
\end{equation}
If each component of $\mathbf{y}$ is distributed according to a univariate Gaussian distribution, then $T$, which consists of a sum of squares of such variables, has a $\chi^2$ distribution with degrees of freedom equal to the number of variables, namely $\mathrm{rank}(\mathbf{\Sigma}_\mathrm{F}^{+})$. This is indeed the case, as the covariance matrix equals the identity matrix
\begin{align}
\langle \mathbf{y} \mathbf{y}^{T} \rangle &= \langle  (\mathbf{\tilde{D}}^{1/2})^T \mathbf{U}^{T} \mathbf{g} \mathbf{g}^{T}  \mathbf{U} \mathbf{\tilde{D}}^{1/2}    \rangle\textrm{,}
\end{align}
and the expectation values may be taken to exclude the constant parts so that
\begin{align}
\langle \mathbf{y} \mathbf{y}^{T} \rangle &=  (\mathbf{\tilde{D}}^{1/2})^T \mathbf{U}^{T} \langle  \mathbf{g} \mathbf{g}^{T} \rangle \mathbf{U} \mathbf{\tilde{D}}^{1/2},
\end{align}
and evaluated to give the frequentist covariance matrix $\mathbf{\Sigma}_\mathrm{F}$:
\begin{align}
 \langle \mathbf{y} (\mathbf{y})^{T} \rangle &=  (\mathbf{\tilde{D}}^{1/2})^{T} \mathbf{U}^{T} \mathbf{\Sigma}_\mathrm{F} \mathbf{U} \mathbf{\tilde{D}}^{1/2}.
\end{align}
Expressing this in the diagonalised form (\ref{eq:diagd}) and reordering we get
\begin{align}
\langle \mathbf{y} \mathbf{y}^{T} \rangle &=  (\mathbf{\tilde{D}}^{1/2})^{T} \mathbf{U}^{T} (\mathbf{U} \mathbf{D} \mathbf{U}^{T}) \mathbf{U} \mathbf{\tilde{D}}^{1/2}
\\ &=  (\mathbf{\tilde{D}}^{1/2})^{T} (\mathbf{U}^{T} \mathbf{U}) \mathbf{D} (\mathbf{U}^{T} \mathbf{U}) \mathbf{\tilde{D}}^{1/2} = (\mathbf{\tilde{D}}^{1/2})^{T} \mathbf{D} \mathbf{\tilde{D}}^{1/2} = \mathbf{I},
\end{align}
since $\mathbf{U}^{T} \mathbf{U} = \mathbf{I}$ ($\mathbf{U}$ is orthogonal as $\mathbf{\Sigma}_\mathrm{F}$, the matrix being diagonalised, is symmetric).

\section{The Anderson-Darling test} 
\label{sec:adtest}

The Anderson-Darling test \cite{adarling} checks if a particular sample is likely to have come from a given distribution. Consider a sample $x_1,x_2,\ldots,x_n$ of a stochastic variable $X$. When this is sorted in ascending order $x_{(1)},x_{(2)},\ldots,x_{(n)}$, the empirical distribution function $F_n(x)$ can be constructed as the fraction of the sample that has values less than or equal to $x$. In asking if a sample is likely to come from a probability distribution $P(x,\boldsymbol\theta)$, where $\boldsymbol\theta$ denotes the set of parameters specifying the probability distributions, the empirical distribution function $F_n(x)$ may be compared with the cumulative distribution function $F(x,\boldsymbol\theta) = \int_{-\infty}^{x} \mathrm{d}y \, P(y,\boldsymbol\theta)$ as a weighted square error in the test statistic
\begin{equation}
	W^{\ast} = n \int_{-\infty}^{\infty} \mathrm{d}F(x) \, (F_n(x)-F(x,\boldsymbol\theta))^2 \psi(x),
\end{equation}
where, for the Anderson-Darling test, $\psi(x)= (F(x,\boldsymbol\theta) (1- F(x,\boldsymbol\theta)) )^{-1}$. For the Gaussian distribution $\boldsymbol \theta = (\mu,\sigma)$ and $F(x,\boldsymbol\theta) = (2 \pi \sigma^2)^{-1/2} \int_{-\infty}^{x} \exp({-(x-\mu)^2/(2 \sigma^2)})$.

In practice, another form is used. Upon calculating $z_{(i)} = F(x_{(i)},\boldsymbol\theta)$, the Anderson-Darling statistic is
\begin{equation}
A_n^2 = - n^{-1} \sum_{x=1}^{n} (2i - 1) (\log z_i + \log(1-z_{n+1-i})) - n\textrm{.} \label{eq:adt}
\end{equation}
Rejection at a given significance level is then based on the value of $A_n^2$ being greater than a tabulated value.
If the parameters of the distribution $\boldsymbol \theta$ are unknown they are replaced by the average and the sample standard deviation of the sample. It is then necessary to modify the test statistic in order to take this additional uncertainty into account so that $A^{\ast} = A_n^2 (1 + b_0/n + b_1/n^2 )$ and then compare with tabulated values \cite{techand}.

Alternatively, the significance level may be assessed by generating a large number of samples taken from a Gaussian distribution with zero mean and unit variance. For each, the \emph{sample} standard deviation is used as standard deviation and the \emph{mean} as average. Then the test statistic (\ref{eq:adt}) is computed. The test statistic obtained for the original is then compared with the distribution of test statistics for the simulated samples to assess the significance.

\section{Diagonal approximation} 
\label{sec:diagap}

When we adopt the common approximation of truncating the covariance matrix $\mathbf{\Sigma}$ \eqref{eq:sigmamat} to the diagonal only, there is no evidence for a constant quadrupole modulation nor one scaling with wave number as a power law. The best-fit constant \eqref{eq:best-fit} and power-law \eqref{eq:genepowlaw} quadrupole modulations and their uncertainties are listed in Table~\ref{tab:dmain} and Table~\ref{tab:dpowlawquad}, respectively. The p-values for these spectra obtained using the test statistic $T_{g_2}$ \eqref{eq:thetg2} are listed in Table~\ref{tab:dg2}. When assuming just one preferred direction \eqref{eq:newform} we find the best-fit amplitudes and their uncertainties to be those listed in Table~\ref{tab:dastpowlaws}. The non-parametric reconstructions for $\lambda=100$ are shown in Fig.~\ref{fig:dmainres}. With the test statistic $T_{g_2}$ they are found to be unusual at a $2.28 \sigma$ significance level and at $2.21\sigma$ for intermediate wave number ranges, a marginal significance. This is illustrated in Fig.~\ref{fig:dg2p}. With the test statistic $T$ \eqref{eq:t1}, they are found to be unusual at $4.63\sigma$ and $2.72\sigma$ for the full and intermediate wave number range, respectively. This is illustrated in Fig.~\ref{fig:dtwodisfigs} and Fig.~\ref{fig:dpvaltests}. When fitting a constant quadrupole modulation to the discrepancy in the wave number range $0.005$-$0.008 \, \mathrm{Mpc}^{-1}$ noted in $\mathrm{Im}\, g_{22}$ of Fig.~\ref{fig:dmainres}, we find that it is perpendicular to the CMB dipole and aligned with the direction of hemispherical asymmetry in the case of a negative amplitude and perpendicular to the direction of hemispherical asymmetry in the case of a positive amplitude. The posterior distributions of the direction for positive and negative amplitude are shown in Fig.~\ref{fig:dfeatdir} and the angular distances to the CMB dipole and direction of hemispherical asymmetry are listed in Table~\ref{tab:dmodes}.

\begin{table}[ht]
\begin{center}
\begin{tabular}{|l|r|r|}
\hline
$M$ & $ 10^2 \times \mathrm{Re} \, {g}_{2M \star}$ & $10^2 \times \mathrm{Im} \, {g}_{2M \star}$ \\
\hline
$0$ & $0.34\pm 0.34$ &  \\
\hline
$1$ & $0.04\pm 0.21$ & $-0.23 \pm 0.21$  \\
\hline
$2$ & $0.15 \pm 0.21$ & $0.09 \pm 0.21$ \\
\hline
\end{tabular}
\caption{The best-fit constant quadrupole modulation $g_{2M \star}$ \eqref{eq:best-fit}, under the diagonal approximation.}
\label{tab:dmain}
\end{center}
\end{table}

\begin{table}[h!]
\begin{center}
\begin{tabular}{|l|r|r| l| r| r|}
\hline
$M$ & $ 10^2 \times \mathrm{Re} \, {g}_{2M \star}$ & $10^2 \times \mathrm{Im} \, {g}_{2M \star}$ & $M$ & $ 10^2 \times \mathrm{Re} \, {g}_{2M \star}$ & $10^2 \times \mathrm{Im} \, {g}_{2M \star}$ \\
\hline
\multicolumn{3}{|c|}{$q=-2$} & \multicolumn{3}{|c|}{$q=2$} \\
\hline
$0$ & $0.04\pm 0.14$ & & $0$ & $0.02\pm 0.12$ &\\
\hline
$1$ & $0.09\pm 0.11$ & $-0.06 \pm 0.10$ & $1$ & $-0.03\pm 0.07$ & $-0.09 \pm 0.07$ \\
\hline
$2$ & $0.03 \pm 0.11$ & $0.21 \pm 0.11$ & $2$ & $0.15 \pm 0.07$ & $-0.05 \pm 0.07$ \\
\hline
\multicolumn{3}{|c|}{$q=-1$} & \multicolumn{3}{|c|}{$q=1$} \\
\hline
$0$ & $0.44\pm 0.40$ & & $0$ & $0.12\pm 0.21$ & \\
\hline
$1$ & $0.14\pm 0.27$ & $-0.32 \pm 0.27$ & $1$ & $-0.01\pm 0.13$ & $-0.15 \pm 0.13$ \\
\hline
$2$ & $0.07 \pm 0.27$ & $0.23 \pm 0.27$ & $2$ & $0.17 \pm 0.13$ & $0.00 \pm 0.13$ \\
\hline
\end{tabular}
\caption{Power-law quadrupole modulations of the form $g_{2M} \propto g_{2M \star} (k/k_{\ast})^{q} \mathcal{P}(k)$, under the diagonal approximation.}
\label{tab:dpowlawquad}
\end{center}
\end{table}

\begin{table}[ht]
\begin{center}
\begin{tabular}{|l|r|r|r|r|r|}
\hline
$q$ & -2 & -1 & 0 & 1 & 2 \\
\hline
p-value & 0.40 & 0.59 & 0.68 & 0.66 & 0.33 \\
\hline
\end{tabular}
\caption{P-value test of $g_2$ \eqref{eq:g2} for power law quadrupole modulations from data compared with $10^4$ noise realisations, under the diagonal approximation.}
\label{tab:dg2}
\end{center}
\end{table}

\begin{table}[ht]
\begin{center}
\begin{tabular}{|r|r|r|r|r|r|}
\hline
$q$ & $-2$ & $-1$ & $0$ & $1$& $2$ \\
\hline
 $ 10^2 {g}_{\ast}$ & $-0.03^{+0.22}_{-0.23}$ & $0.11^{+0.54}_{-0.57}$ & $0.10^{+0.40}_{-0.41}$ & $0.00^{+0.25}_{-0.25}$ & $-0.05^{+0.19}_{-0.18}$ \\
\hline
\end{tabular}
\end{center}
\caption{Best-fit power-law quadrupole modulations of the form \eqref{eq:newform} for different power law indices, under the diagonal approximation.}
\label{tab:dastpowlaws}
\end{table}

\begin{figure}[ht]
\centering
\subfigure{\includegraphics[width=0.48\textwidth]{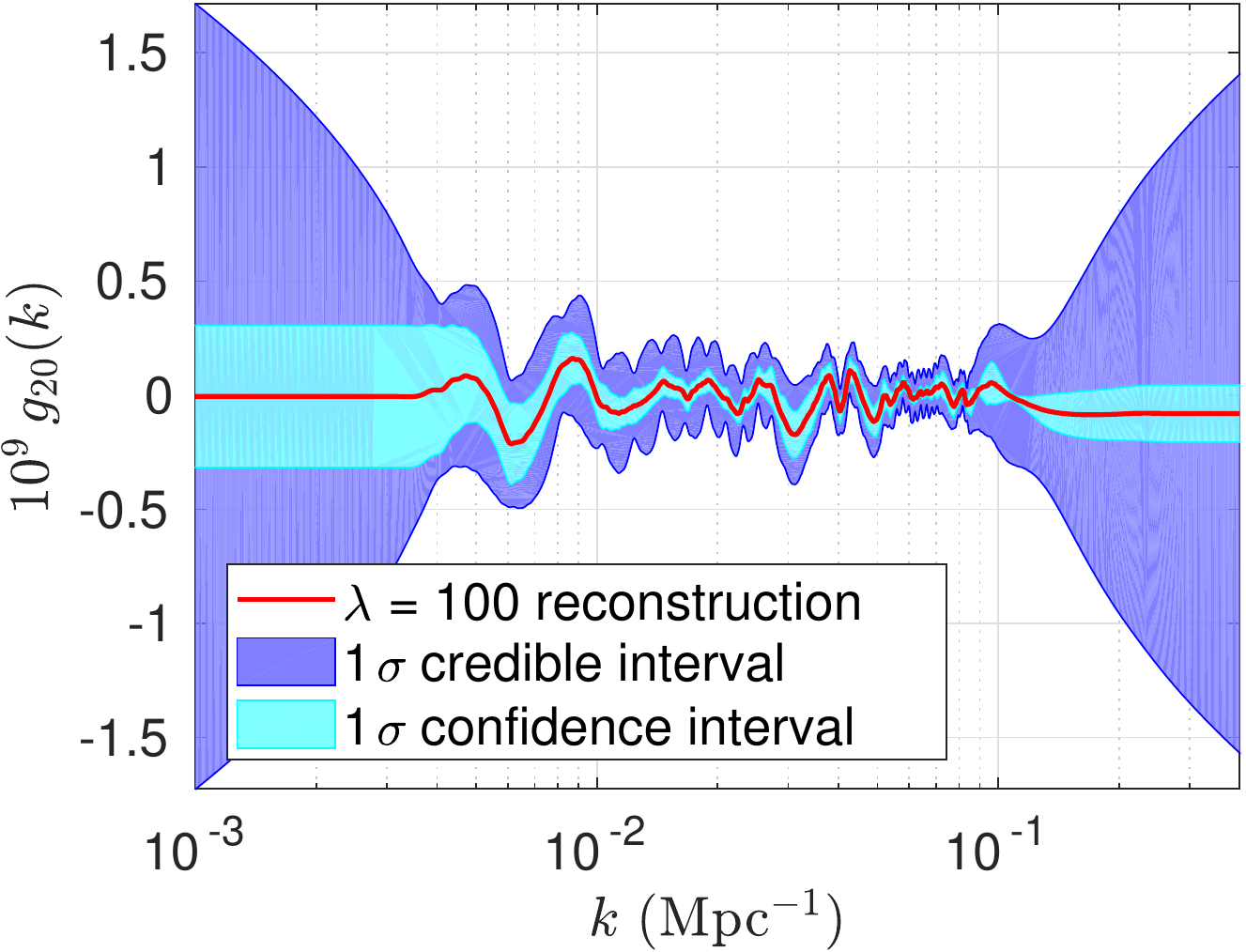}
\label{fig:m0}
} \\
\subfigure{\includegraphics[width=0.48\textwidth]{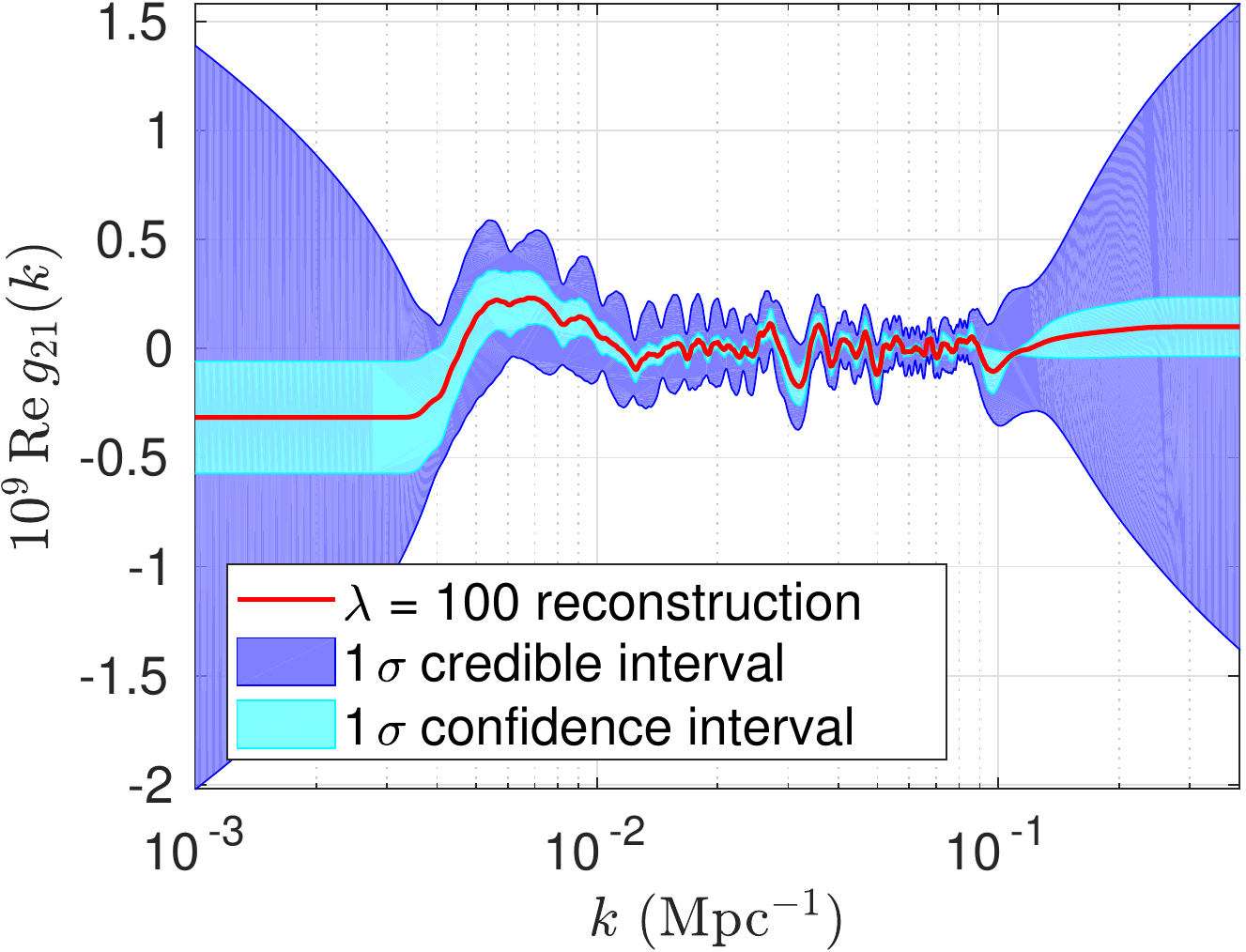}
\label{fig:m1}
}
\subfigure{\includegraphics[width=0.48\textwidth]{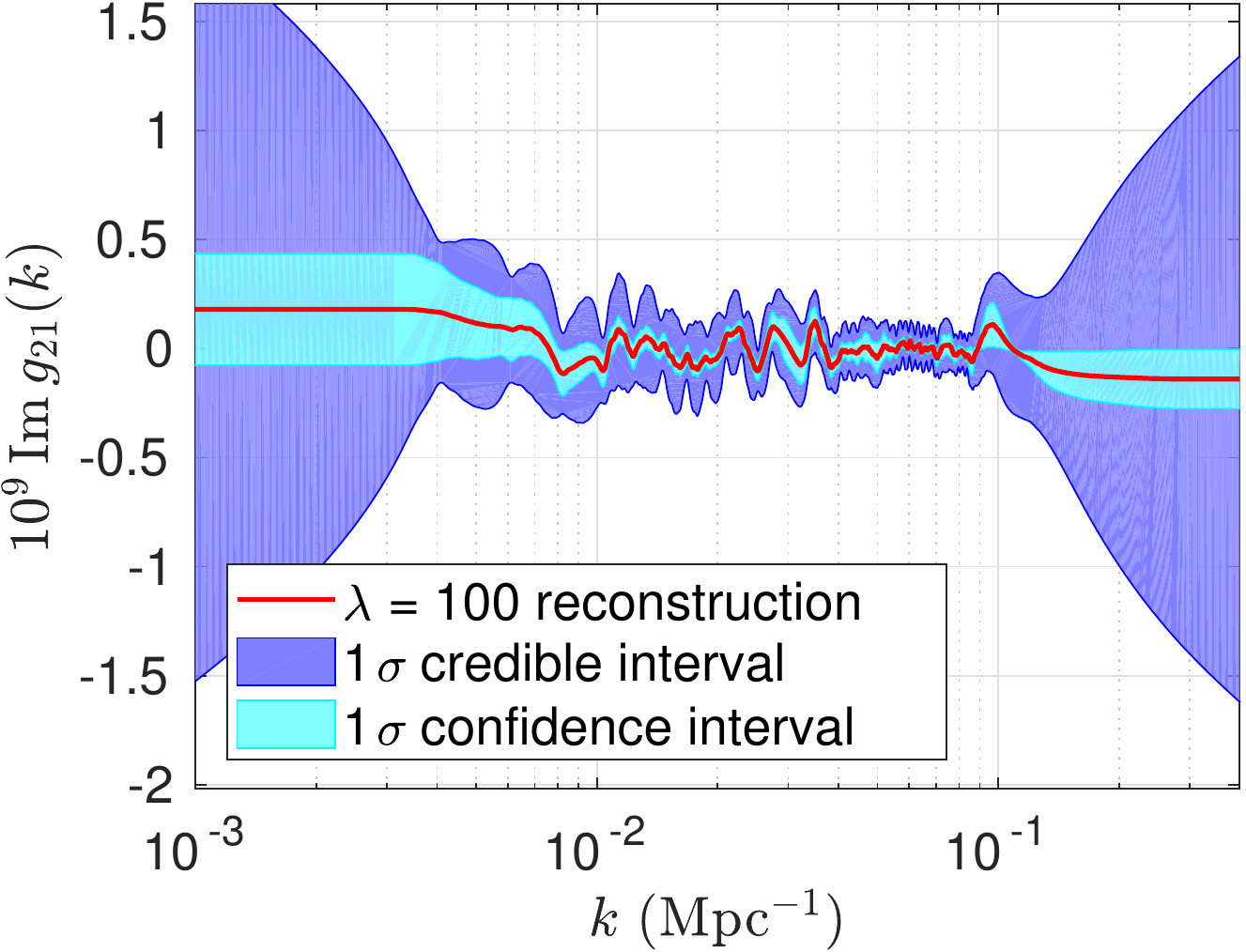}
\label{fig:m1im}
}
\subfigure{\includegraphics[width=0.48\textwidth]{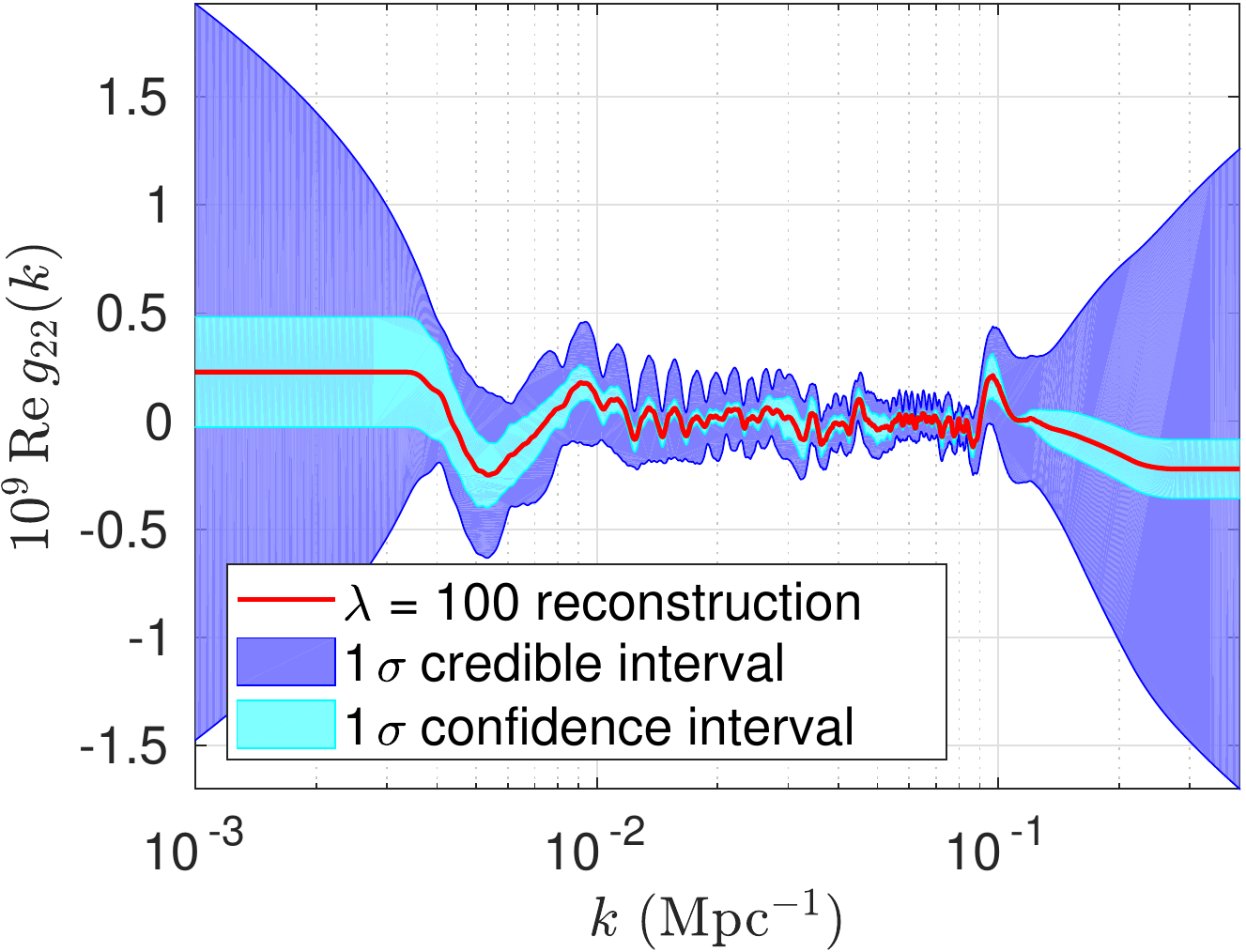}
\label{fig:m2}
}
\subfigure{\includegraphics[width=0.48\textwidth]{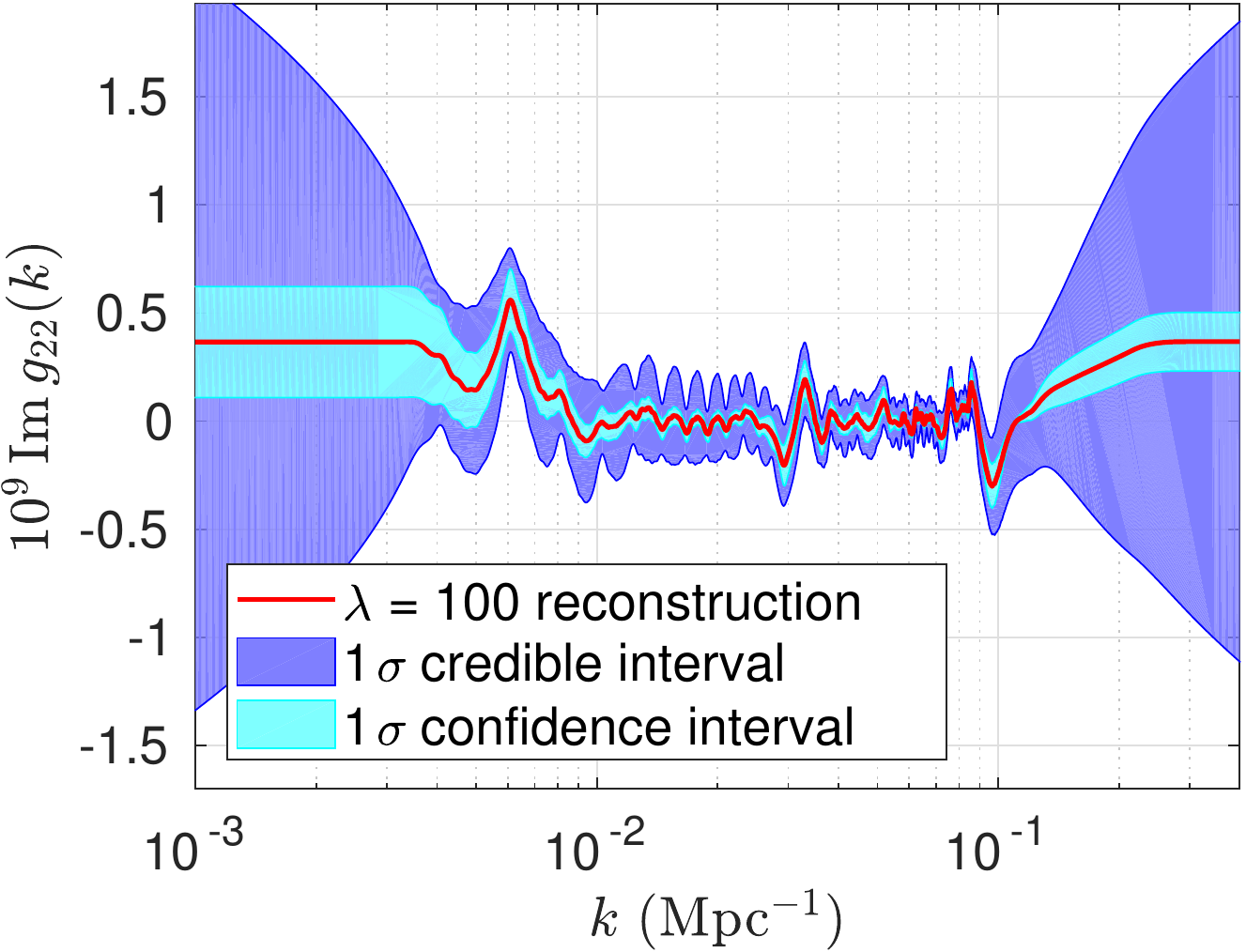}
\label{fig:m2im}
}
\caption{Reconstruction (full red line) of the quadrupole modulation from masked and binned ($\Delta \ell = 30$) PR2--2015 SMICA temperature data in the multipole range $30\leq \ell \leq 1200$, assuming a diagonal covariance matrix $\mathbf{\Sigma}$. Purple bands and cyan bands indicate the $1\sigma$ credible intervals and $1\sigma$ confidence intervals, respectively. The regularisation parameter was set to $\lambda = 100$. Note the feature at $k \sim 6 \times 10^{-3} \, \mathrm{Mpc}^{-1}$ in $\mathrm{Im}\, g_{22}$.}
\label{fig:dmainres}
\end{figure}

\begin{figure}[ht]
\centering
\includegraphics[width=0.48\textwidth¤]{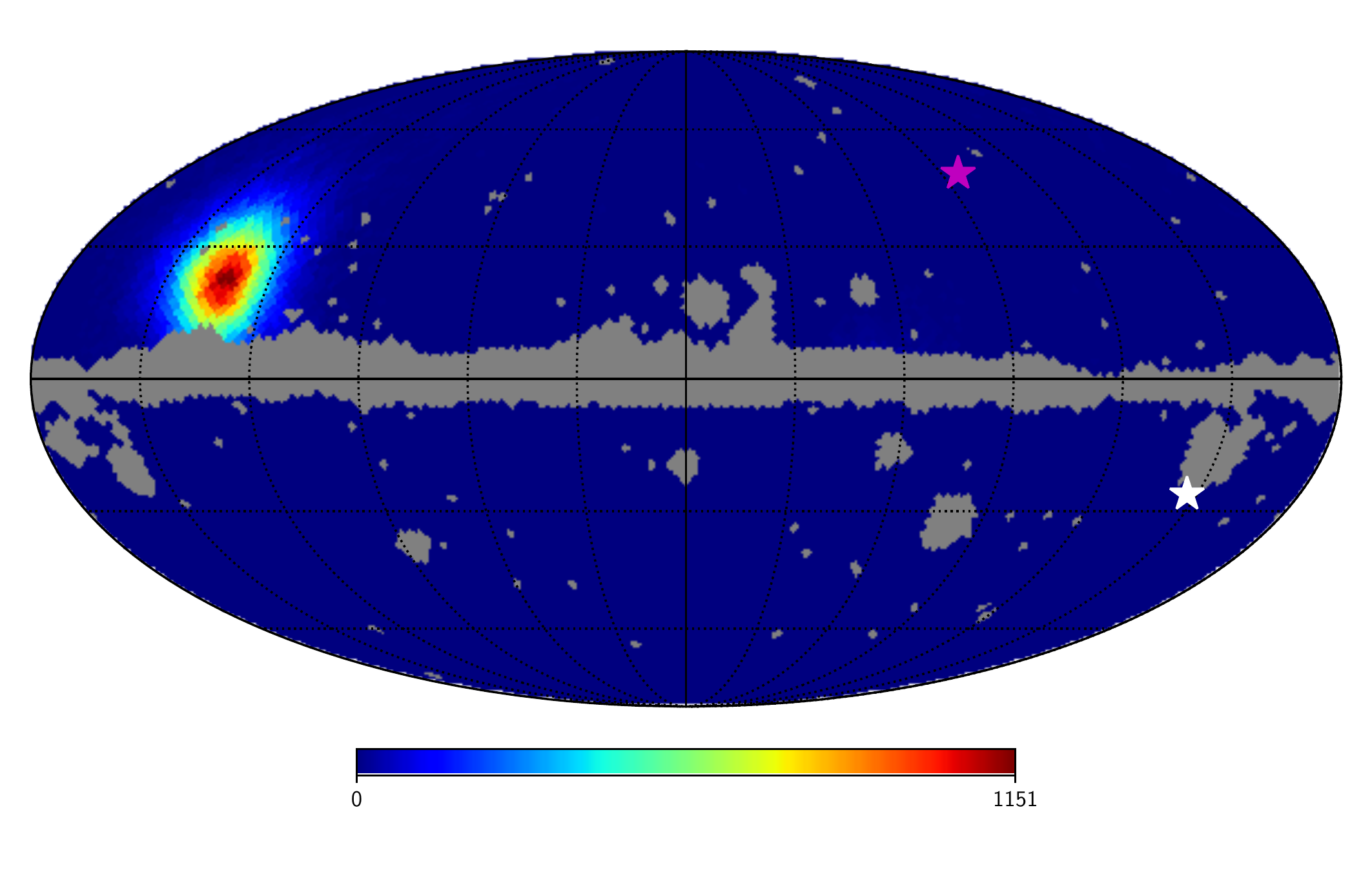}
\includegraphics[width=0.48\textwidth]{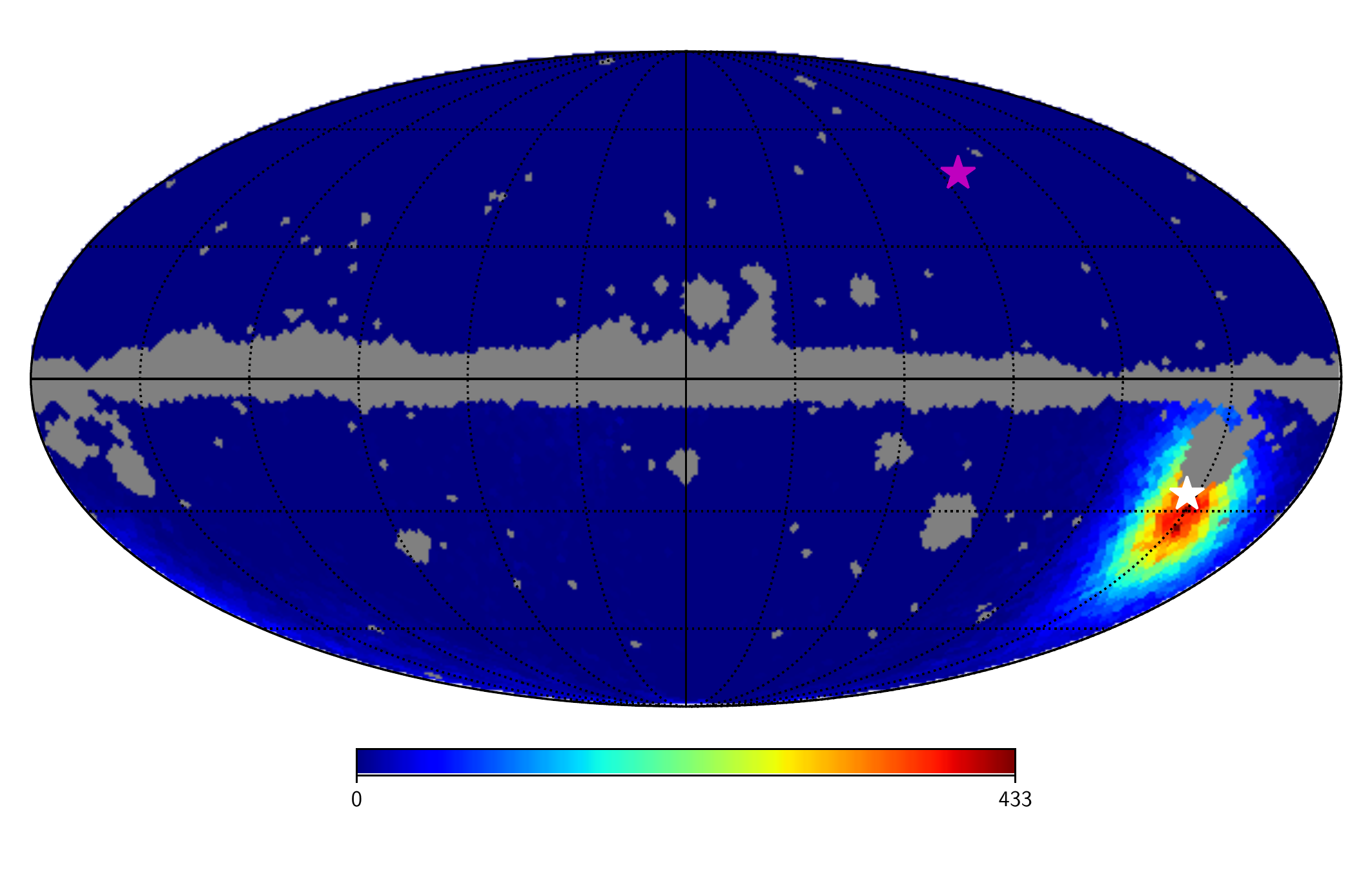}
\caption{The posterior distribution of the direction $\hat{\mathbf{n}}$ for the case of a hot ($10^{9} g_{A} = 0.51\pm 0.16$) constant quadrupole modulation (left panel) and a cold ($10^{9} g_{A} = -0.42 \pm 0.18$) modulation (right panel) in the wave number range $0.005 \leq k/\mathrm{Mpc}^{-1}\leq 0.008$, assuming a diagonal covariance matrix $\mathbf{\Sigma}$. The SMICA mask (in grey) is superimposed for comparison. The magenta and white stars indicate the directions of the CMB dipole and the hemispherical asymmetry, respectively.}
\label{fig:dfeatdir}
\end{figure}

\begin{table}[ht]
\begin{center}
\begin{tabular}{|l|c|c|c|}
\hline
\multicolumn{2}{|l|}{Range of $k=0.005$-$0.008 \, \mathrm{Mpc}^{-1}$ } & \multicolumn{2}{|c|}{Angular distances to} \\
\hline
Amp. $10^{9} g_A$ & Direction $(l,b)$ & CMB dipole $(264^{\circ},48^{\circ})$ & Hemisph. asym. $(213^{\circ},-26^{\circ})$ \\
\hline
$0.51\pm 0.16$  & $({136^{\circ}}^{+13}_{-12},{22^{\circ}}^{+10}_{-11})$ & $ 96^{\circ}$ & $89^{\circ}$ \\
\hline
$-0.42 \pm 0.18$ & $({212^{\circ}}^{+13}_{-12},{-30^{\circ}}^{+14}_{-15})$ & $91^{\circ}$ & $4^{\circ}$ \\
 \hline
\end{tabular}
\caption{Details of the two modes of the posterior distribution of the quadrupole modulation direction shown in Fig.~\ref{fig:featdir}, under the diagonal approximation. The first mode is for a hot modulation. Angular distances to the CMB dipole and the hemispherical asymmetry directions are also indicated.}
\label{tab:dmodes}
\end{center}
\end{table}

\begin{figure}[h]
 \centering
 \subfigure{\includegraphics[width=0.48\textwidth]{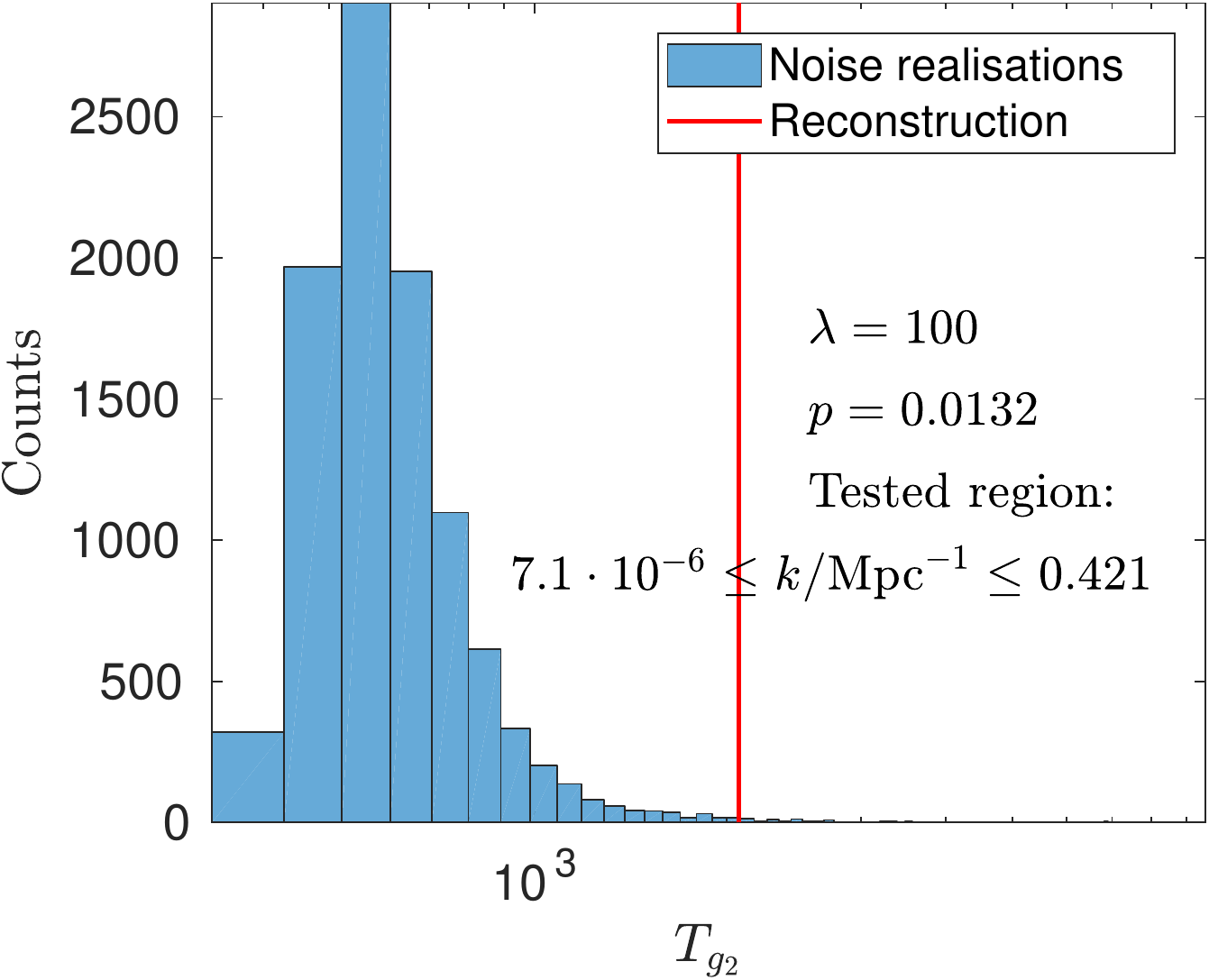}}
 \subfigure{\includegraphics[width=0.48\textwidth]{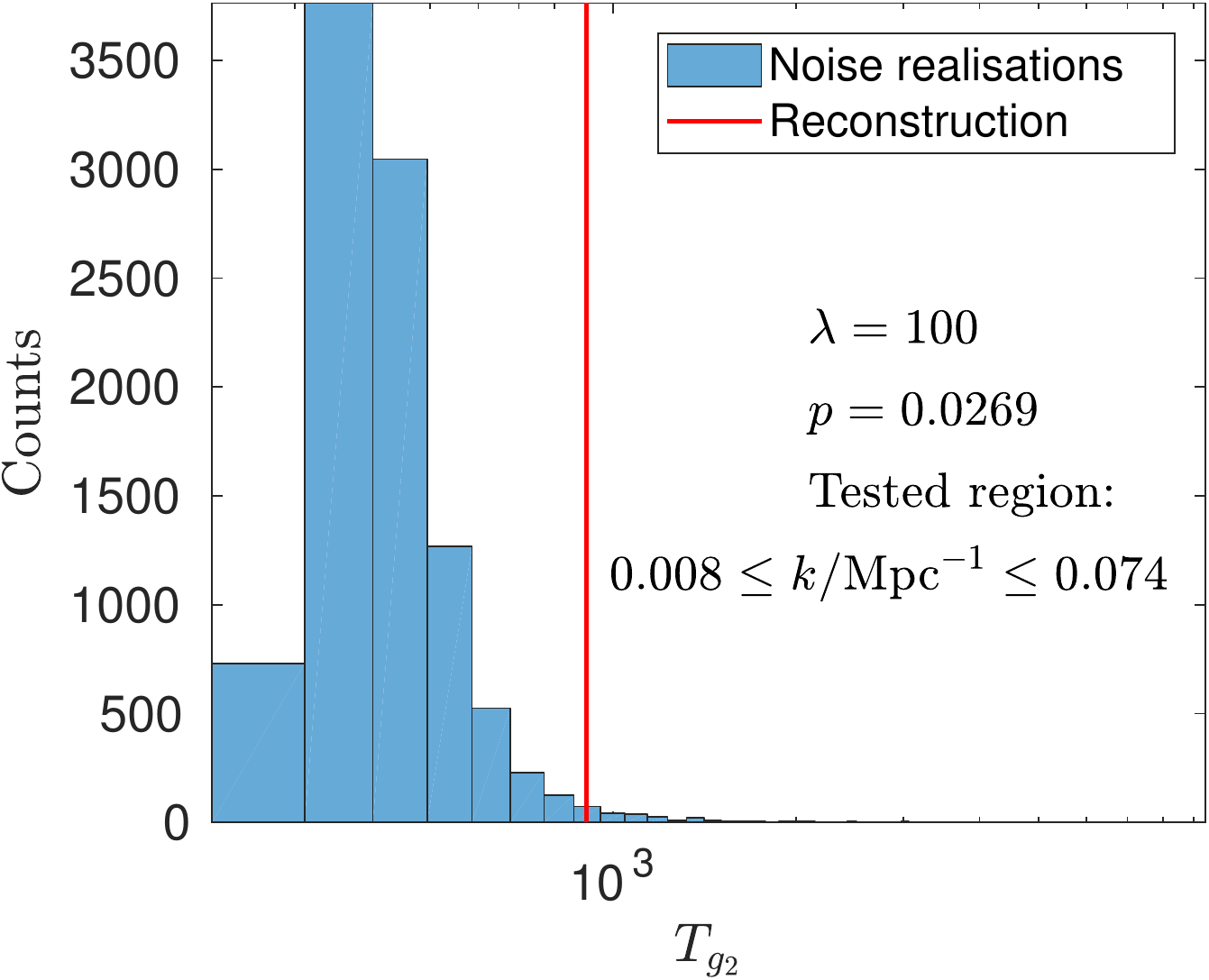}}
 \caption{P-value test using $T_{g_2}$ under the diagonal approximation. The distribution of $T_{g_2}$ for noise realisations and its value for the reconstruction from data (red line) are shown. The right panel shows the same test, but for a more restricted range of wave numbers.}
 \label{fig:dg2p}
 \end{figure}

\begin{figure}[ht]
\centering
 \subfigure{\includegraphics[width=0.48\textwidth]{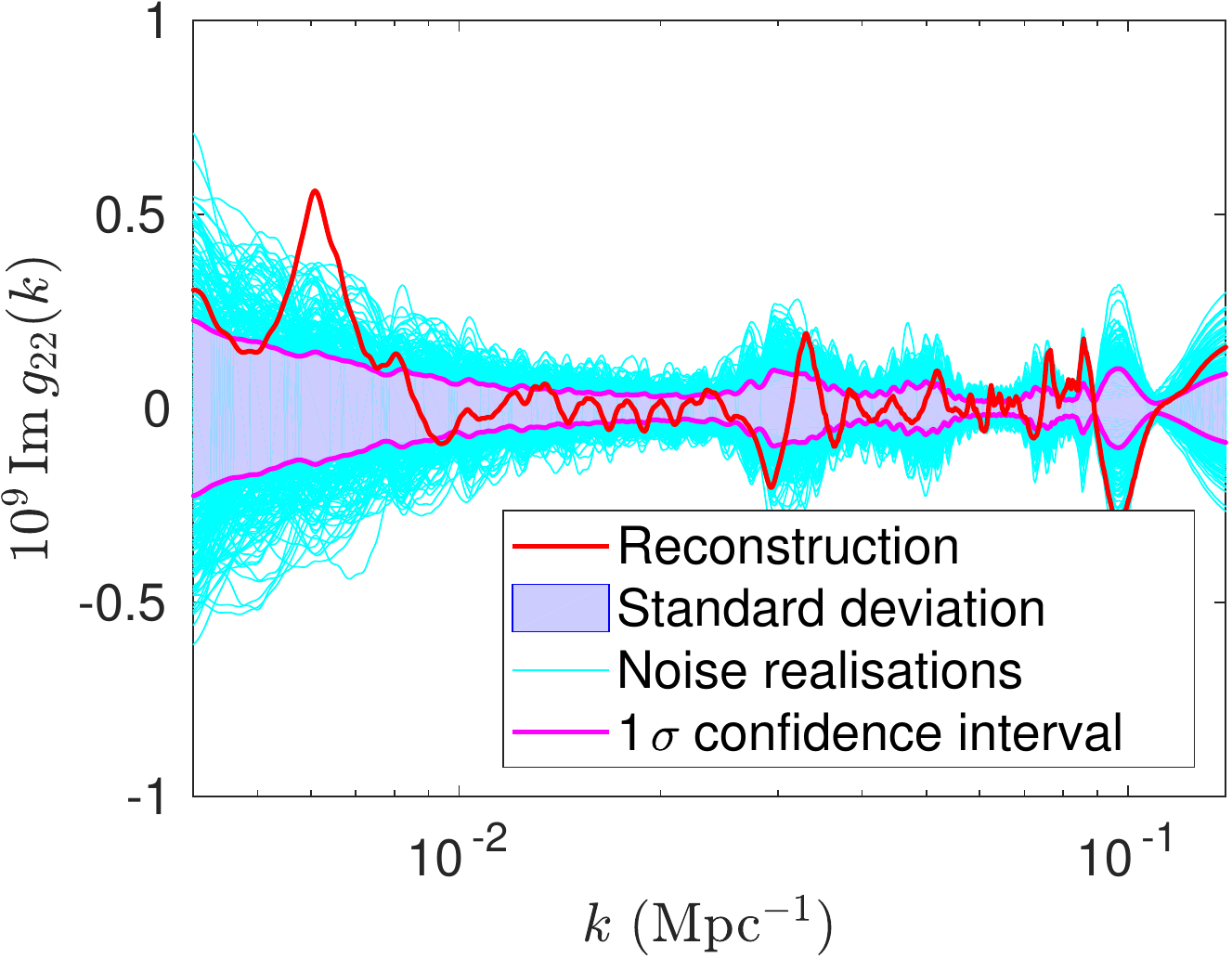}
 }
\subfigure{\includegraphics[width=0.48\textwidth]{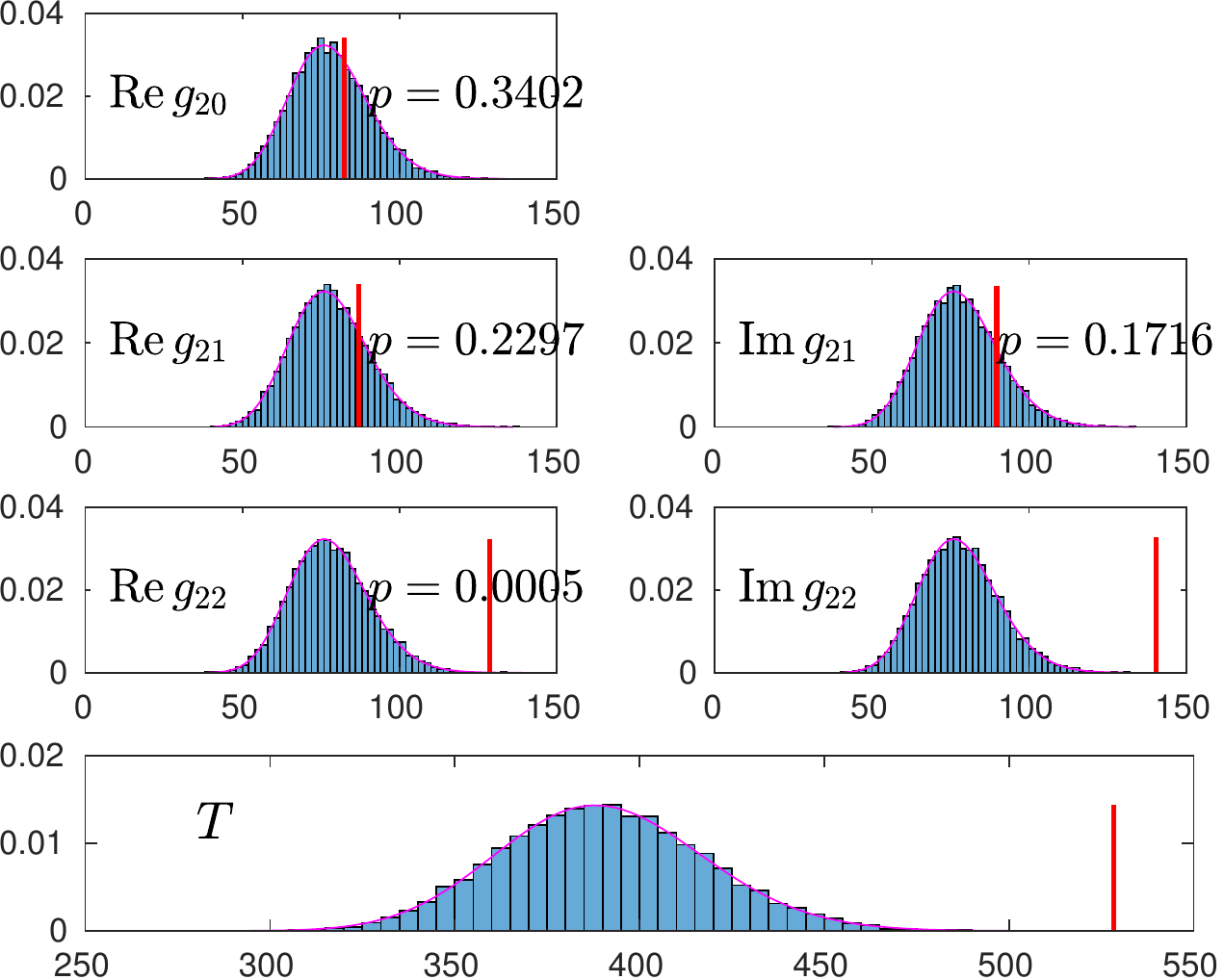}}
\caption{The left panel shows the noise realisations (thin cyan lines), their standard deviation (purple band) as well as the reconstruction from data (red line), under the diagonal approximation. The $1\sigma$ confidence interval (magenta lines) derived from the frequentist covariance matrix exactly matches the standard deviation. The right panel shows the results of the p-value test of $T(g_{2M})$ and the final test statistic $T$ --- with a red line indicating their values for the reconstruction from data. The purple lines are the theoretically expected distributions of the test statistic.}
\label{fig:dtwodisfigs}
\end{figure}
\begin{figure}[ht]
\centering
\subfigure{\includegraphics[width=0.48\textwidth]{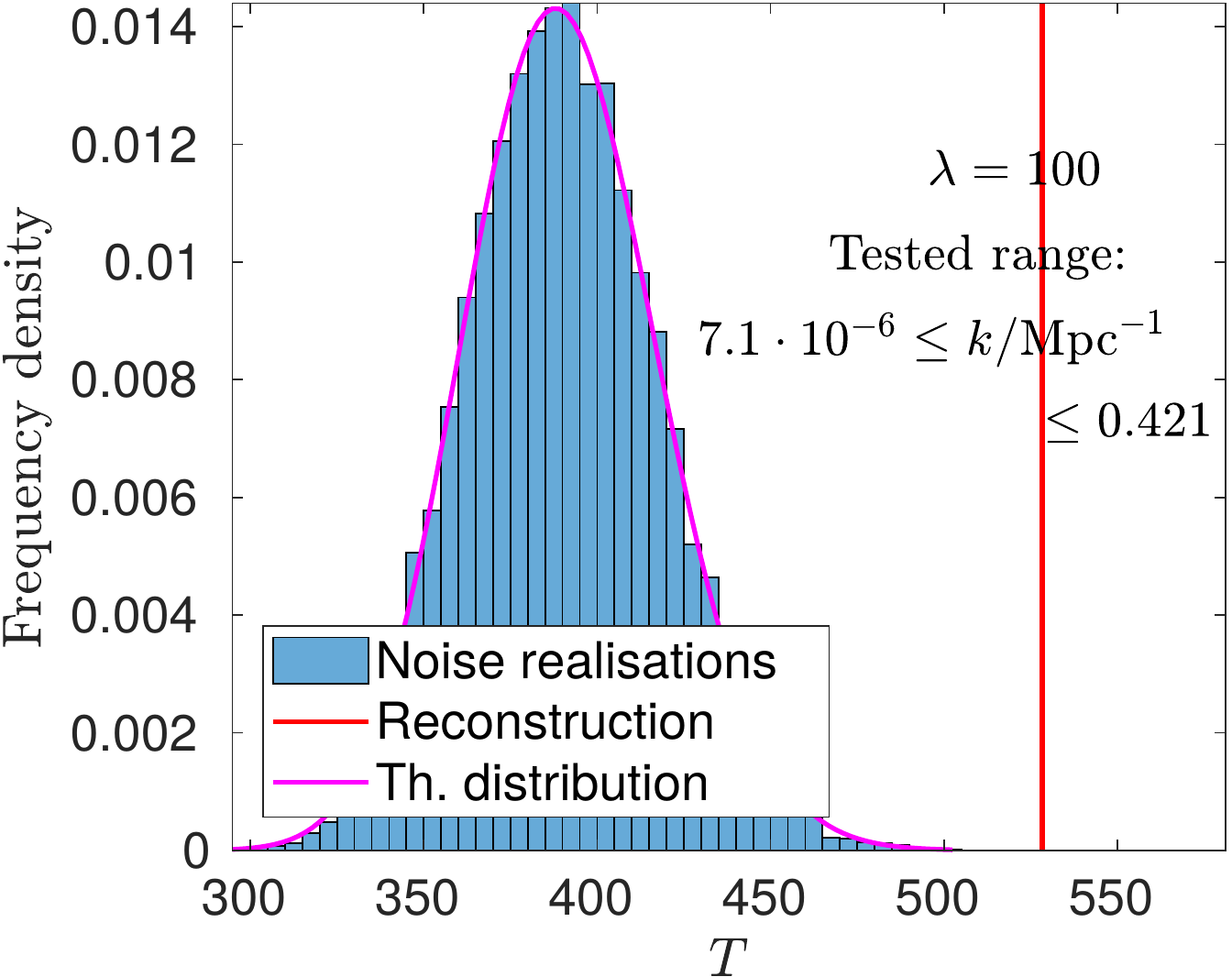}}
\subfigure{\includegraphics[width=0.48\textwidth]{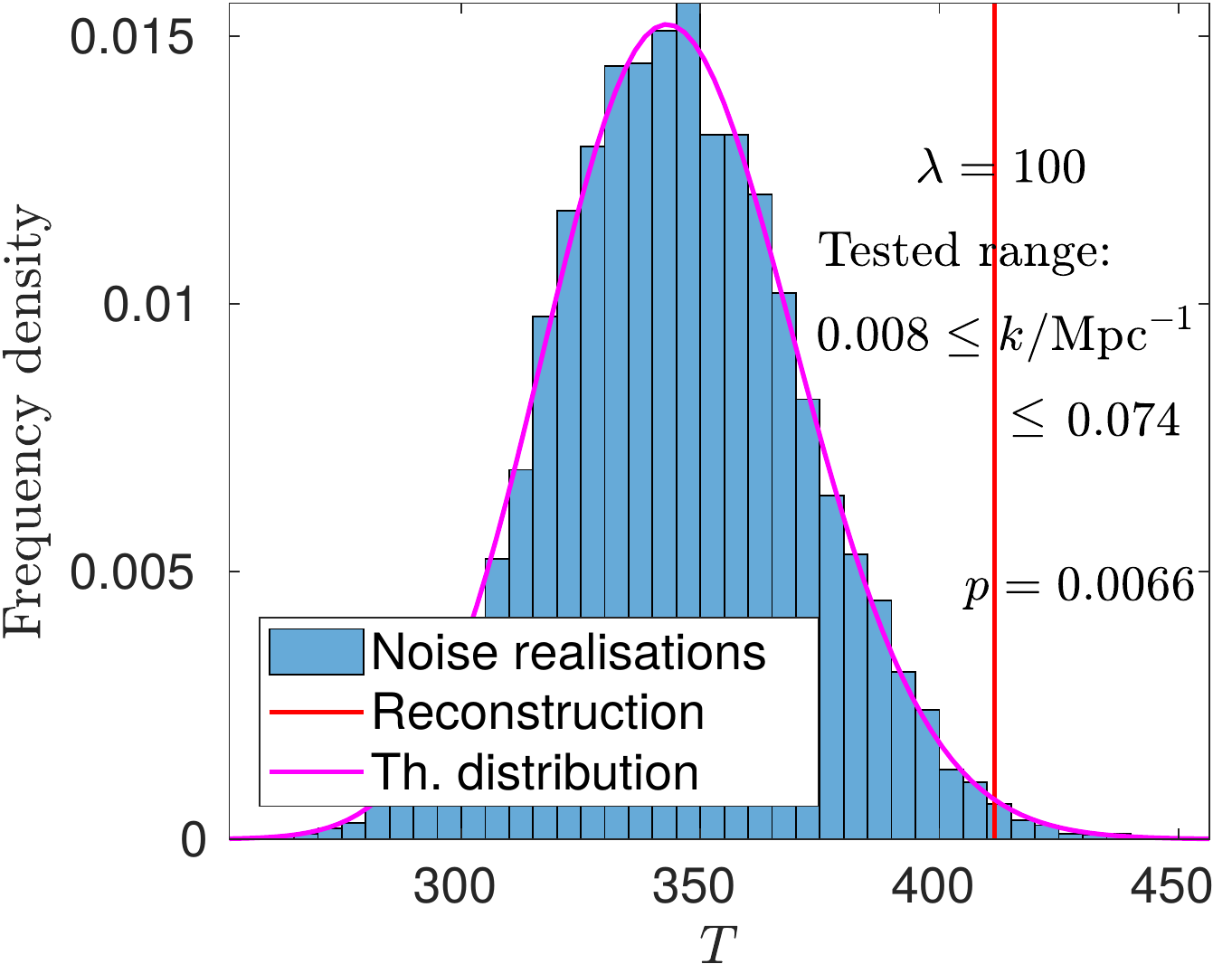}}
\caption{P-value tests of $T$ for the full range of wave numbers (left panel) and a more limited range (right panel), under the diagonal approximation. The value of $T$ for the reconstruction from data is indicated with a red line.}
\label{fig:dpvaltests}
\end{figure}



\end{document}